\RequirePackage{snapshot}
\pdfoutput=1
\documentclass[10pt,a4paper,onecolumn]{amsart}

\usepackage{etex}
 
\usepackage[utf8]{inputenc}
\usepackage[left=2.8cm,right=2.8cm,top=2.8cm,bottom=2.8cm]{geometry}

\usepackage{mathtools}
\usepackage{amssymb}
\usepackage{amsfonts}
\usepackage{amsthm}
\usepackage{tikz}
\usepackage{subcaption}
\expandafter\def\csname ver@subfig.sty\endcsname{}

\usepackage{subfig}

\usepackage{caption}
\usepackage{tensorNotation}
\usepackage{booktabs}

\usepackage[
colorlinks=true,
linkcolor=black, % einfache interne Verkn�pfungen
anchorcolor=black,% Ankertext
citecolor=blue, % Verweise auf Literaturverzeichniseintr�ge im Text
filecolor=magenta, % Verkn�pfungen, die lokale Dateien �ffnen
menucolor=red, % Acrobat-Men�punkte
urlcolor=cyan,
plainpages=false,% zur korrekten Erstellung der Bookmarks
pdfpagelabels,% zur korrekten Erstellung der Bookmarks
hypertexnames=false,% zur korrekten Erstellung der Bookmarks
linktocpage % Seitenzahlen anstatt Text im Inhaltsverzeichnis verlinken
]{hyperref}

\title[A numerical stabilization framework for viscoelastic fluid flow]{A numerical stabilization framework for viscoelastic fluid flow using the finite volume method on general unstructured meshes}

\author[M.\ Niethammer]{Matthias Niethammer}
\address{Mathematical Modeling and Analysis Group, Technische \mbox{Universit{\"a}t} Darmstadt, Germany}
\email{niethammer@mma.tu-darmstadt.de}
\author[H.\ Marschall]{Holger Marschall}
\address{Mathematical Modeling and Analysis Group, Technische \mbox{Universit{\"a}t} Darmstadt, Germany}
\email{marschall@mma.tu-darmstadt.de}

\author[C.\ Kunkelmann]{Christian Kunkelmann}
\address{GCP/RC - BASF SE, Ludwigshafen, Germany}

\author[D.\ Bothe]{Dieter Bothe}
\address{Mathematical Modeling and Analysis Group, Technische \mbox{Universit{\"a}t} Darmstadt, Germany}
\email{bothe@mma.tu-darmstadt.de}

\subjclass{flu-dyn}
\date{February 8, 2017.}


\begin{document}

\begin{abstract}
%% Text of abstract
A robust finite volume method for viscoelastic flow analysis on general unstructured meshes is developed. It is built upon a general-purpose stabilization framework for high Weissenberg number flows. The numerical framework provides full combinatorial flexibility between different kinds of rheological models on the one hand, and effective stabilization methods on the other hand. A special emphasis is put on the velocity-stress-coupling on co-located computational grids. Using special face interpolation techniques, a semi-implicit stress interpolation correction is proposed to correct the cell-face interpolation of the stress in the divergence operator of the momentum balance. Investigating the entry-flow problem of the 4:1 contraction benchmark, we demonstrate that the numerical methods are robust over a wide range of Weissenberg numbers and significantly alleviate the high Weissenberg number problem. The accuracy of the results is evaluated in a detailed mesh convergence study.
\end{abstract}

\maketitle

%% main text
\section{Introduction}
\label{Introduction}
%%%%%%%%%%%%%%%%%%%%%%%%%%%%%%%%%%%%%%%%%%%%%%%%%%%%%%%%%%%%%
The numerical simulation of viscoelastic flows in complex geometries is a challenging problem due to the presence of geometric singularities and sharp stress boundary layers which occur in the high Weissenberg number limit \cite{Renardy2000}. Indeed, the high Weissenberg number problem (HWNP) \cite{Joseph1985, Keunings1986} has been a major stumbling block in computational rheology for more than four decades (see, e.g.\ the reviews \cite{Keunings2000, Keunings2001, Walters2003} or the book \cite{Owens2002}). It refers to the incapability of numerical methods to resolve these geometric singularities and boundary layers, causing all computations to break down at relatively low critical Weissenberg numbers. The observed critical Weissenberg number varies not only with the numerical method and the geometry, but also with the rheological model. One example is the entry flow problem \cite{Boger1972, Boger1987} in a contraction geometry, where the stresses and deformation rates are infinite at the re-entrant corner singularity. However, the fact that in such a problem the high Weissenberg number limit is particularly pronounced for certain rheological models (e.g.,\ UCM/Oldroyd-B), should not be taken as an argument for relating the origin of the HWNP to rheological modelling.

Ever since the stability analysis by Fattal and Kupferman \cite{Fattal2004}, it is evident that the HWNP is a solely numerical issue of stiffness due to strong spatial gradients. Referring to this work, the HWNP is related to the inappropriateness of numerical advection schemes to transport the steep spatial stress profiles, that are locally of exponential type (e.g.,\ at geometric singularities or stagnation points). Since the numerical advection schemes, which are used in finite volume methods (FVM), finite element methods (FEM), etc. are based on polynomial interpolation, the approximation of the transported stress profiles is decidedly poor, which gives rise to local multiplicative accumulation of the stress in space over time, resulting in a numerical blowup \cite{Fattal2005}.

Because of the reasons given above, a numerical stabilization is required for the development of robust numerical methods and algorithms. There have been considerable advances in the development of effective stabilization methods in the most recent decade, which will be discussed in detail in section \ref{Stabilization}. The present work develops a novel unified mathematical and numerical framework which generalizes existing specific stabilization methods on the one hand and covers a large number of different rheological models on the other hand. We build a general-purpose library for the numerical simulation of viscoelastic fluid flow on top of a widely used open source CFD (computational fluid dynamics) package, which has been designed for the flow analysis in complex geometries and is fully parallelized. For this purpose, the OpenFOAM \cite{weller1998, jasak1996, Darwish2015} package appears to be suitable, since it provides a FVM on general unstructured meshes and, being written in C++, it offers the capability of a highly flexible and modular code design for our developments by using object-oriented and generic programming techniques (C++ template programming).

We shall remark that a viscoelastic fluid flow solver has been included previously into the OpenFOAM package \cite{Favero2010} that includes various rheological models. However, in our computational benchmark tests the solver is found to be unstable at moderate and higher Weissenberg numbers, since it does not use an effective stabilization approach to alleviate the HWNP. Moreover, we observe a decoupling between the velocity and stress, giving rise to unphysical checkerboard flow fields. This is related to an issue on the formulation of the stress at the faces of the control-volumes, within the context of a cell-centered FVM. The present contribution eliminates such deficiencies and represents a substantial improvement regarding robustness, accuracy, the consistent implementation in the FV framework and the high degree of generality regarding the use of rheological models.

The objectives of this work are:
\begin{enumerate}
\item Based on the derivation of a unified mathematical stabilization framework which is both independent from specific stabilization approaches and from certain rheological models, we deduce a novel generic numerical FV framework for viscoelastic flow analysis on general unstructured meshes. The generic numerical framework provides full combinatorial flexibility between different kinds of rheological models on the one hand, and effective stabilization methods on the other hand. We demonstrate that many widely used rheological models can be incorporated in a highly flexible and efficient way by means of a set of scalar coefficients which depend on the invariants of the conformation tensor. Moreover, we demonstrate that constitutive equations including Gordon-Schowalter convected derivatives are covered.
\item We develop a consistent velocity-stress-coupling approach for the FVM on co-located computational grids, using special face interpolation techniques. We propose a new correction for the cell-face interpolation of the stress in the momentum balance to remove the decoupling between the velocity and the stress fields.
\item We study the convergence, robustness and accuracy of the numerical methods by means of a detailed quantitative mesh convergence study in the computational benchmark of an entry flow in a 4:1 contraction. For the first time, we compare the predictions from a set of different stabilization approaches over a wide range of Weissenberg numbers. In particular, above the critical Weissenberg number we obtain novel benchmark results: We show that the convergence properties depend strongly on the particular stabilization approach used. Given the same spatial resolution, the root conformation representations with small root functions yield the most accurate predictions. However, with an increasing degree of resolution all representations converge towards one mesh-independent solution. Although being robust, we show that for an increasing fluid elasticity the numerical methods are subject to severe requirements to obtain accurate predictions of the flow fields. For characteristic quantities, like the corner vortex size and its intensity, we provide new benchmark results for highly elastic flows of Oldroyd-B and PTT fluids in the 4:1 contraction benchmark.
\end{enumerate}

\section{Mathematical modeling framework}
\label{Mathmod}
%%%%%%%%%%%%%%%%%%%%%%%%%%%%%%%%%%%%%%%%%%%%%%%
The balance equations for laminar flows of incompressible and isothermal viscoelastic fluids are the mass conservation
\begin{equation}
\label{contiEq}
\div \velocity = 0,
\end{equation}
where $\velocity$ denotes the fluid velocity and the momentum conservation
\begin{equation}
\label{momentumEq}
\rho \ddt \velocity + \rho \left( \velocity \dprod \nabla \right) \velocity = \nabla \dprod \stress,
\end{equation}
where $\rho$ is the fluid density, which is assumed to be constant. We decompose the total stress $\stress$ into an isotropic pressure part $p \IT$, where $\IT$ is the unit tensor, and into the extra stress tensor $\stressE$, that vanishes in equilibrium,
\begin{equation}
\label{stressdef}
\stress = - p \IT + \stressE.
\end{equation}
The extra stress tensor consists of two contributions: the Newtonian solvent contribution and, assuming a multimode approximation of the relaxation spectrum, the combination of all $K$ modal polymer stress contributions, i.e.\
\begin{equation}
\label{stresssplit}
{\stressE} = \stressS + \sum_{k = 1}^{K} \stressPk.
\end{equation}
Note that (\ref{stresssplit}) is sometimes called the solvent-polymer stress splitting \cite{Bird1977}.
The Newtonian solvent contribution $\stressS$ is a linear
function of the rate-of-deformation tensor
$\DT = \frac{1}{2} \left( \grad \velocity + \trans{\grad \velocity} \right)$,
\begin{equation}
\label{solventstress}
\stressS = 2\eta_S\DT,
\end{equation}
where $\eta_S$ is the solvent viscosity. The polymer stress tensor of the $k$-th relaxation mode $\stressPk$ depends, in general, nonlinearly on the local deformation and on the flow history of the fluid. A mathematical expression for the material-dependent modal polymer stress tensor $\stressPk$ needs to be obtained from rheological constitutive equations.
%%%%%%

For a macroscopic fluid element, the evolution of the stress is governed by a statistical distribution of its internal microstructure, viz.\ the conformation of the macro-molecules; cf. \cite{Bird1977, Larson1988, Keunings2000}. Therefore, the modal polymer stress is directly related to a
conformation tensor $\Ck$, which may be regarded as the average distribution of all microscopic configurations of the polymer macromolecules in the macroscopic fluid element. In particular, the conformation tensor is usually related to the second moment of a distribution function of all molecular configurations \cite{Larson1988} which implies that, by definition, it is a symmetric positive definite (\textsc{spd}) second-rank tensor. Indeed, the \textsc{spd} property of the conformation tensor is a universal requirement for the derivation of a class of numerical stabilization methods, as will be discussed in section \ref{Stabilization}. Moreover, the positive definiteness can be exploited in any numerical algorithm to avoid unphysical results. Therefore, an evolution of the conformation tensor rather than the stress tensor should be generally preferred in the development of numerical algorithms.

Consequently, our purpose at this point is twofold: to determine a representation for the tensor-valued tensor function $\stressPk = \cvec{T}(\Ck)$, and to determine an evolution equation for the conformation tensor $\Ck$ that shall comply with the following requirements: in this work, we shall restrict our considerations to partial differential conformation tensor constitutive equations. Our particular purpose is to consider one tensorial constitutive equation for each modal conformation $\Ck$, that shall be generic by means of representing all rheological models that are compatible with such requirements. Moreover, regarding the derivation of the stabilization method, we shall impose an assumption on the isotropy of the material.
Let us, therefore, consider the generic equations given by
\begin{align}
\label{defConstEqC1}
\ddt \Ck + \left( \velocity \dprod \nabla \right)  \Ck  - \LT \dprod \Ck - \Ck \dprod \trans{\LT}
&=
\frac{1}{\tau_{1,k}} \cvec{P}\left(\Ck \right),\\
\label{defConstEqC1b}
\stressPk &= \cvec{T}(\Ck),
\end{align}
where $\tau_{1,k}$ is the relaxation time,
$\LT := \trans{\grad \velocity} - {\zeta}_k \DT, \ {\zeta}_k \in [0, 2]$, and let
${\zeta}_k$ be a material parameter (sometimes called ''slip parameter``) which characterizes the degree of nonaffine response of the polymer chains to an imposed deformation.
Note that the l.h.s. of (\ref{defConstEqC1}) is called the convected time-derivative of \Ck,
defined by
\begin{equation}
\label{defConDeriv}
\overset{\square}{\CT}_k = \DDt{\Ck} - \LT \dprod \Ck - \Ck \dprod \trans{\LT},
\end{equation}
where $\DDt{\Ck} = \ddt \Ck + \left( \velocity \dprod \nabla \right)\Ck$ is the substantial time-derivative of \Ck. Note further that the motion becomes affine for ${\zeta}_k = 0$ in which case the left-hand side of (\ref{defConstEqC1}) reduces to the upper-convected derivative, first proposed by Oldroyd \cite{Oldroyd1950}.

The formulation of constitutive equations in a convected coordinate system that is deformed with the fluid elements is convenient under continuum mechanical considerations, namely the principle of frame-invariance \cite{Truesdell1965}. On the other hand, an identical form is obtained in the derivations from molecular theories; cf. \cite{Larson1988, Huilgol1997}. It is worthwhile to mention that some of the most successful theories for the derivation of macroscopic differential constitutive equations of the form (\ref{defConstEqC1}) and (\ref{defConstEqC1b}) are the configuration space and phase space kinetic theories proposed by Kirkwood \cite{Kirkwood1967} and generalized by Bird et al. \cite{Bird1987b}, the network theories by Lodge \cite{Lodge1956} and Yamamoto \cite{Yamamoto1956} and, to some extent, the reptation theories, based on Doi and Edwards \cite{Doi1978} and de Gennes \cite{deGennes1979}.
Furthermore, by introducing flow-induced anisotropic mobility of the molecular configuration, Giesekus \cite{Giesekus1966} proposed another theory from which such type of constitutive equations have been derived.

In this work, we assume the modal stress function to be linear in $\Ck$ and write a general relation for the modal stress in the form
\begin{equation}
\label{conversion}
{\stressPk} = \frac{{\eta_{P,k}}}{{\tau_{1,k}} (1 - {\zeta}_k)} \left( h_{0,k} \IT + h_{1,k} \Ck \right),
\end{equation}
where $h_{j,k}$ are scalar functions. One way to derive the conversion between the macroscopic quantities ${\stressPk}$ and $\Ck$ in (\ref{conversion}) is from a closed form of the Kramers expression for the extra stress tensor \cite{Bird1987b, Larson1988}. Closure approximations have to be included into the microscopic model to avoid dependencies of higher order moments. For most rheological models, $h_{j,k}$ are either functions of the first invariant of $\Ck$, i.e.\ $h_{j,k} = \hat{h}_{j,k}(\tr \Ck), \ j = 0,1$ or constants (cf.\ Table \ref{tab:constitutiveEx}). The exact expression depends on both, the connector force in the molecular model and the closure approximation. For the most simple case of a Hookean spring force law both $h_{0,k} = -1$ and $h_{1,k} = 1$ are constants.

Suppose now that on the right-hand side of (\ref{defConstEqC1}), $\cvec{P} \left(\Ck \right)$ is a real analytic tensor function of the argument $\Ck$, and let
\begin{equation}
\label{defisotropicf}
\OT \dprod \cvec{P} \left(\Ck \right) \dprod \trans{\OT} = \cvec{P} \left(\OT \dprod \Ck \dprod \trans{\OT} \right)
\end{equation}
for every orthogonal tensor $\OT$. Then, $\cvec{P} \left(\Ck \right)$ is said to be an \textit{isotropic} tensor-valued function of the \textsc{spd} second-rank tensor $\Ck$. Rivlin and Ericksen \cite{RivlinEricksen1955} and Rivlin \cite{Rivlin1955} proved that every real analytic and isotropic tensor-valued function $\cvec{P} \left(\Ck \right)$ of one second-rank tensor $\Ck$ has a representation of the form
\begin{equation}
\label{Pfunc}
\cvec{P}\left(\Ck \right) = g_{0,k} \IT + g_{1,k} \Ck + g_{2,k} \Ck^2,
\end{equation}
where $g_{0,k}$, $g_{1,k}$ and $g_{2,k}$ are isotropic invariants (isotropic scalar functions) of $\Ck$ and hence can be expressed as functions of its three scalar principal invariants, i.e.\ $g_{i,k} = \hat{g}_{i,k}(I_{1,k}, I_{2,k}, I_{3,k}), \ i = 0,1,2$. The principal invariants $I_{1,k}, I_{2,k}, I_{3,k}$ are defined by
\begin{equation}
I_{1,k} = \textnormal{tr} \: \Ck, \ I_{2,k} = \frac{1}{2} \left[ I_{1,k}^2 - \tr{\Ck^2} \right]\!, \ I_{3,k} = \det \Ck.
\end{equation}
A large class of viscoelastic materials is represented by (\ref{contiEq}) to (\ref{defConstEqC1}), (\ref{conversion}), and (\ref{Pfunc}). The complexity of the closure problem is now reduced to the determination of expressions for the scalar functions $g_{0,k}$, $g_{1,k}$ and $g_{2,k}$, as well as $h_{1,k}$ and $h_{2,k}$ in terms of the principal invariants of $\Ck$, instead of determining $2 \times 6$ independent components of the two symmetric tensors ${\PT}_k = \cvec{P} \left(\Ck \right)$ and ${\TT}_k = \cvec{T}(\Ck)$. On the other hand, some generality is lost with the restriction (\ref{defisotropicf}) to isotropic forms in $\Ck$, though the framework could be extended to some anisotropic representations in order to include further rheological models. But this would go beyond the scope of the present work. Moreover, we exploit the isotropy of $\cvec{P}\left(\Ck \right)$ for the derivation of the stabilization method in section \ref{Stabilization}. In Table \ref{tab:constitutiveEx}, we present the model-specific expressions for some widely used rheological models that fit into this framework.
%%%%%%%%%%%%%%%%%%%%%%%%%%%%%%%%%%%%%%%%%%%%%%%%%%%%%%%%%%%%%%%%%%%%%

\begin{table}[h!]
\begin{center}
\begin{tabular}{@{}lcccccc@{}}
\toprule
constitutive model & \ ${\zeta}_k$ \ & \ $g_{0,k}$ \ & \ $g_{1,k}$ \ & \ $g_{2,k}$ \ & \ $h_{0,k}$ \ & \ $h_{1,k}$ \ \\ 
\midrule
Maxwell/Oldroyd-B & $0$ & $1$ & $-1$ & $0$ & $-1$ & $1$ \\
%%%%%%%%%%%%%
Leonov & $0$ & $ \frac{1}{2} $ & $  \frac{1}{6} \left( I_{1,k} - I_{2,k} \right)  $ & $- \frac{1}{2}$ & $-1$ & $1$  \\
%%%%%%%%%%%%
Giesekus & $0$ & $1 - \alpha_k$ & $2 \alpha_k - 1$ & $ - \alpha_k$ & $-1$ & $1$  \\
%%%%%%%%%%%%%
Johnson-Segalman & $\in \left[0, 2 \right]$ & $1$ & $-1$ & $0$ & $-1$ & $1$  \\
%%%%%%%%%%%%%
LPTT & $\in \left[0, 2 \right]$ & $1 + \frac{\varepsilon_k}{1 - {\zeta}_k} \left( \tr \Ck - 3 \right) $ & $ - g_{0,k} $ & $0$ & $-1$ & $1$  \\  
%%%%%%%%%%%%%
EPTT & $\in \left[0, 2 \right]$ & $ \exp \left[  \frac{\varepsilon_k}{1 - {\zeta}_k} \left( \tr \Ck - 3 \right) \right] $ & $ - g_{0,k} $ & $ 0 $ & $-1$ & $1$  \\
%%%%%%%%%%%%%%
%%%%%%%%%%%%%
FENE-P & 0 & 1 & $- \frac{1}{1 - {\tr \left( \Ck \right) }/b_k}$ & $0$ & $-1$ & $ \frac{1}{1 - {\tr \left( \Ck \right) }/b_k}$  \\
%%%%%%%%%%%%
FENE-CR & 0 & $ \frac{1}{1 - {\tr \left( \Ck \right) }/b_k}$ & $-g_{0,k}$ & $0$ & $-1$ & $ \frac{1}{1 - {\tr \left( \Ck \right) }/b_k}$ \\
%%%%%%%%%%%%
%%%%%%%%%%%%
%%%%%%%%%%%%
\bottomrule
\end{tabular}
\end{center}
\caption{Model-dependent scalar-valued tensor functions for the generic constitutive equation (\ref{defConstEqC1}) and (\ref{Pfunc}).}
\label{tab:constitutiveEx}
\end{table}
For later use, let us also give an equivalent formulation of (\ref{defConstEqC1}) for the polymer stress tensor $\stressPk$:
\begin{equation}
\label{pStressConst}
\overset{\square}{\stressE}_{p, k} = - 2 G_k h_{0,k} \DT + \frac{{1}}{{\tau_{1,k}}} \cvec{P} \left(\stressPk \right),
\end{equation}
where $G_k = \frac{\eta_{P,k}}{\tau_{1,k}}$ and $\cvec{P} \left(\stressPk \right)$ is uniquely determined by the representation
\begin{equation}
\label{pStressConst2}
\cvec{P} \left(\stressPk \right) = g^{*}_{0,k} \IT + g^{*}_{1,k} \stressPk + g^{*}_{2,k} \stressPk^{2},
\end{equation}
%%%
where the relations for $g^{*}_{0,k}, g^{*}_{1,k}$ and $g^{*}_{2,k}$ are given by
\begin{equation}
\label{pStressConst3}
g^{*}_{0,k} = g_{0,k} - \frac{h_{0,k}}{h_{1,k}} g_{1,k} + \frac{h^{2}_{0,k}}{h^{2}_{1,k}} g_{2,k},
\end{equation}
\begin{equation}
\label{pStressConst4}
g^{*}_{1,k} = \frac{1-\zeta_{k}}{G_{k} h_{1,k}} \left( g_{1,k} - 2 g_{2,k} \frac{h_{0,k}}{h_{1,k}}\right),
\end{equation}
\begin{equation}
\label{pStressConst5}
g^{*}_{2,k} = \frac{(1-\zeta_{k})^{2}}{G^{2}_{k} h^{2}_{1,k}} g_{2,k}.
\end{equation}
Equations (\ref{pStressConst}) to (\ref{pStressConst5}) can be regarded as the direct counterpart of the generic conformation tensor constitutive framework in terms of the stress tensor. We will need this form in the derivation of our finite volume method on general unstructured meshes.
%%%%%%%%%%%%%%%%%%%%%%%%%%%%%%%%%%%%%%%%%%%%%

\section{Stabilization}
\label{Stabilization}
The objective of the present section is to present a unified mathematical stabilization framework for the general conformation tensor constitutive equations (\ref{defConstEqC1}).
Special care will be taken on the derivation of a form which is convenient to be implemented into a novel generic C++ library. The consistent implementation of the mathematical framework into a FVM is addressed in section \ref{Numerical Method}.

\subsection{Literature survey}
\label{sec:ltrv}
The numerical stabilization approaches for high Weissenberg number flows can be classified into three categories.

In the first category, an additional diffusion term is added on both sides of the momentum equation. One of them is treated implicitly and the other one is treated as an explicit source term. The matrix coefficients that arise from the implicit diffusion term might improve the convergence of the iterative solution algorithm. This is referred to as the both-sides diffusion (BSD) approach \cite{Xue1995}. However, since this approach does not address the essential causes of the HWNP, i.e.\ the advection of the steep spatial stress profiles, the stabilizing effect in computational benchmarks has been found to be relatively poor compared to approaches of the other two categories \cite{Chen2013}. Moreover, it has been reported that in transient simulations the BSD tends to generate overdiffusion \cite{Xue2004}, hence being not suitable to accurately capture the transient nature of the flow.

The stabilization methods of the second category focus on preserving the positive definiteness of the conformation tensor by means of the numerical schemes. By definition, the conformation tensor is a symmetric positive definite tensor; cf.\ \cite{Grmela1987, Hulsen1988, Hulsen1990}.
Any numerical algorithm must guarantee this positivity-preserving property during time evolution in order to obtain physically reliable results. The first-order accurate positive definiteness preserving scheme (PDPS) \cite{Stewart2008} has been proven to keep the conformation tensor positive definite as long as some stability constraint is satisfied. However, one should clearly distinguish between the stability of the numerical method and the mathematical constraint of a positive definite conformation tensor, since the loss of positive definiteness is not necessarily the cause of the HWNP, but only a consequence of the numerical failure to advect steep profiles.

The objective of the stabilization approaches of the third category is to reduce the stiffness of the problem by certain tensor-transformed representations of the conformation tensor constitutive equations. We will call such methods the change-of-variable representations (CVR). The most popular approach is the logarithm conformation representation (LCR), first proposed by Fattal and Kupferman in \cite{Fattal2004} and applied to a FEM in \cite{Hulsen2005}, where the conformation tensor equations are reformulated with logarithmic variables. The representation of the conformation tensor equations with logarithmic functions has two major advantages: the first one is the linearization of the locally steep spatial conformation tensor profiles which removes the previously mentioned numerical advection issue completely. Secondly, the LCR inherently preserves the positive definiteness of the conformation tensor. Therefore, CVRs are most effective in defeating the HWNP, yet they do not represent a rigorous mathematical solution of the problem. We emphasize that such methods need a diagonalization of the conformation tensor which generally makes them computationally more expensive than those of the previous categories. 
Balci et al.\ \cite{Balci2011} proposed a square root conformation representation (SRCR), where the diagonalization of the conformation tensor can be avoided. Moreover, Afonso et al.\ \cite{Afonso2012} proposed a generic kernel conformation tensor transformation framework, which includes different transformation functions.

Because of their benefits regarding the stability at high Weissenberg numbers, we restrict our considerations to the change-of-variable representations. In the next sections, we present a general mathematical framework which has been implemented into our numerical stabilization library (stabLib).

\subsection{Diagonalization of the conformation tensor}
Let $\CT = \trans{\CT}$ be a \textsc{spd} second rank tensor.
Then, 
$\CT$ has a diagonalization
\begin{equation}
\label{diagonalize1}
\CT
=
\QT \dprod \LambdaT \dprod \trans{\QT},
\end{equation}
where the diagonal tensor $\LambdaT$ contains the 3 real eigenvalues $\lambda_1, \lambda_2, \lambda_3$ of $\CT$ on its diagonal in ascending order. Note that the eigenvalues must be strictly positive, because $\CT$ is a positive definite tensor. The corresponding set of the 3 orthonormalized eigenvectors (orthonormal basis) are the columns of the tensor $\QT$. Recall that $\QT$ satisfies $\trans{\QT} \dprod \QT = \QT \dprod \trans{\QT} = \IT$, i.e.\ ${\QT}^{-1} = \trans{\QT}$.

%%%%%%%%%%%%%%%%%%%%%%%%%%%%%%%%%%%%%%
\subsection{Local decomposition of the deformation terms}
Let us define $\OmegaT := \DDt {\QT} \dprod \trans{\QT}$, hence $\trans{\OmegaT} = \QT \dprod \DDt {\trans{\QT}}$. Then, plugging in the diagonalization $\left( \ref{diagonalize1} \right)$ into the material derivative and applying the {chain rule}, we obtain
\begin{equation}
%\label{diagonalizeCr1a}
\DDt {\CT}
%\DDt {\left( \QT \dprod \LambdaT \dprod \trans{\QT} \right)}
%&=
%\DDt {\QT} \dprod \LambdaT \dprod \trans{\QT}
%+ \QT \dprod \DDt {\LambdaT} \dprod \trans{\QT}
%+ \QT \dprod \LambdaT \dprod \DDt {\trans{\QT}}
%\\
%&=
\label{diagonalizeCr1b}
=
\QT \dprod \DDt {\LambdaT} \dprod \trans{\QT}
+ \OmegaT \dprod \CT + \CT \dprod \trans{\OmegaT}.
\end{equation}
We exploit the symmetry properties of the previously defined tensors $\OmegaT$ and $\trans{\OmegaT}$ in $\left( \ref{diagonalizeCr1b} \right)$. Differentiating the orthogonality condition ${\QT \dprod \trans{\QT} = \IT}$
shows that ${\OmegaT}$ is a skew-symmetric tensor, i.e.\ ${\OmegaT} = - \trans{{\OmegaT}}$.
%%%%%%%%%%%%%%%%%
Therefore, the material derivative $\left(\ref{diagonalizeCr1b}\right)$ can be written in the form
\begin{equation}
\label{diagonalizeCr1c}
\DDt {\CT}
=
\QT \dprod \DDt {\LambdaT} \dprod \trans{\QT}
+ \OmegaT \dprod \CT - \CT \dprod \OmegaT.
\end{equation}
%%%%%%%%%%%%%%%
Inserting equation $\left(\ref{diagonalizeCr1c}\right)$ into the evolution equation of the conformation tensor $\left( \ref{defConstEqC1} \right)$, we obtain
\begin{equation}
\label{defConstEqC2}
\QT \dprod \DDt {\LambdaT} \dprod \trans{\QT}
+ \OmegaT \dprod \CT - \CT \dprod \OmegaT
=
\LT \dprod \CT + \CT \dprod \trans{\LT}
+ \frac{1}{\tau_1} \cvec{P} \left(\CT \right).
\end{equation}
Our objective is now to determine an expression for $\OmegaT$. First, we write equation $\left( \ref{defConstEqC2} \right)$ in a more convenient form along the principal axes of $\CT$. Let us define $\tilde{\OmegaT} := \trans{\QT} \dprod \DDt {\QT}$, $\LT =: \QT \dprod \tilde{\LT} \dprod \trans{\QT}$
and recall that $\cvec{P} \left(\CT \right)$ is an isotropic symmetric tensor-valued tensor function of the symmetric second-order tensor $\CT$. From the isotropy condition (\ref{defisotropicf}), we have $\cvec{P} \left(\CT \right) = \cvec{P} \left(\QT \dprod \LambdaT \dprod \trans{\QT} \right) = \QT \dprod \cvec{P} \left(\LambdaT \right) \dprod \trans{\QT}$. Hence equation $\left( \ref{defConstEqC2} \right)$ can be written in the form
\begin{equation}
\label{defConstEqC3}
{
{\DDt {\LambdaT}}
}
+
{
{\tilde{\OmegaT} \dprod \LambdaT
- \LambdaT \dprod \tilde{\OmegaT}}
}
=
{
{\LambdaT \dprod \tilde{\LT}
+ \trans{\tilde{\LT}} \dprod \LambdaT}
}
+ 
{
{\frac{1}{\tau_1} 
\cvec{P} \left(\LambdaT \right)}
}.
\end{equation}
At this point, we take advantage of the structure of equation $\left( \ref{defConstEqC3} \right)$. On the left-hand side it consists of one diagonal term ${\DDt {\LambdaT}}$ and two skew-symmetric terms $\tilde{\OmegaT} \dprod \LambdaT$ and $- \LambdaT \dprod \tilde{\OmegaT}$. However, note that the sum of both skew-symmetric terms $\tilde{\OmegaT} \dprod \LambdaT - \LambdaT \dprod \tilde{\OmegaT}$ is in turn symmetric. On the right-hand side $\cvec{P} \left(\LambdaT \right)$ is a diagonal tensor. Only the two terms $\tilde{\LT}$ and $\trans{\tilde{\LT}}$ on the right-hand side are neither diagonal nor skew-symmetric. Consequently, a sensible next step is to decompose these two terms. For this purpose, let $\tilde{\BT}$ be a diagonal tensor that contains the diagonal elements ${\tilde{l}}_{ii}$ of the tensor $\tilde{\LT}$, i.e.\
\begin{align}
\label{tensorCVD}
\tilde{\BT} =
\left\lbrace
{
\begingroup
\renewcommand*{\arraystretch}{1.66}
\begin{array}{lll}
{\tilde{l}}_{ij} \ & \textnormal{if} \ & i = j,
\\
0 \ & \textnormal{if} \ & i \neq j
\end{array}
\endgroup
}
\right.
\end{align}
and let $\tilde{\OT}$ be a tensor with all diagonal elements equal to zero and containing all off-diagonal elements ${\tilde{l}_{{ij}, \; {i \neq j}}}$ of the tensor $\tilde{\LT}$, i.e.\
\begin{align}
\label{tensorOT}
\tilde{\OT} =
\left\lbrace
{
\begingroup
\renewcommand*{\arraystretch}{1.66}
\begin{array}{lll}
0 \ & \textnormal{if} \ & i = j,
\\
{\tilde{l}}_{ij} \ & \textnormal{if} \ & i \neq j .
\end{array}
\endgroup
}
\right.
\end{align}
Then, splitting equation $\left( \ref{defConstEqC3} \right)$ into its diagonal and off-diagonal contributions, we have
\begin{align}
\label{diagonalizeCr81}
\left\lbrace
{
\begingroup
\renewcommand*{\arraystretch}{1.66}
\begin{array}{l}
{
{\DDt {\LambdaT}}
}
=
{
{2 \tilde{\BT} \dprod \LambdaT}
}
+ 
{
\frac{1}{\tau_1} 
\cvec{P} \left(\LambdaT \right)
},
\\
{
{\tilde{\OmegaT} \dprod \LambdaT
- \LambdaT \dprod \tilde{\OmegaT}}
}
=
{\LambdaT \dprod \tilde{\OT}
+ \trans{\tilde{\OT}} \dprod \LambdaT},
\end{array}
\endgroup
}
\right.
\end{align}
where we use the fact that the two diagonal tensors $\LambdaT$ and $\tilde{\BT}$ commute, so that $\LambdaT \dprod \tilde{\BT} + \trans{\tilde{\BT}} \dprod \LambdaT = 2 \tilde{\BT} \dprod \LambdaT$. However, from the off-diagonal contribution in $\left( \ref{diagonalizeCr81} \right)$ it follows that
\begin{equation}
\label{diagonalizeCr82}
\tilde{\omega}_{{{ij}, \; {i \neq j}}} = \frac{\lambda_{ii} {\tilde{l}_{{ij}, \; {i \neq j}}} + \lambda_{jj} {\tilde{l}_{{ji}, \; {j \neq i}}}}{\lambda_{jj} - \lambda_{ii}}, \ i, j = 1, 2, 3 .
\end{equation}
By the relation (\ref{diagonalizeCr82}), the tensor ${\tilde{\OmegaT}}$ in (\ref{defConstEqC3}) is determined as a function of the known tensor $\tilde{\LT}$.
Keeping this in mind, we write the material derivative of the conformation tensor in the form
\begin{align}
\label{evolutionConf1}
\DDt \CT
&=
{\QT} \dprod \left[
{
2 \tilde{\BT} \dprod {\LambdaT}
}
+ \tilde{\OmegaT} \dprod {\LambdaT}
- {\LambdaT} \dprod \tilde{\OmegaT}
+ 
{
\frac{1}{\tau_1} 
\cvec{P} \left(\LambdaT \right)
}
\right] 
\dprod \trans{\QT}
\\
\label{evolutionConf2}
&=
2 \BT \dprod \CT
+ \OmegaT \dprod \CT
- \CT \dprod {\OmegaT}
+ {\frac{1}{\tau_1} 
\cvec{P} \left(\CT \right)},
\end{align}
where the definition $\BT := \QT \dprod \tilde{\BT} \dprod \trans{\QT}$ was used. Equation (\ref{evolutionConf2}) is the final result for the local decomposition of the tensors $\LT$ and $\trans{\LT}$ in the evolution equation for the conformation tensor. The final form is very similar to the decomposition theorem which has been proposed by Fattal and Kupferman \cite{Fattal2004}. However, (\ref{evolutionConf2}) is slightly more general, since non-affine motion has been included into our framework, i.e. $\LT = \trans{\grad \velocity} - {\zeta}_k \DT$, which implies that it covers a larger range of constitutive models.

Note that if we would solve (\ref{evolutionConf2}) at this stage, the numerical computation of the tensor $\tilde{\OmegaT}$ would generate numerical issues if $\epsilon = |\lambda_{jj} - \lambda_{ii}|$ was very small. Floating point exception errors could be avoided by evaluating the off-diagonal elements of the tensor $\tilde{\KT} := \tilde{\OmegaT} \dprod \LambdaT - \LambdaT \dprod {\tilde{\OmegaT}}$ directly as 
\begin{equation}
\label{diagonalizeCr83}
\tilde{k}_{{{ij}, \; {i \neq j}}} = {\lambda_{ii} {\tilde{l}_{{ij}, \; {i \neq j}}} + \lambda_{jj} {\tilde{l}_{{ji}, \; {j \neq i}}}}, \ i, j = 1, 2, 3
\end{equation}
instead of explicitly computing $\tilde{\OmegaT}$. Nevertheless, solving (\ref{evolutionConf2}) instead of (\ref{defConstEqC1}) doesn't make much sense, since we have only decomposed the deformation terms which does not necessarily affect the numerical robustness. The local decomposition is rather important for the derivation of CVRs using tensor transformation functions, since it is in general required for applying tensor transformation functions to the conformation tensor constitutive equations. The tensor transformation of (\ref{evolutionConf2}) will be described in section \ref{sec:changeofvariable}. 

%%%%%%%%%%%%%%%%%%%%%%%%%%%%%%%%%%%%%%%%%%%%%%%%%%%%%%%%
%%%%%%%%%%%%%%%%%%%%%%%%%%%%%%%%%%%%%%%%%%%%%%%%%%%%%%%%
\subsection{Tensor transformation (change-of-variable)}
\label{sec:changeofvariable}
%%%%%%%%%%%%%%%%%%%%%%%%%%%%%%%%%%%%%%%%%%%%%%%%%%%%%%%%
%%%%%%%%%%%%%%%%%%%%%%%%%%%%%%%%%%%%%%%%%%%%%%%%%%%%%%%%
Let $\cvec{F}(\CT)$ be a real analytic tensor function of the \textsc{spd} second-rank tensor $\CT$, such that
\begin{equation}
\label{diagonalizef0}
\OT \dprod \cvec{F}(\CT) \dprod \trans{\OT}
=
\cvec{F}(\OT \dprod \CT \dprod \trans{\OT}),
\end{equation}
for every orthogonal tensor ${\OT}$, i.e.\ let $\cvec{F}(\CT)$ be an \textit{isotropic} tensor-valued function of $\CT$. Then the diagonalization of $\CT$ satisfies
\begin{equation}
\label{diagonalizef1}
\cvec{F}(\CT)
=
\cvec{F}(\QT \dprod \LambdaT \dprod \trans{\QT})
=
\QT \dprod \cvec{F}(\LambdaT) \dprod \trans{\QT}.
\end{equation}
Applying the chain rule to the material derivative of $\cvec{F}(\CT)$, we have
\begin{equation}
\label{materialderivativef1}
\DDt {\cvec{F}(\CT)}
=
\QT \dprod \DDt {\cvec{F}(\LambdaT)} \dprod \trans{\QT} +
\OmegaT \dprod {\cvec{F}(\CT)} - {\cvec{F}(\CT)} \dprod \OmegaT.
\end{equation}
Regarding the first term on the right-hand side of (\ref{materialderivativef1}), we exploit the property that $\LambdaT$
is a diagonal second rank tensor, i.e.\ recall that analytical functions ${\cvec{F}\left(\LambdaT(t)\right)}$ of a diagonal tensor $\LambdaT(t)$ can generally be evaluated by computing each diagonal component ${\mathcal{F}}_{ii}\left(\lambda_{ii}(t)\right)$ according to
\begin{equation}
\label{diagonalF01}
{\cvec{F}\left(\LambdaT(t)\right)} = \operatorname{diag} \left({\mathcal{F}}_{ii}\left(\lambda_{ii}(t)\right)\right), \ i = 1, 2, 3.
\end{equation}
Thus, we can write any derivative of the non-zero diagonal elements in (\ref{diagonalF01}) as a composition
\begin{equation}
\label{diagonalF02}
%{\cvec{F}_{ii}^{'}\left(\LambdaT_{ii}(t)\right)} 
\left({\mathcal{F}}_{ii}\left(\lambda_{ii}(t)\right)\right)^{'}
= {\mathcal{F}}_{ii}^{'}\left(\lambda_{ii}(t)\right) \lambda_{ii}^{'}(t),
\end{equation}
hence
\begin{equation}
\label{diagonalF02b}
\left({\cvec{F}\left(\LambdaT(t)\right)}\right)^{'}
=
{\cvec{F}^{'}\left(\LambdaT(t)\right)} \dprod \LambdaT^{'}(t),
\end{equation}
for every diagonal tensor $\LambdaT(t)$, where ${\cvec{F}^{'}\left(\LambdaT(t)\right)} = \operatorname{diag} \left({\mathcal{F}_{ii}^{'}\left(\lambda_{ii}(t)\right)}\right)$ and $\LambdaT^{'}(t) = \operatorname{diag} \left(\lambda_{ii}^{'}(t)\right)$, respectively.
Substituting (\ref{diagonalF02b}) into the material derivative (\ref{materialderivativef1}), we obtain
\begin{equation}
\DDt {\cvec{F}\left(\LambdaT(t)\right)} = \tilde{\UpsilonT}\left(\LambdaT(t)\right) \dprod \DDt {\LambdaT(t)},
% \tilde{\UpsilonT}\left(\LambdaT(t)\right),
%\frac{D \cvec{F}(\LambdaT)}{D \LambdaT}
\end{equation}
where the diagonal tensor $\tilde{\UpsilonT}$ is defined as
\begin{equation}
\label{diagonalF03}
\tilde{\UpsilonT}\left(\LambdaT(t)\right) := {\cvec{F}^{'}\left(\LambdaT(t)\right)} = \operatorname{diag}
\left(
{\mathcal{F}}_{ii}^{'}\left(\lambda_{ii}(t)\right)\right).
\end{equation}
Note that $\tilde{\UpsilonT}$ is known analytically for certain functions $\cvec{F}$; cf.\ table \ref{tab:JJ}.
From (\ref{diagonalizeCr81}), we have an expression for the derivative $\DDt {\LambdaT}$. Using all this information, and defining $\UpsilonT := \QT \dprod \tilde{\UpsilonT} \dprod \trans{\QT}$ the generic constitutive equation for a function $\cvec{F}(\CT)$ has the form
\begin{equation}
\label{evolutionConf3}
\DDt {\cvec{F}(\CT)}
=
%\QT \dprod \left( {2 \tilde{\BT} \dprod \LambdaT} + {\frac{1}{\tau_1} \cvec{P} \left(\LambdaT \right)} \right) \dprod \UpsilonT \dprod \trans{\QT}
{2 \BT \dprod {\UpsilonT} \dprod \CT}
+
\OmegaT \dprod {\cvec{F}(\CT)} - {\cvec{F}(\CT)} \dprod \OmegaT
+
\frac{1}{\tau_1} \UpsilonT \dprod \cvec{P} \left(\CT \right).
\end{equation}
It is important to note that (\ref{evolutionConf3}) is generic and does not inherently influence the robustness of the numerical solution. Whether the robustness is increased or not, depends strongly on which particular function $\cvec{F}(\CT)$ is introduced. Therefore, (\ref{evolutionConf3}) may serve as a common basis for the formulation of certain stabilized versions of the conformation tensor constitutive laws, though additional considerations about the function type are to be incorporated into the framework.

We impose two fundamental requirements that $\cvec{F}(\CT)$ must satisfy in order to obtain both, a numerically robust, but also a physically meaningful result: firstly, $\cvec{F}(\CT)$ must flatten the locally steep spatial conformation tensor profiles such that the stiffness of the advection problem is substantially reduced. Secondly, $\cvec{F}(\CT)$ must be chosen such that the positivity of the conformation tensor is always guaranteed.

In this work, we use two types of functions that satisfy both conditions: the $k$-th root of $\CT$, and the logarithm of $\CT$ to base $a$. Table \ref{tab:JJ} shows the corresponding $\tilde{\UpsilonT}$ and $\UpsilonT$ terms (see equations (\ref{defRT}) and (\ref{defST}) for the definitions of $\RT$ and $\ST$, respectively). The details are discussed further in \ref{sec:RCR} and \ref{sec:LCR}.
\begin{table}[htbp]
\renewcommand{\arraystretch}{1.5}
\begin{center}
\linespread{1.2}
\begin{tabular}{rlll}
\hline
\ & \cvec{F}(\CT) & \ $\tilde{\UpsilonT}$ \ & \ $\UpsilonT$ \ \\
\hline
conformation tensor \ & \CT & \ $\IT$ \ & \ $\IT$ \ \\
$k$-th root of $\CT$ \ & \RT & \ $\frac{1}{k} \LambdaT^{\frac{1-k}{k}}$ \ & \ $\frac{1}{k} \RT^{1-k}$ \ \\
logarithm of $\CT$ to base $a$ \ & \ST & \ $\frac{1}{\operatorname{ln}(a)} \LambdaT^{-1}$ \ & \ $\frac{1}{\operatorname{ln}(a)} a^{-\ST}$ \ \\
%%%%%%%%%%%%
\hline 
\end{tabular}
\end{center}
\renewcommand{\arraystretch}{1.0}
\caption{$\tilde{\UpsilonT}$ and $\UpsilonT$ terms for the conformation tensor $\CT$, the $k$-th root of $\CT$, and the logarithm of $\CT$ to base $a$}
\label{tab:JJ}
\end{table}

Note that the computation of the symmetric tensor $\FT := \OmegaT \dprod {\cvec{F}(\CT)} - {\cvec{F}(\CT)} \dprod \OmegaT$ needs to be further analyzed to avoid numerical issues if $\epsilon = |\lambda_{jj} - \lambda_{ii}|$ is very small. Recalling (\ref{diagonalizeCr82}), the off-diagonal elements of the tensor $\tilde{\FT} := \trans{\QT} \dprod \FT \dprod \QT = \tilde{\OmegaT} \dprod {\cvec{F}(\LambdaT)} - {\cvec{F}(\LambdaT) \dprod \tilde{\OmegaT}}$ can be computed by
\begin{equation}
\label{diagonalizeCr84}
\tilde{f}_{{{ij}, \; {i \neq j}}} = \frac{\mathcal{F}_{jj}(\lambda_{jj}) - \mathcal{F}_{ii}(\lambda_{ii})}{\lambda_{jj} - \lambda_{ii}} \left({\lambda_{ii} {\tilde{l}_{{ij}, \; {i \neq j}}} + \lambda_{jj} {\tilde{l}_{{ji}, \; {j \neq i}}}} \right), \ i, j = 1, 2, 3
\end{equation}
while all diagonal elements of ${\tilde{\FT}}$ are zero. In the limit as $\epsilon \rightarrow 0$, we have
\begin{equation}
\label{diagonalizeCr85}
\tilde{f}_{{{ij}, \; {i \neq j}}} = \tilde{l}_{{ij}, \; {i \neq j}} + \tilde{l}_{{ji}, \; {j \neq i}} , \ i, j = 1, 2, 3. %\textnormal{for} \ \epsilon \rightarrow 0
\end{equation}
We have stabilized our code by implementing the numerical decision to use (\ref{diagonalizeCr85}) if $\epsilon$ falls below some threshold and (\ref{diagonalizeCr84}) otherwise.

%%%%%%%%%%%%%%%%%%%%%%%%%%%%%%%%%%%%%%%%%%%%%%%%%%%%%%%%%%%%%%%%%%%%%%%%
\subsubsection{The $k$-th root conformation representation (RCR)}
\label{sec:RCR}
Let $\CT$ be a symmetric positive definite tensor and let $\RT$ be the symmetric and positive definite
$k$-th root of $\CT$, such that $\RT^{k} = \CT$, for any positive number $k$. Then, we write
\begin{equation}
\label{defRT}
\RT = \CT^{\frac{1}{k}} = \QT \dprod \LambdaT^{\frac{1}{k}} \dprod \trans{\QT}.
\end{equation}
Moreover, we substitute $\RT$ for ${\cvec{F}(\CT)}$ in the constitutive equation (\ref{evolutionConf3}) and obtain, after a few straightforward manipulations, the $k$-th root conformation representation (RCR)
\begin{equation}
\label{evolutionConfRT}
\DDt \RT
=
\frac{2}{k} \BT \dprod \RT
+
\OmegaT \dprod {\RT} - {\RT} \dprod \OmegaT
+
\frac{1}{k \tau_1} \RT^{1-k} \dprod \cvec{P} \left({\RT}^{k} \right).
\end{equation}
Additionally, considering the representation (\ref{Pfunc}), we obtain
\begin{equation}
\label{evolutionConfRTb}
\DDt \RT
=
\frac{2}{k} \BT \dprod \RT
+
\OmegaT \dprod {\RT} - {\RT} \dprod \OmegaT
+
\frac{1}{k \tau_1} \left(g_0 \RT^{1-k} + g_1 \RT + g_2 \RT^{1+k}\right).
\end{equation}
For $k = 1$, (\ref{evolutionConfRT}) and (\ref{evolutionConfRTb}) are obviously identical to the conformation tensor equation (\ref{evolutionConf2}). For $k = 2$, we obtain the square root conformation representation (SRCR)
\begin{equation}
\label{evolutionConfSRT}
\DDt \SigmaT
=
\BT \dprod \SigmaT
+
\OmegaT \dprod {\SigmaT} - {\SigmaT} \dprod \OmegaT
+
\frac{1}{2 \tau_1} \SigmaT^{-1} \dprod \cvec{P} \left({\SigmaT}^{2} \right),
\end{equation}
or
\begin{equation}
\label{evolutionConfSRTb}
\DDt \SigmaT
=
\BT \dprod \SigmaT
+
\OmegaT \dprod {\SigmaT} - {\SigmaT} \dprod \OmegaT
+
\frac{1}{2 \tau_1}  \left(g_0 \SigmaT^{-1} + g_1 \SigmaT + g_2 \SigmaT^{3}\right),
\end{equation}
where $\SigmaT = \CT^{\frac{1}{2}}$, which was first proposed by Balci et al.\ \cite{Balci2011}. However, we have not applied the symmetrization procedure that has been presented in \cite{Balci2011} for the special case of the SRCR. Instead, we use the more general decomposition of the deformation according to equation (\ref{evolutionConf3}) also in our SRCR, being aware of the fact that the diagonalization might lead to higher computational costs.

%%%%%%%%%%%%%%%%%%%%%%%%%%%%%%%%%%%%%%%%%%%%%%%%%%%%%%%%%%%%%%%%%%%%%%%%%%%%%%%%%%%%%%%%
\subsubsection{The $\loga{}$-logarithm conformation representation (LCR)}
\label{sec:LCR}
Let $\CT$ be a symmetric positive definite tensor and let $\ST$ be the symmetric and positive definite logarithm of $\CT$ to base $a$, such that $a^{\ST} = \CT$, for any positive real number $a \neq 1$. Then, we write
\begin{equation}
\label{defST}
\ST = \loga{\CT} = \QT \dprod \loga{(\LambdaT)} \dprod \trans{\QT}.
\end{equation}
Moreover, we substitute $\ST$ for ${\cvec{F}(\CT)}$ in the constitutive equation (\ref{evolutionConf3}) and obtain, after a few straightforward manipulations, the $\loga{}$ conformation representation (LCR)
\begin{equation}
\label{evolutionConfST}
\DDt \ST
=
\frac{2}{\operatorname{ln}(a)} \BT
+
\OmegaT \dprod {\ST} - {\ST} \dprod \OmegaT
+
\frac{1}{\operatorname{ln}(a) \tau_1} a^{-\ST} \dprod \cvec{P} \left(a^{\ST} \right).
\end{equation}
Additionally, considering the representation (\ref{Pfunc}), we obtain
\begin{equation}
\label{evolutionConfSTb}
\DDt \ST
=
\frac{2}{\operatorname{ln}(a)} \BT
+
\OmegaT \dprod {\ST} - {\ST} \dprod \OmegaT
+
\frac{1}{\operatorname{ln}(a) \tau_1} \left(g_0 a^{-\ST} + g_1 \IT + g_2 a^{\ST}\right).
\end{equation}
For $a = e$, (\ref{evolutionConfST}) and (\ref{evolutionConfSTb}) reduce to the natural logarithm conformation representation (NLCR)
\begin{equation}
\label{nevolutionConfST}
\DDt \PsiT
=
{2} \BT
+
\OmegaT \dprod {\PsiT} - {\PsiT} \dprod \OmegaT
+
\frac{1}{\tau_1} e^{-\PsiT} \dprod \cvec{P} \left(e^{\PsiT} \right),
\end{equation}
or
\begin{equation}
\label{nevolutionConfSTb}
\DDt \PsiT
=
{2} \BT
+
\OmegaT \dprod {\PsiT} - {\PsiT} \dprod \OmegaT
+
\frac{1}{\tau_1} \left(g_0 e^{-\PsiT} + g_1 \IT + g_2 e^{\PsiT}\right),
\end{equation}
where $\PsiT = \ln{\CT}$, which was first proposed by Fattal and Kuferman \cite{Fattal2004}.

%%%%%%%%%%%%%%%%%%%%%%%%%%%%%%%%%%%%%%%%%%%%%%%%%%%%%%%%%%%%%%%%%%%%%%%%%%%%%%%%%%%%%%%%
\subsubsection{The kernel conformation representation (KCR)}
\label{sec:KCR}
The kernel conformation representation (KCR) is the first generic analytical framework, proposed by Afonso et al. \cite{Afonso2012, Martins2015}. The KCR has been proposed for any continuous, invertible and differentiable matrix transformation function. In section \ref{sec:changeofvariable} we present the derivarion of a generic framework for $\cvec{F}(\CT)$, given by (\ref{evolutionConf3}), where $\cvec{F}(\CT)$ is a real analytic tensor-valued function of the second-rank tensor $\CT$ and is necessarily isotropic. We demonstrate that imposing isotropy (\ref{diagonalizef0}) on $\cvec{F}(\CT)$ is indeed the crucial point for the derivation of the generic form (\ref{evolutionConf3}). However, this is not mentioned in \cite{Afonso2012}.
Moreover, non-affine motion is considered in our derivation, providing more generality regarding the rheological models. Finally, we assume that $\cvec{P} \left(\CT \right)$ has a unique representation (\ref{Pfunc}). Therefore, it is convenient to substitute (\ref{Pfunc}) into (\ref{evolutionConf3}). From this it follows that the RCR is given by (\ref{evolutionConfRTb}), and the LCR is given by (\ref{evolutionConfSTb}). This reduces the complexity of the closure problem, while maintaining sufficient generality to cover a wide range of rheological models; cf. Table \ref{tab:constitutiveEx}. The model-independent representations (\ref{evolutionConfRTb}) and (\ref{evolutionConfSTb}) are very convenient to be implemented as specializations of (\ref{evolutionConf3}) into an object-oriented numerical code. On the other hand, the constitutive models remain completely separated from the transport equation of $\cvec{F}(\CT)$. The three scalar cofficients $g_{0,k}$, $g_{1,k}$ and $g_{2,k}$ represent the relation between certain constitutive models (cf. table \ref{tab:constitutiveEx}) and the general transport equation of $\cvec{F}(\CT)$. This framework is used in the present work for the development of the stabLib (see also section \ref{subsec:solutionprocedure}).

%%%%%%%%%%%%%%%%%%%%%%%%%%%%%%%%%%%%%%%%%%%%%%%%%%%%%%%%%%%%%%%%%%%%%%%%%%%%%%%%%%%%%%%%
\subsection{Further remarks}
\label{sec:FurtherRemarks}
We shall emphasize that although the change-of-variable representations have proven to significantly alleviate the HWN instability, there remain some issues.
CVRs do not guarantee \textit{accurate} solutions at high Weissenberg numbers (see section \ref{sec:results}). Such representations can be viewed as an approximation to the real problem, which is not directly solved. Therefore, the critical evaluation of the solution, i.e.\ the mesh convergence in computational benchmarks is a matter of considerable importance. Having all representations available on the same computational platform, we are able to investigate and compare the accuracy of different representations in a rigorous way. In section \ref{sec:results} we present a detailed mesh convergence study of various CVRs in the computational benchmark of a 4:1 contraction flow.

\section{Numerical method}
\label{Numerical Method}
This section addresses the development of a second-order accurate cell-centered FV method on a general unstructured mesh.
The description of the FV method is subdivided into four parts: In section \ref{subsec:domaindiscretization} we discuss the discretization of the computational domain, which is subdivided into a set of control volumes (CVs) that can be of general polyhedral shape. Section \ref{subsec:eqdiscretization} addresses the equation discretization. The inter-equation coupling between the velocity and the stress is a non-trivial task, as shall be detailed in section \ref{subsec:velStressCoupling}. One reason is related to the fact that we are using a co-located or non-staggered variable arrangement in our computational grid, where all dependent variables are stored in the center of the CVs. It is well known, that the co-located variable arrangement requires special cell-face interpolation techniques, as proposed by Rhie and Chow \cite{RhieChow1983}, for the pressure-velocity coupling in order to prevent unphysical checkerboard pressure fields. It is evident, that special interpolation methods are also needed for the cell-face interpolation of the stresses in the momentum equation. Indeed, Oliveira \cite{Oliveira1998} demonstrated that the velocity-stress coupling in a segregated approach can be addressed very similar to the pressure-velocity coupling, though we are now dealing with tensors that are increased by one in rank. Our particular aim is to derive a cell-face interpolation correction term that removes the decoupling between the velocity and stress fields within our FV method on a general unstructured mesh. The derivation depends on the form of the transport equations for momentum and the constitutive variable, as well as on the FV discretization on unstructured meshes. To our knowledge, such a derivation has not been carried out for the model-independent equations of the generic framework, viz. (\ref{contiEq}) to (\ref{defConstEqC1}), (\ref{conversion}), (\ref{Pfunc}), (\ref{pStressConst}) to (\ref{pStressConst5}) and (\ref{evolutionConf3}). Finally, section \ref{subsec:solutionprocedure} addresses the numerical solution of the equation system.

\subsection{Discretization of the computational domain}
\label{subsec:domaindiscretization}
\begin{figure}[h]
\centering
%{\resizebox{0.37\textwidth}{!}{\input{poly/FVM-polyhedralCell-FVM.tex}}}
\includegraphics[width=.37\textwidth]{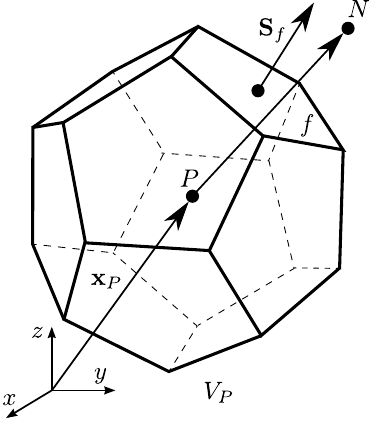}
\caption{A control volume $V_P$ around the computational point $P$}
\label{fig:polyCV}
\end{figure}

The discretization of space is carried out by a subdivision of the computational domain into a set of convex CVs, that do no overlap and completely fill the computational domain. The CVs shall be of general polyhedral shape. Fig. \ref{fig:polyCV} shows a CV $V_P$ around a computational point $P$ that is located in its centroid. $V_P$ is bounded by a fixed number of flat faces $f$. $\vec{s}_f$ is defined for each face as the outward-pointing vector normal to $f$ with the magnitude of the surface area of $f$, $\vec{s}_f = \vec{n}_f\vert \vec{s}_f \vert$, where $\vec{n}_f$ is the face unit normal vector. Each face of $V_P$ is shared with one neighboring CV around the computational point $N$.

\subsection{Equation discretization}
\label{subsec:eqdiscretization}
Let us consider the integral form of the model-independent equations (\ref{contiEq}), (\ref{momentumEq}) and (\ref{evolutionConf3}), viz.
\begin{equation}
\label{integral01}
\oint_{\partial V_P} \vec{n} \dprod \velocity \dS = 0
\end{equation}
\begin{align}
\nonumber
&\frac{\dop}{\dop t} \int_{V_P} \velocity \dV + \oint_{\partial V_P} \vec{n} \dprod  \left( \velocity \otimes \velocity \right) \dS
- \oint_{\partial V_P} \vec{n} \dprod \left({\frac{\eta_s}{\rho}} \nabla {\velocity} \right) \dS \\ 
&=
\label{integral02}
- \int_{V_P} {\frac{1}{\rho}} \nabla p \dV + \oint_{\partial V_P} \vec{n} \dprod \left({\frac{1}{\rho}} \stressE \right) \dS
\end{align}
\begin{equation}
\label{integral03}
\frac{\dop}{\dop t} \int_{V_P} \GT \dV + \oint_{\partial V_P} \vec{n} \dprod \left( \velocity \otimes \GT \right) \dS = \int_{V_P} \cvec{S}_{\GT_P} \dV,
\end{equation}
where $\GT := \cvec{F}(\Ck), k = 1, 2, ..., K$ and $\stressE = \sum_{k=1}^{K} \stressPk$. It is evident that the direct discretization of this set of integral balance equations ensures both local and global volume and momentum conservation on the discrete level. Since each control volume $V_P$ is bounded by a set of flat faces $f$, the surface integrals can be transformed into a sum of integrals over all faces that shall be approximated in this work to second order accuracy. The results will be discussed individually for each equation.

\subsubsection{Momentum equation}
The FV discretization of (\ref{integral02}) yields
\begin{align}
\label{momentumeqdiscr01}
\frac{ V_P}{\vartriangle \! t} \left(\velocity^n - \velocity^o\right)
&+ \sum_f \vec{s}_f \dprod \left(\velocity \otimes \velocity \right)_f^{\beta}
- \sum_f {\frac{\eta_s}{\rho}} \vec{s}_f \dprod \left(\nabla {\velocity}\right)_f^{\beta}
=
-{\frac{V_P}{\rho}} \left(\nabla p\right)^{\beta}
+ \sum_f {\frac{1}{\rho}} \vec{s}_f \dprod \stressE_f^{\beta},
%&+ \frac{1}{1+\beta} \sum_f \vec{s}_f \dprod \velocity_f^n \velocity_f^n 
%+ \frac{1}{1+\beta} \sum_f {\frac{\eta_s}{\rho}} \vec{s}_f \dprod \nabla {\velocity}_f^n \\
%&+ \frac{\beta}{1+\beta} \sum_f \vec{s}_f \dprod \velocity_f^o \velocity_f^o 
%+ \frac{\beta}{1+\beta} \sum_f {\frac{\eta_s}{\rho}} \vec{s}_f \dprod \nabla {\velocity}_f^o \\
%&= -{\frac{1}{\rho}} \nabla p^n V_P + {\frac{1}{\rho}} \stressE^n V_P
\end{align}
where $\beta$ is the off-centering coefficient in the interpolation between the old time steps $o$ and the new time steps $n$, i.e. for any scalar, vector or tensor quantity $\PhiT_f^{\beta}$ we have
\begin{equation}
\label{beta1}
\PhiT_f^{\beta} = \frac{1}{1+\beta} \PhiT_f^n + \frac{\beta}{1+\beta} \PhiT_f^o.
\end{equation}
Note that $\beta=1$ gives a second-order Crank-Nicholson scheme, while $\beta=0$ results in a first-order Euler scheme. In this work, we use a small amount of off-centering, i.e. $\beta=0.95$ to increase the stability of the Crank-Nicholson scheme.

We emphasize that some components of (\ref{momentumeqdiscr01}) are lagged in time due to the fact that we are using a segregated solution approach to handle the inter-equation coupling, and due to the equations being solved componentwise for any vector or tensor variable. Let us therefore denote a lagged quantity at the current time level $n$ as $l(n)$, and let
\begin{equation}
\label{beta2}
\PhiT_f^{l(\beta)} = \frac{1}{1+\beta} \PhiT_f^{l(n)} + \frac{\beta}{1+\beta} \PhiT_f^o.
\end{equation}
Equation (\ref{beta2}) is used for those terms which cannot be discretized fully implicitly in this approach, such as the non-linear advection term and the source terms which depend on the current values of $p$ or $\stressE$. We assume that in the segregated solution approach the system of equations is solved multiple times within one time step (the procedure is described in detail in section \ref{subsec:solutionprocedure}). Therefore, we have $\PhiT_f^{l(n)} = \PhiT_f^{o}$ when the equations are solved for the first time after the time index is increased by one. But as the number of outer iterations on the same time-level is increased, $\PhiT_f^{l(n)}$ is updated with the latest values, subsequently to the solution of the respective equations, such that it converges successively towards $\PhiT_f^{n}$. This iterative approach may improve the accuracy in transient simulations considerably. 
Using (\ref{beta2}), the non-linear advection term in (\ref{momentumeqdiscr01}) is linearised by replacing the cell-face flux $\phi_f^{\beta} = \vec{s}_f \dprod \velocity^{\beta}$ with $\phi_f^{l(\beta)} = \vec{s}_f \dprod \velocity^{l(\beta)}$. Since the inter-equation coupling is treated in an explicit manner, the source terms on the r.h.s. of (\ref{momentumeqdiscr01}) are replaced, too. Therefore, (\ref{momentumeqdiscr01}) becomes
\begin{align}
\label{momentumeqdiscr02}
\frac{ V_P}{\vartriangle \! t} \left(\velocity_P^n - \velocity_P^o\right)
&+ \sum_f \phi_f^{l(\beta)} \velocity_f^{\beta}
- \sum_f {\frac{\eta_s}{\rho}} \vec{s}_f \dprod \left(\nabla {\velocity}\right)_f^{\beta}
=
-{\frac{V_P}{\rho}} \left(\nabla p\right)_P^{l(\beta)}
+ \sum_f {\frac{1}{\rho}} \vec{s}_f \dprod \stressE_f^{l(\beta)}
\end{align}
The face-centroid velocity $\velocity_f^{n}$ in the advection term of (\ref{momentumeqdiscr02}) is handled implicitly by interpolating from the cell-centroids $P$ and $N$, using the high-resolution scheme Gamma \cite{Jasak1999}. This interpolation scheme guarantees the boundedness of the solution by combining second-order accurate linear interpolation with unconditionally bounded upwind interpolation. The blending factor between linear and upwind interpolation is influenced by the scheme's constant $\beta_m$. We use the value $\beta_m = 1/10$, which is recommended in \cite{Jasak1999}, in order to preserve second-order accuracy and to avoid convergence problems. The lagged quantities $\PhiT_f^{l(n)}$ are handled explicitly by linear interpolation according to
\begin{equation}
\label{linearinterpolation}
\PhiT_f^{l(n)} = \xi_f \PhiT_P^{l(n)} + \left(1 - \xi_f \right) \PhiT_N^{l(n)},
\end{equation}
where the linear interpolation factor to interpolate between the cell-centroids $P$ and $N$ is defined as $\xi_f = {\overline{f N}}/{\overline{P N}}$.

%%%%%%%%%%%%%%%%%%%%%%%%%%%%%%%%%%
The derivative $\vec{s}_f \dprod \left(\nabla {\velocity}\right)_f^{n}$ in the diffusion term of (\ref{momentumeqdiscr02}) is discretized according to Mathur and Murthy \cite{Mathur1997} as
\begin{equation}
\label{laplacian}
\vec{s}_f \dprod \left(\nabla {\velocity}\right)_f^{n} 
= 
\underbrace{\frac{\left\vert \vec{s}_f \right\vert}{\vec{n}_f \dprod \vec{d}_f} \left(\velocity_N^n - \velocity_P^n \right)}_{\textnormal{orthogonal contribution}}
+ 
\underbrace{\left(\vec{s}_f - \frac{\left\vert \vec{s}_f \right\vert}{\vec{n}_f \dprod \vec{d}_f}\vec{d}_f \right) \dprod \left(\nabla \velocity \right)_f^{l(n)}
}_{\textnormal{non-orthogonal correction}},
\end{equation}
where $\vec{d}_f$ represents the connection vector of the two computational points $P$ and $N$, $\vec{d}_f = \vec{x}_N - \vec{x}_P$. The orthogonal contribution is handled implicitly by evaluating the current second order accurate central difference of the neighboring cells directly at the face $f$. The non-orthogonal correction or cross-diffusion term is evaluated explicitly by linear interpolation of the cell-centroid gradients at $P$ and $N$ to the face $f$. The cell-centroid gradients are computed using the Gauss integral theorem. From (\ref{laplacian}) it is obvious, that for an orthogonal mesh the discretization is fully implicit (i.e. the non-orthogonal correction term becomes zero) while the explicit non-orthogonal correction increases as the angle between $\vec{n}_f$ and $\vec{d}_f$ is increased and the evaluation of this term becomes more important.

Substituting (\ref{laplacian}) into (\ref{momentumeqdiscr02}), the discretized momentum equation can be written as a linear algebraic equation in the form
\begin{equation}
\label{algebraicform01}
a_P \velocity_P^n + \sum_N a_N \velocity_N^n = \vec{r}_P - {\frac{1}{\rho}}\left(\nabla p \right)_P^{l(n)},
\end{equation}
where the coefficients $a_P$ and $a_N$ read
\begin{align}
\label{coeffap01a}
a_P &= \frac{1 + \beta}{\vartriangle \! t} + \frac{\eta_s}{\rho V_P} \sum_f \frac{\left\vert \vec{s}_f \right\vert}{\vec{n}_f \dprod \vec{d}_f} + a_P^{adv}\\
\label{coeffan02a}
a_N &= - \frac{\eta_s}{\rho V_P} \frac{\left\vert \vec{s}_f \right\vert}{\vec{n}_f \dprod \vec{d}_f} + a_N^{adv}
\end{align}
and where the source term vector $\vec{r}_P$ contains the remaining old-time level terms $o$ and the lagged terms $l(n)$ of (\ref{momentumeqdiscr02}). The coefficients $a_P^{adv}$ and $a_N^{adv}$ represent the contributions from the discretization of the advection term, using the Gamma discretization scheme.

%%%%%%%%%%%%%%%%%%%%%%%%%%%%%%%%%%
\subsubsection{Continuity equation}
The FV discretized continuity equation (\ref{integral01}) reads
\begin{equation}
\label{contidiscretization01}
\sum_{f} \vec{s}_f \dprod \velocity_f^{n} = 0.
\end{equation}
In a segregated solution approach, the continuity equation (\ref{contidiscretization01}) and the momentum equation (\ref{algebraicform01}) are combined to yield a pressure equation. The pressure equation is obtained as follows. From (\ref{algebraicform01}) the cell-centroid velocity can be expressed in the form
\begin{equation}
\label{vel01}
\velocity_{P}^{n} = \frac{\vec{h}_{P}}{a_P} - \frac{1}{a_P} \left(\nabla p^* \right)_P^{n},
\end{equation}
where the vector $\vec{h}_{P}$ is defined as
\begin{equation}
\label{vel02}
\vec{h}_{P} := \vec{r}_P - \sum_N a_N \velocity_N^{l(n)}
\end{equation}
and where $p^* = p/\rho$. Equation (\ref{vel01}) can be used to construct a cell-face velocity equation according to
\begin{equation}
\label{vel03}
\velocity_f^{n} = \left(\frac{\vec{h}}{a}\right)_f - \left(\frac{1}{a}\right)_f \left(\nabla p^* \right)_f^{n},
\end{equation}
where $\left(\frac{\vec{h}}{a}\right)_f$ and $\left(\frac{1}{a}\right)_f$ are computed by linear interpolation (\ref{linearinterpolation}) of the cell-centroid expressions at $P$ and $N$, next to the face $f$. The pressure equation is obtained from substituting (\ref{vel03}) into (\ref{contidiscretization01})
\begin{equation}
\label{contidiscretization02}
\sum_{f} \left(\frac{1}{a}\right)_f \vec{s}_f \dprod \left(\nabla p^* \right)_f^{n} = \sum_{f} \vec{s}_f \dprod \left(\frac{\vec{h}}{a}\right)_f,
\end{equation}
where the term $\vec{s}_f \dprod \left(\nabla p^* \right)_f^{n}$ is discretized analogous to (\ref{laplacian}) as follows
\begin{equation}
\label{laplacianpeq}
\vec{s}_f \dprod \left(\nabla p^*\right)_f^{n} 
= 
\underbrace{\frac{\left\vert \vec{s}_f \right\vert}{\vec{n}_f \dprod \vec{d}_f} \left({p^*_N}^n - {p^*_P}^n \right)}_{\textnormal{orthogonal contribution}}
+ 
\underbrace{\left(\vec{s}_f - \frac{\left\vert \vec{s}_f \right\vert}{\vec{n}_f \dprod \vec{d}_f}\vec{d}_f \right) \dprod \left(\nabla {p^*} \right)_f^{l(n)}
}_{\textnormal{non-orthogonal correction}}.
\end{equation}
After solving the pressure equation (\ref{contidiscretization02}), the reconstructed cell-face fluxes
\begin{equation}
\label{contidiscretization03}
\phi_f^n = \vec{s}_f \dprod \left(\frac{\vec{h}}{a}\right)_f - \left(\frac{1}{a}\right)_f \vec{s}_f \dprod \left(\nabla p^* \right)_f^{n}
\end{equation}
are conservative, which means that they satisfy the continuity equation (\ref{contidiscretization01})
\begin{equation}
\label{contidiscretization04}
\sum_{f} \phi_f^n = 0.
\end{equation}
Note that using the cell-face velocity equation (\ref{vel03}) together with the linear interpolation of $\left(\frac{\vec{h}}{a}\right)_f$ and $\left(\frac{1}{a}\right)_f$ to construct the pressure equation (\ref{contidiscretization02}) and (\ref{laplacianpeq}) to discretize the pressure gradient at the cell-face is formally equivalent to applying the original Rhie and Chow momentum interpolation \cite{RhieChow1983}. This approach effectively prevents decoupling of the pressure fields, because the orthogonal contribution is discretized on a compact stencil that depends only on the pressure in adjacent cell-centers. However, in the implementation further modifications have to be considered, e.g.\ regarding the consistency of the temporal discretization schemes; cf. \cite{Yu2002}.

%%%%%%%%%%%%%%%%%%%%%%%%%%%%%%%%%%
\subsubsection{Constitutive equation}
Applying the discretization described above also to the constitutive equation (\ref{integral03}), we have
\begin{equation}
\label{constdiscretization01}
\frac{V_P}{\vartriangle \! t} \left( \GT_P^n - \GT_P^o \right) + \sum_{f} \phi_f^{l(\beta)} \GT^{\beta} = V_P \cvec{S}_{\GT_P}^{l(\beta)}.
\end{equation}
The corresponding linear algebraic equation has the form
\begin{equation}
\label{algebraicform02}
a_P \GT_P^n + \sum_N a_N \GT_N^n = \RT_P,
\end{equation}
where the coefficients $a_P$ and $a_N$ read
\begin{align}
\label{coeffap01b}
a_P &= \frac{1 + \beta}{\vartriangle \! t} + a_P^{adv},\\
\label{coeffan02b}
a_N &= a_N^{adv}
\end{align}
and where the symmetric tensor $\RT_P$ contains the remaining old-time level terms $o$ and the lagged terms $l(n)$.

%%%%%%%%%%%%%%%%%%%%%%%%%%%%%%%%%%%%%%%%%%%%%%%%%%%%%%%%%%%%%%%%%%%%%%%%%%%%%%%%%%%%%%%%%%%%%%%%%%%%%%%%%%%%%%%%%%
\subsection{Velocity-stress coupling}
\label{subsec:velStressCoupling}
This section addresses the velocity-stress coupling of the discretized equations using a co-located FV discretization method on general unstructured meshes. The purpose of this approach is to prevent checkerboarding in the flow fields by expressing the cell-face stresses in the momentum equation in terms of the adjacent cell-centroid velocities. In particular, the cell-face gradients in the Laplacian operators are discretized directly at the cell-faces on a compact computational molecule, rather than interpolating the cell-centroid gradients to the cell-faces.
Furthermore, by subtracting the linear interpolant of the cell-face stresses from the discretization at the cell-face, we derive a correction for the linear interpolation of the cell-face stresses. This correction shall be added to the linear interpolation of the cell-face stresses in the momentum equation to prevent checkerboarding.

First, let us consider the discretized form of (\ref{pStressConst}), viz.
\begin{equation}
\label{stressdiscretization01}
\frac{V_P}{\vartriangle \! t} \left( \stressE_P^n - \stressE_P^o \right) + \sum_{f} \phi_f^{l(\beta)} \stressE_f^{\beta}
= V_P \cvec{S}_{\stressE_P}^{l(\beta)} + V_P \left(\LT \dprod \stressE + \stressE \dprod \trans{\LT} - 2 G h_0 \DT \right)_P^{l(\beta)}.
\end{equation}
Let us write (\ref{stressdiscretization01}) as a linear algebraic equation in the form
\begin{equation}
\label{salgebraicform01}
a_P \stressE_P^{n} + \sum_N a_N \stressE_N^{n} = \RT_P + \left(\LT \dprod \stressE + \stressE \dprod \trans{\LT} - 2 G h_0 \DT \right)_P^{l(n)},
\end{equation}
where the coefficients $a_P$ and $a_N$ read
\begin{align}
\label{coeffaps01a}
a_P &= \frac{1 + \beta}{\vartriangle \! t} + a_P^{adv},\\
\label{coeffans02a}
a_N &= a_N^{adv}
\end{align}
and where the symmetric tensor $\RT_P$ contains the remaining old-time level terms $o$ and the lagged terms $l(n)$.

Our purpose is now to express the cell-centroid stress $\stressE_P^{l(n)}$ by rearranging (\ref{salgebraicform01}) as follows
\begin{equation}
\label{salgebraicform02}
\stressE_{P}^{l(n)} = \frac{\HT_{P}}{a_P} + \frac{1}{a_P} \left(\LT \dprod \stressE + \stressE \dprod \trans{\LT} - 2 G h_0 \DT \right)_P^{l(n)},
\end{equation}
where the symmetric tensor $\HT_{P}$ is defined as
\begin{equation}
\label{defH}
\HT_{P} = \RT_P - \sum_N a_N \stressE_N^{l(n)}.
\end{equation}
Moreover, the cell-face stress $\stressE_f^{l(n)}$ can be reconstructed with (\ref{salgebraicform02}) according to
\begin{equation}
\label{salgebraicform03}
\stressE_f^{l(n)} = \left(\frac{\HT}{a}\right)_f + \left(\frac{1}{a}\right)_f \left(\LT \dprod \stressE + \stressE \dprod \trans{\LT} - 2 G h_0 \DT \right)_f^{l(n)},
\end{equation}
where the terms $\left(\frac{\HT}{a}\right)_f$ and $\left(\frac{1}{a}\right)_f$ can be obtained by cell-face interpolation of the $P$ and $N$ cell-centroid expressions.
Recalling the definition $\LT = \trans{\grad \velocity} - {\zeta} \DT$ , we write (\ref{salgebraicform03}) in the form
\begin{equation}
\label{salgebraicform04}
\stressE_f^{l(n)} = \left(\frac{\HT}{a}\right)_f + \GammaT_{1,f}^{l(n)} \dprod \left(\nabla \velocity\right)_f^{l(n)} + \left(\trans{\nabla \velocity}\right)_f^{l(n)} \dprod \GammaT_{1,f}^{l(n)} - \GammaT_{2,f}^{l(n)} \dprod \left(\trans{\nabla \velocity}\right)_f^{l(n)} - \left(\nabla \velocity\right)_f^{l(n)} \dprod \GammaT_{2,f}^{l(n)},
\end{equation}
where the symmetric tensors $\GammaT_{1,f}^{l(n)}$ and $\GammaT_{2,f}^{l(n)}$ are defined as
\begin{align}
\label{coeffgamma01}
\GammaT_{1,f}^{l(n)} &:= \left(\frac{1}{a}\right)_f \left[\left(1 - \frac{\zeta}{2} \right) \stressE - h_0 G \IT \right]_f^{l(n)},\\
\label{coeffgamma02}
\GammaT_{2,f}^{l(n)} &:= \left(\frac{1}{a}\right)_f \left[\frac{\zeta}{2} \stressE \right]_f^{l(n)}.
\end{align}
Note that for affine motion $(\zeta = 0)$, we have
\begin{align}
\label{coeffgamma01b}
\GammaT_{1,f}^{l(n)} &:= \left(\frac{1}{a}\right)_f \left[\stressE - h_0 G \IT \right]_f^{l(n)},\\
\label{coeffgamma02b}
\GammaT_{2,f}^{l(n)} &:= \vec{0}.
\end{align}
Let us now consider the stress term in the momentum balance (\ref{momentumeqdiscr02}), viz.
\begin{equation}
\label{rci01}
\sum_f {\frac{1}{\rho}} \vec{s}_f \dprod \stressE_f^{l(n)}.
\end{equation}
The face-centroid stress $\stressE_f^{l(n)}$ could be evaluated by inserting (\ref{salgebraicform03}) or (\ref{salgebraicform04}), where all terms on the r.h.s.\ are obtained by cell-face interpolation of the corresponding $P$ and $N$ cell-centroid expressions. However, this is problematic, since cell-face interpolation of the cell-centroid velocity gradients generally leads to an extended stencil. To make this obvious, let us insert (\ref{salgebraicform04}) into (\ref{rci01}), viz.
\begin{equation}
\label{rci02}
{\frac{1}{\rho}} \sum_f \vec{s}_f \dprod \left(\frac{\HT}{a}\right)_f 
+ {\frac{1}{\rho}} \sum_f \vec{s}_f \dprod \left(\MT_{f}^{l(n)} + {\trans{\MT_{f}}}^{l(n)} \right) 
- {\frac{1}{\rho}} \sum_f \vec{s}_f \dprod \left(\NT_{f}^{l(n)} + {\trans{\NT_{f}}}^{l(n)} \right),
\end{equation}
where the tensors $\MT_{f}^{l(n)}$ and $\NT_{f}^{l(n)}$ are defined as
\begin{align}
\label{coeffM}
\MT_{f}^{l(n)} &:= \GammaT_{1,f}^{l(n)} \dprod \left(\nabla \velocity \right)_f^{l(n)},\\
\label{coeffN}
\NT_{f}^{l(n)} &:= \left(\nabla \velocity \right)_f^{l(n)} \dprod \GammaT_{2,f}^{l(n)}.
\end{align}
If we hypothetically assume that the cell-centroid gradients $\left(\nabla \velocity \right)_P^{l(n)}$ and $\left(\nabla \velocity \right)_N^{l(n)}$ were evaluated using Gauss' theorem, the linear cell-face interpolation of the cell-centroid gradients 
\begin{equation}
\label{lint02}
\overline{\left(\nabla \velocity \right)}_f^{l(n)} = \xi_f^{l(n)} \left(\nabla \velocity \right)_P^{l(n)} + \left(1 - \xi_f^{l(n)} \right) \left(\nabla \velocity \right)_N^{l(n)}
\end{equation}
would add the stencils of the cell-centroid gradients to yield a cell-face gradient extended stencil giving rise to checkerboard velocity fields. In order to avoid this, we use a special cell-face interpolation 
equivalent to the Rhie and Chow \cite{RhieChow1983}. The key aspect of the method is that the cell-face gradient along the $PN$ direction is discretized on a compact stencil in terms of the adjacent cell-centroid variables only. For this purpose, we write the discretization of the term $\sum_f \vec{s}_f \dprod \MT_{f}^{l(n)}$ in (\ref{rci02}) in the form
\begin{equation}
\label{rci03}
\sum_f \vec{s}_f \dprod \MT_{f}^{l(n)} = 
\sum_f \left[\vec{s}_f \dprod \GammaT_{1,f}^{l(n)} - \gamma _{1,f}^{l(n)} \vec{s}_f \right] \dprod \overline{\left(\nabla \velocity \right)}_f^{l(n)}
+
\sum_f
\gamma _{1,f}^{l(n)} \vec{s}_f \dprod \left(\nabla \velocity \right)_f^{l(n)},
\end{equation}
where $\gamma _{1,f}^{l(n)}$ is defined as
\begin{equation}
\label{rci03b}
\gamma _{1,f}^{l(n)} := \vec{n}_f \dprod \GammaT_{1,f}^{l(n)} \dprod \vec{n}_f
\end{equation}
and where the contribution $\vec{s}_f \dprod \left(\nabla \velocity \right)_f^{l(n)}$ in last term on the r.h.s.\ is discretized using (\ref{laplacian}) to give
\begin{equation}
\label{laplacian2}
\vec{s}_f \dprod \left(\nabla {\velocity}\right)_f^{l(n)}
= 
\underbrace{\frac{\left\vert \vec{s}_f \right\vert}{\vec{n}_f \dprod \vec{d}_f} \left(\velocity_N^{l(n)} - \velocity_P^{l(n)} \right)}_{\textnormal{orthogonal contribution}}
+ 
\underbrace{\left(\vec{s}_f - \frac{\left\vert \vec{s}_f \right\vert}{\vec{n}_f \dprod \vec{d}_f}\vec{d}_f \right) \dprod \overline{\left(\nabla \velocity \right)}_f^{l(n)}
}_{\textnormal{non-orthogonal correction}}.
\end{equation}
The orthogonal contribution in (\ref{laplacian2}) obviously reduces the stencil of the cell-face gradient along the $PN$ direction to the adjacent cell-centroids.
From (\ref{rci03}) - (\ref{laplacian2}) we obtain
\begin{equation}
\label{rci04}
\sum_f \vec{s}_f \dprod {\MT}_{f}^{l(n)}
=
\sum_f \vec{s}_f \dprod \overline{\MT}_{f}^{l(n)} 
+
\sum_f
\gamma _{1,f}^{l(n)} \left[\frac{\left\vert \vec{s}_f \right\vert}{\vec{n}_f \dprod \vec{d}_f} \left(\velocity_N^{l(n)} - \velocity_P^{l(n)} \right) - \frac{\left\vert \vec{s}_f \right\vert}{\vec{n}_f \dprod \vec{d}_f}\vec{d}_f \dprod \overline{\left(\nabla \velocity \right)}_f^{l(n)}\right],
\end{equation}
where the last term on the r.h.s.\ represents the correction for the linear cell-face interpolation of $\sum_f \vec{s}_f \dprod \overline{\MT}_{f}^{l(n)}$. Moreover, the orthogonal contribution in (\ref{rci04}) can be evaluated implicitly at the current time step by replacing $\velocity^{l(n)}$ with $\velocity^{n}$ to give
\begin{equation}
\label{rci04b}
\sum_f \vec{s}_f \dprod {\MT}_{f}^{n}
=
\sum_f \vec{s}_f \dprod \overline{\MT}_{f}^{l(n)} 
+
\sum_f
\gamma _{1,f}^{l(n)} \left[\frac{\left\vert \vec{s}_f \right\vert}{\vec{n}_f \dprod \vec{d}_f} \left(\velocity_N^{n} - \velocity_P^{n} \right) - \frac{\left\vert \vec{s}_f \right\vert}{\vec{n}_f \dprod \vec{d}_f}\vec{d}_f \dprod \overline{\left(\nabla \velocity \right)}_f^{l(n)}\right].
\end{equation}
Therefore, in (\ref{rci04b}), the cell-face gradient along the $PN$ direction depends on the cell-centroid velocities of the two adjacent cells to the face $f$ at the current time step, $\velocity_P^{n}$ and $\velocity_N^{n}$, respectively.
%%%%%%%%%%%%%%%%%%%%%%%%%%%%%%%%%%%%%%%%%%%%%%%%%%%%%%%%%%%%%%%%%%%%%%%%%%%%%%%%%
% Transpose term
%%%%%%%%%%%%%%%%%%%%%%%%%%%%%%%%%%%%%%%%%%%%%%%%%%%%%%%%%%%%%%%%%%%%%%%%%%%%%%%%%
Additionally, considering the transpose in (\ref{rci02}), we obtain
\begin{align}
\label{rci04c}
\sum_f \vec{s}_f \dprod {\trans{\MT_{f}}}^{l(n)}
&=
\sum_f \vec{s}_f \dprod \overline{\trans{\MT_{f}}}^{l(n)}
+
\sum_f \frac{\left\vert \vec{s}_f \right\vert}{\vec{n}_f \dprod \vec{d}_f} \gamma_{2,f}^{l(n)}
\left(
\GammaT_{1,f}^{l(n)} \dprod \vec{n}_f
\right),
%\\
%\label{rci04c}
%&+
%\sum_f
%\vec{n}_f \dprod
%\left\lbrace
%\left[
%\frac{\left\vert \vec{s}_f \right\vert}{\vec{n}_f \dprod \vec{d}_f}
%\left(
%\velocity_N^{l(n)} - \velocity_P^{l(n)}
%\right)
%-
%\frac{\left\vert \vec{s}_f \right\vert}{\vec{n}_f \dprod \vec{d}_f} \vec{d}_f
%\dprod \overline{\left(\nabla \velocity \right)}_f^{l(n)}
%\right]
%\otimes
%\left(
%\GammaT_{1,f}^{l(n)} \dprod \vec{n}_f
%\right)
%\right\rbrace.
%&-
%\vec{n}_f \dprod
%\left[
%\left(
%\frac{\left\vert \vec{s}_f \right\vert}{\vec{n}_f \dprod \vec{d}_f} \vec{d}_f
%\dprod \overline{\left(\nabla \velocity \right)}_f^{l(n)}
%\right)
%\otimes
%\left(
%\GammaT_{1,f}^{l(n)} \dprod \vec{n}_f
%\right)
%\right]
\end{align}
where the scalar coefficient $\gamma_{2,f}^{l(n)}$ in the correction term is defined as
\begin{equation}
\label{rci04d}
\gamma_{2,f}^{l(n)} :=
\vec{n}_f \dprod
\left[
%\frac{\left\vert \vec{s}_f \right\vert}{\vec{n}_f \dprod \vec{d}_f}
\left(
\velocity_N^{l(n)} - \velocity_P^{l(n)}
\right)
-
%\frac{\left\vert \vec{s}_f \right\vert}{\vec{n}_f \dprod \vec{d}_f} 
\vec{d}_f
\dprod \overline{\left(\nabla \velocity \right)}_f^{l(n)}
\right].
\end{equation}
Combining (\ref{rci04}) to (\ref{rci04d}), we write the discretization of the second term in (\ref{rci02}) as
\begin{equation}
\label{rci05a}
\sum_f \vec{s}_f \dprod \left({\MT}_{f}^{n} + {\trans{\MT_{f}}}^{l(n)}\right)
=
\sum_f \vec{s}_f \dprod \left(\overline{{\MT}_{f}}^{l(n)} + \overline{\trans{\MT_{f}}}^{l(n)}\right)
+
\sum_f
\frac{\left\vert \vec{s}_f \right\vert}{\vec{n}_f \dprod \vec{d}_f}
\vec{c}_{\MT, f}^{n}
+
\sum_f
\frac{\left\vert \vec{s}_f \right\vert}{\vec{n}_f \dprod \vec{d}_f}
\vec{c}_{\MT, f}^{l(n)},
\end{equation}
where the correction vectors for the linear cell-face interpolation $\vec{c}_{\MT, f}^{n}$ and $\vec{c}_{\MT, f}^{l(n)}$ are defined as:
\begin{itemize}
\item[(i)] \textit{semi-implicit} discretization
\begin{align}
\label{rci05b1}
\vec{c}_{\MT, f}^{n}
&:=
\gamma _{1,f}^{l(n)}
%\frac{\left\vert \vec{s}_f \right\vert}{\vec{n}_f \dprod \vec{d}_f} 
\left(\velocity_N^{n} - \velocity_P^{n} \right),\\
%%%%%%%%%%%%%%%%%%%%%%%%%%
\label{rci05b2}
\vec{c}_{\MT, f}^{l(n)}
&:=
%\frac{\left\vert \vec{s}_f \right\vert}{\vec{n}_f \dprod \vec{d}_f}
%\left(
\gamma _{2,f}^{l(n)}
\left(
\GammaT_{1,f}^{l(n)} \dprod \vec{n}_f
\right)
-
\gamma _{1,f}^{l(n)}
\left(
\vec{d}_f \dprod \overline{\left(\nabla \velocity \right)_f}^{l(n)}
\right),
%\right)
\end{align}
\item[(ii)] \textit{explicit} discretization
\begin{align}
\label{rci05b3}
\vec{c}_{\MT, f}^{n}
&:= \vec{0},\\
%%%%%%%%%%%%%%%%%%%%%%%%%%
\label{rci05b4}
\vec{c}_{\MT, f}^{l(n)}
&:=
%\frac{\left\vert \vec{s}_f \right\vert}{\vec{n}_f \dprod \vec{d}_f}
%\left(
\gamma _{1,f}^{l(n)}
\left[
\left(\velocity_N^{l(n)} - \velocity_P^{l(n)} \right)
-
\vec{d}_f \dprod \overline{\left(\nabla \velocity \right)_f}^{l(n)}
\right]
+
\gamma _{2,f}^{l(n)}
\left(
\GammaT_{1,f}^{l(n)} \dprod \vec{n}_f
\right).
%\right)
\end{align}
\end{itemize}
%%%%%%%%%%%%%%%%%%%%%%%%%%%%%%%%%%%%%%%%%%%%%%%%%
%%%%%%%%%%%%%%%%%%%%%%%%%%%%%%%%%%%%%%%%%%%%%%%%%
From similar analysis of the term $\sum_f \vec{s}_f \dprod {\NT}_{f}^{l(n)}$ in (\ref{rci02}), we obtain
\begin{equation}
\label{rci06}
\sum_f \vec{s}_f \dprod {\NT}_{f}^{l(n)}
= 
\sum_f \vec{s}_f \dprod \overline{{\NT}_{f}}^{l(n)}
+
\sum_f
\frac{\left\vert \vec{s}_f \right\vert}{\vec{n}_f \dprod \vec{d}_f}
\left[
\left(
\velocity_N^{l(n)} - \velocity_P^{l(n)}
\right)
-
\vec{d}_f
\dprod \overline{\left(\nabla \velocity \right)}_f^{l(n)}
\right]
\dprod
\GammaT_{2,f}^{l(n)}
%\sum_f \vec{s}_f \dprod {\NT}_{f}^{l(n)}
%=
%\sum_f \vec{s}_f \dprod \overline{\NT}_{f}^{l(n)} 
%+ 
%\gamma _{2,f}^{l(n)} \left[\frac{\left\vert \vec{s}_f \right\vert}{\vec{n}_f \dprod \vec{d}_f} \left(\velocity_N^{l(n)} - \velocity_P^{l(n)} \right) - \frac{\left\vert \vec{s}_f \right\vert}{\vec{n}_f \dprod \vec{d}_f}\vec{d}_f \dprod \overline{\left(\nabla \velocity \right)}_f^{l(n)}\right]
\end{equation}
and for $\sum_f \vec{s}_f \dprod {\trans{\NT_{f}}}^{l(n)}$, we have
\begin{equation}
\label{rci06b}
\sum_f \vec{s}_f \dprod {\trans{\NT_{f}}}^{l(n)}
=
\sum_f \vec{s}_f \dprod \overline{\trans{\NT_{f}}}^{l(n)}
+
\sum_f
\frac{\left\vert \vec{s}_f \right\vert}{\vec{n}_{f} \dprod \vec{d}_f}
\gamma_{3,f}^{l(n)} \vec{n}_f,
\end{equation}
where the scalar coefficient $\gamma_{3,f}^{l(n)}$ in the correction term is defined as
\begin{equation}
\label{rci06c}
\gamma_{3,f}^{l(n)}
:=
\vec{n}_{f}
\dprod
\left\lbrace
\left[
\left(
\velocity_N^{l(n)} - \velocity_P^{l(n)}
\right)
-
\vec{d}_f
\dprod \overline{\left(\nabla \velocity \right)}_f^{l(n)}
\right]
\dprod
\GammaT_{2,f}^{l(n)}
\right\rbrace.
\end{equation}
From (\ref{rci06}) - (\ref{rci06c}) the discretization of the second term in (\ref{rci02}) reads
\begin{equation}
\label{rci07}
\sum_f \vec{s}_f \dprod \left({\NT}_{f}^{n} + {\trans{\NT_{f}}}^{l(n)}\right)
=
\sum_f \vec{s}_f \dprod \left(\overline{{\NT}_{f}}^{l(n)} + \overline{\trans{\NT_{f}}}^{l(n)}\right)
+
\sum_f
\frac{\left\vert \vec{s}_f \right\vert}{\vec{n}_f \dprod \vec{d}_f}
\vec{c}_{\NT, f}^{l(n)},
\end{equation}
where the correction vector for the linear cell-face interpolation $\vec{c}_{\NT, f}^{l(n)}$ is defined as
\begin{equation}
\label{rci07b}
\vec{c}_{\NT, f}^{l(n)}
:=
\gamma_{3,f}^{l(n)} \vec{n}_f +
\left[
\left(
\velocity_N^{l(n)} - \velocity_P^{l(n)}
\right)
-
\vec{d}_f
\dprod \overline{\left(\nabla \velocity \right)}_f^{l(n)}
\right]
\dprod
\GammaT_{2,f}^{l(n)}.
\end{equation}
%%%%%%%%%%%%%%%%%%%%%%%%%%%%%%%%%%%%%%%%%%%%%%%%%%%%%%%%%%%%%%%%%%%%%%%%%%%%%%%%
Finally, from substituting (\ref{rci05a}) and (\ref{rci07}) into (\ref{rci02}), we obtain
\begin{equation}
\label{rci08}
{\frac{1}{\rho}}
\sum_f
\vec{s}_f \dprod {\stressE}_f^{n}
=
{\frac{1}{\rho}}
\sum_f
\vec{s}_f \dprod \overline{\stressE}_f^{l(n)}
+
{\frac{1}{\rho}}
\sum_f
\frac{\left\vert \vec{s}_f \right\vert}{\vec{n}_f \dprod \vec{d}_f}
\vec{c}_{\MT, f}^{n}
+
{\frac{1}{\rho}}
\sum_f
\frac{\left\vert \vec{s}_f \right\vert}{\vec{n}_f \dprod \vec{d}_f}
\left(
\vec{c}_{\MT, f}^{l(n)}
+
\vec{c}_{\NT, f}^{l(n)}
\right).
\end{equation}
The last two terms on the r.h.s\ of (\ref{rci08}) remove the velocity-stress checkerboarding of the linear cell-face interpolation.
%%%%%%%%%%%%%%%%%%%%%%%%%%%%%%%%%%
When (\ref{rci05b1}) is used to compute $\vec{c}_{\MT, f}^{n}$, we call (\ref{rci08}) the finite-volume implicit stress interpolation correction (FV-ISIC). The discretization of the correction term $\vec{c}_{\MT, f}^{n}$ formally equals that of an implicit Laplace operator with scalar diffusion coefficients $\gamma_{1,f}^{l(n)}$. We emphasize that other than in the BSD, the implicit diffusion term in the FV-ISIC is not artificial. It has been derived from the FV-discretized form of the equations. The FV-ISIC (\ref{rci05b1}) and the explicit discretization (\ref{rci05b3}) should give the same result, if the system of equations is solved multiple times within each time step. However, since the implicit term (\ref{rci05b1}) gives rise to the algebraic matrix coefficients of the momentum equation, the convergence properties of the iterative algorithm are influenced by the FV-ISIC. Depending on the local values of $\gamma_{1,f}^{l(n)}$, the implicit Laplace operator potentially increases the convergence, similar to the BSD.

%%%%%%%%%%%%%%%%%%%%%%%%%%%%%%%%%%%%%%%%%%%%%%%%%%%%%%%%%%%%%%%%%%%%%%%%%%%%%%%%%%%%%%%%%%%%%%%%%%%%%%%%
\subsection{Solution algorithm}
\label{subsec:solutionprocedure}
The pressure-velocity coupled system is solved using a segregated approach, based on the iterative pressure-implicit with splitting of operators (PISO) algorithm \cite{Issa1986}. The PISO algorithm involves an iterative pressure correction strategy, where the pressure equation (\ref{contidiscretization02}) is solved implicitly, and then
the velocity is corrected with the new pressure at time-level $n$, according to
\begin{align}
\label{corrector01}
\velocity_{P}^{n} = \frac{\vec{h}_P}{a_P} + \frac{1}{a_P} \left(\nabla p^{*} \right)_P^{n}.
\end{align}
Without modifying the iterative PISO corrector strategy for the pressure-velocity coupling problem, let us establish an extended segregated approach to incorporate the inter-equation coupling between the velocity and the stress. In this procedure, the PISO algorithm and the solution of the constitutive equations are performed multiple times within each time step. This iterative strategy allows all lagged terms, such as the inter-equation coupling terms and the non-linear advection terms, to converge successively to the solution of the new time step $n$, as the number of outer iterations within each time step is increased. We use three of such iterations per time step to obtain good accuracy in our transient flow calculations. The solution procedure can be summarized in 4 steps:
\begin{itemize}
\item[1.] \textit{Initialization.} For given initial fields of $p^{l(\beta)}$, $\velocity^{l(\beta)}$, $\stressP^{l(\beta)}$ and the generic tensor transport variable $\GT^{l(\beta)}$, compute a cell-centroid velocity estimate $\frac{\vec{h}_{P}}{a_{P}}$ using (\ref{algebraicform01}) and (\ref{vel02}) without considering the pressure gradient.
\item[2.] \textit{PISO algorithm.} With $\frac{\vec{h}_{P}}{a_{P}}$, compute the new pressure ${p^*_P}^n$ by solving the pressure-equation (\ref{contidiscretization02}) implicitly and, subsequently, correct the face-fluxes according to (\ref{contidiscretization03}). Update the velocity as (\ref{corrector01}). Repeat the pressure-velocity correction step sequentially to improve the accuracy of the transient solution. For this, use the new velocity as an estimate to update the lagged terms in (\ref{contidiscretization02}).
\item[3.] \textit{Constitutive equations.} Assemble the constitutive equations (\ref{constdiscretization01}), using the new velocity $\velocity_{P}^{n}$. Compute the new constitutive transport variable $\GT^n$ by solving the constitutive equations implicitly and, subsequently, update the stress $\stressP^n$ by transforming the new transport variable $\GT^n$ back. A detailed description of the procedure for step 3 is given below.
\item[4.] Repeat steps 1, 2 and 3 within each time step to increase the accuracy of the transient solution. For this, the new ${p^*_P}^n$, $\velocity_{P}^{n}$, $\stressP^n$ and $\GT^n$ become ${p^*_P}^{l(n)}$, $\velocity_{P}^{l(n)}$, $\stressP^{l(n)}$ and $\GT^{l(n)}$, respectively. Use the latter to update all lagged terms in (\ref{contidiscretization02}), (\ref{contidiscretization03}), (\ref{corrector01}), and (\ref{constdiscretization01}).
\end{itemize}
Let us now consider the detailed procedure for assembling and solving the constitutive equations (step 3). In the extended segregated approach, the constitutive equations are assembled and solved using the novel stabilization framework (stabLib). This generic numerical framework provides a runtime selective combinatorial flexibility between different kinds of rhelogical models on the one hand, and the previously derived stabilization methods on the other hand. The generic procedure for assembling and solving the constitutive equations can be summarized in 4 steps:
\begin{itemize}
%%%%%%%%%%
\item[3.1.] \textit{Construction of the generic tensor transport variable $\GT_{k} := \cvec{F}(\Ck)$.} First, transform the modal polymer stress tensor $\stressPk^{l(n)}$ using (\ref{conversion}) according to
\begin{align}
\label{transformC}
\Ck^{l(n)} = \frac{{\tau_{1,k}} (1 - {\zeta}_k)}{{\eta_{P,k}} h_{1,k}} \stressPk^{l(n)} - \frac{h_{0,k}}{h_{1,k}} \IT.
\end{align}
Note that (\ref{transformC}) contains the model-dependent scalar parameters $\tau_{1,k}$ and ${\zeta}_k$, as well as ${\eta_{P,k}}$, $h_{0,k}$ and $h_{1,k}$ which may be either constant scalars or non-constant scalar functions in the domain\footnote
{
For the special case that $h_{1,k} = \hat{h}_{1,k}(\tr \Ck^{l(n)})$ (e.g.\ FENE-P models), transform the stress to the tensor ${\Ck^{*}}^{l(n)} = h_{1,k}\Ck^{l(n)}$ according to
\begin{align}
\nonumber
\label{transformCb}
{\Ck^{*}}^{l(n)} = \frac{{\tau_{1,k}} (1 - {\zeta}_k)}{{\eta_{P,k}}} \stressPk^{l(n)} - {h_{0,k}} \IT.
\end{align}
}.
Next, diagonalize the conformation tensor and compute the generic tensor transport variable according to
\begin{align}
\GT_{k}^{l(n)} = \QT_{k}^{l(n)} \dprod \cvec{F}(\LambdaT_{k}^{l(n)}) \dprod {\trans{\QT_{k}}}^{l(n)}.
\end{align}
Keep the diagonal tensors $\LambdaT_{k}$ and $\QT_{k}$ in memory in order to compute the decomposition of $\LT$.
%%%%%%%%%%
\item[3.2.] \textit{Decomposition of $\LT$ and rheological model $\cvec{P}(\GT_{k})$}. In a single loop over all cells, compute the r.h.s.\ of (\ref{evolutionConf3}), which includes the decomposition of the tensor $\LT$ into $\BT$ and $\OmegaT$, as well as the model specific symmetric tensor function $\cvec{P}(\GT_{k})$. Note that we use (\ref{evolutionConfRTb}) as RCR specialization and (\ref{evolutionConfSTb}) as LCR specialization. The scalar coefficients $g_{0,k}$, $g_{1,k}$ and $g_{2,k}$ are provided by a class that contains all rheological models according to table \ref{tab:constitutiveEx}. In most cases, the result is an explicit source term $\cvec{S}_{\GT_{k}}$ that is used to construct the linear algebraic equation (\ref{algebraicform02})-(\ref{coeffan02b}). However, a more implicit discretization of $\cvec{S}_{\GT_{k}}$ may additionally contribute to the matrix coefficients in (\ref{algebraicform02}), but this has not been considered in the present work. After the computation of the source term $\cvec{S}_{\GT_{k}}$, clear the tensors $\LambdaT_{k}$ and $\QT_{k}$ from memory.
\item[3.3.] \textit{Solution of the generic transport equations.}
Solve the generic transport equations for each tensor component semi-implicitly with an indirect iterative method. Note that in the segregated approach all coupling terms are treated explicitly that is why we use more than one outer iteration per time step where we update the lagged coupling terms (see step 4). 
\item[3.4.] \textit{Back-transformation of the generic tensor transport variable $\GT_{k}$.}
Transform the generic transport variable $\GT_{k}$ back to the polymer stress tensor $\stressPk$. For the most general case perform a diagonalization of $\GT_{k}$ and apply transformation functions to the eigen-decomposed representation to obtain $\Ck$. Use (\ref{conversion}) to convert $\Ck$ to $\stressPk$. For some specific cases, e.g. the RCR, the diagonalization can optionally be avoided which might reduce the computational costs. Specializations for RCR are implemented in a run-time type selection mechanism, thus leaving the choice between the general case and some representation specific cases to the user.
\end{itemize}
%%%%%%%%%%%%%%%%%%%%%%%%%%%%%%%%%%%%%%%%%%%%

\section{Numerical studies and discussion}
\label{sec:results}
Viscoelastic entry flows through contractions with sharp re-entrant corners are a challenging subject for all numerical methods due to the complex flow field of locally high deformation rates and the presence of geometric singularities. The viscoelastic entry flow through a sudden planar contraction with a contraction ratio of 4:1 is widely used for benchmarking numerical methods; cf.\ the literature reviews \cite{White1987, Owens2002}. We restrict attention to the flow of Oldroyd-B and PTT fluids in the planar 4:1 contraction benchmark for which quantitative numerical results have been provided in \cite{Phillips1999, Aboubacar2001, Alves2003, Edussuriya2004, Kim2005}, using the stress tensor representation (STR). It has been shown that for Oldroyd-B fluids the STR becomes unstable at $\operatorname{Wi} \geq 3$ \cite{Alves2003} and, therefore, mesh-converged solutions at higher $\operatorname{Wi}$ are missing in literature. Nevertheless, change-of-variable representations effectively alleviate the HWNP; some results at higher Weissenberg numbers have been reported more recently \cite{Afonso2011, Comminal2016, Pimenta2017}, using the natural logarithm conformation tensor representation (LCRN). Yet, the question about mesh-convergence of the change-of-variable representations at higher $\operatorname{Wi}$ remains open. In order to demonstrate the mesh-independence of the predictions, several characteristic quantities like the vortex size, the vortex intensity and the stress profiles need to be studied. Regarding the vortex size, the present work confirms that in a very recent work \cite{Pimenta2017} significantly more accurate results were presented as in previous references \cite{Afonso2011, Comminal2016}.
Since in the literature mentioned above only the LCRN was studied, a quantitative verification whether the results of different change-of-variable representations converge to the same solution as the computational grid is refined is a matter of considerable significance in this regard. This has so far not been studied in detail. Indeed, the sensitivity to the spatial resolution varies with the applied representation, which can be easily shown on under-resolved computational grids, where the predictions of different change-of-variable representations differ significantly from each other. Having distinct numerical methods available on the same computational platform, we demonstrate that the stabLib is a powerful tool to investigate the impact of a set of change-of-variable representations on both, mesh-convergence and mesh-sensitivity over a wide range of Weissenberg numbers, thus benchmarking the numerical code and yielding new understanding of viscoelastic flows through planar 4:1 contractions.

The objective of the present section is to provide a detailed quantitative study of the numerical methods described above regarding convergence, mesh-sensitivity and accuracy in the planar 4:1 contraction benchmark problem for moderate and high Weissenberg number flows. We examine the effect of the constitutive equation representation on the corner vortex size, its intensity and the axial profiles of the normal stress differences, by comparing the results of various RCR and LCR representations on a set of six different computational grids. Moreover, we demonstrate the validity of the model-independent framework by performing computations with both Oldroyd-B fluids and exponential PTT fluids. The range of Weissenberg numbers of Oldroyd-B fluids is divided into three parts: for small $\operatorname{Wi}$ up to three, we compare the results to literature data that has been provided using the STR. For $\operatorname{Wi}$ up to six we study the mesh-dependence of the change-of-variable representations in detail, using a series of successively refined orthogonal, non-uniform computational grids with local refinement near the re-entrant corner. Computations on different computational grids reveal that the spatial resolution of the boundary stress near the re-entrant corner wall is crucial to obtain accurate solutions.
Furthermore, we present results on a non-orthogonal and non-uniform mesh with even smaller cell spacing. Finally, for very high $\operatorname{Wi}$ up to twelve, we study the mesh-sensitivity of the methods on the four coarser meshes, without raising a claim to reach mesh-convergent solutions. The reason for this is that the flow becomes more and more unsteady as the Weissenberg number is increased. Similar observations have been previously reported \cite{Afonso2011, Comminal2016, Pimenta2017}. Therefore, the results are obtained from a time-averaged flow field. The time-interval for computing the average flow field is at least twenty times the relaxation time of the fluid, starting from a point in time where the velocity and stress profiles are already developed. For this reason, the computational costs increase dramatically with increasing $\operatorname{Wi}$. On the other hand, the results suggest that the sensitivity to the spatial resolution increases significantly, too. In order to obtain a convergent solution for highly-elastic flows one would need both, a much finer spatial resolution as our finest mesh and long-time averaged flow fields, which is computational expensive and goes beyond the aim of the present work.

%%%%%%%%%%%%%%%%%%%%%%%%%%%%%%%%%%%%%%%%%%%%
\subsection{Problem specification}
%%%%%%%%%%%%%%%%%%%%%%%%%%%%%%%%%%%%%%%%%%%%%%%%%%%%%%%%%%%%%%%%
%%%%%%%%%%%%%%%%%%%%%%%%%%%%%%%%%%%%%%%%%%%%%%%%%%%%%%%%%%%%%%%%
To characterize the flow in the planar 4:1 contraction benchmark, let us first introduce the dimensionless ratio $\beta$ between the solvent viscosity $\eta_S$ and the total viscosity $\eta_0 = \eta_S + \eta_P$, defined by
\begin{equation}
\beta = \frac{\eta_S}{\eta_0}.
\end{equation}
The viscosity ratio is kept constant. For the Oldroyd-B fluid we set $\beta = 1/9$. According to the benchmark convention, we take the half-width $L$ of the small downstream channel as the characteristic length scale and the mean velocity $U$ of the downstream channel as the characteristic velocity. To quantify the dimensionless ratio of elastic forces to viscous forces, let us introduce the Weissenberg number $\operatorname{Wi}$, according to
\begin{equation}
\operatorname{Wi} = \frac{\tau_1 U}{L}.
\end{equation}
For the Oldroyd-B fluid, $\operatorname{Wi}$ is varied from $0.1$ to $12$. The PTT fluid is studied for Weissenberg numbers between $0.1$ and $10000$. Furthermore, let us define the Reynolds number $\operatorname{Re}$ as the dimensionless ratio of momentum forces to viscous forces, given by
\begin{equation}
\label{defre}
\operatorname{Re} = \frac{\rho U L}{\eta_0}.
\end{equation}
In this work, $\operatorname{Re}$ is kept constant at $0.01$. Note that we do not neglect the advection term in the momentum equation, as has been frequently done, e.g. in \cite{Aboubacar2001,Alves2003,Afonso2011}. This assumption of creeping flow is not rigorous, especially at higher $\operatorname{Wi}$, because of the following reason: in general, the flow of viscoelastic fluids is superimposed by local flow structures, formed by time-dependent elastic effects; often referred to as purely elastic instabilities. With increasing $\operatorname{Wi}$ the impact of such instabilities on the flow becomes more and more pronounced. This has been observed in experiments, e.g.\ \cite{McKinley1991, Rodd2005}. In numerical simulations, the local advective transport of such flow structures through the domain is crucial. Thus, the advection term should in general not be neglected, even though the Reynolds number (\ref{defre}) of the flow problem might be rather small.

\begin{figure}[h!]
\centering
%%%%%%%%%%%%%%%%%%%%%%%%%%%%%%%%
\begin{subfigure}{0.333\linewidth-9pt}
\centering
%%%%%%%%%%%%%%%%%%%%%%%%%%%%%%%%
    \begin{tikzpicture}
        \node[inner sep=0pt] (BLVC) at (0,0)
        {
            \includegraphics[width=.99\textwidth]{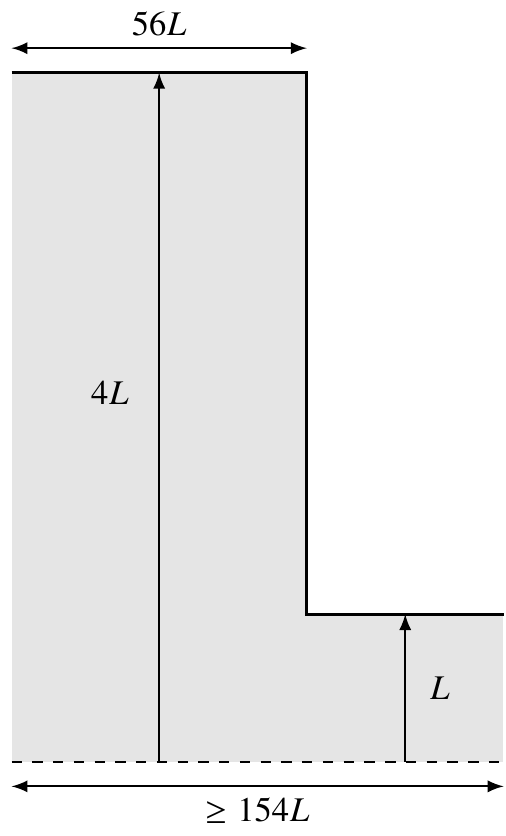}
        };
%        (0,0) at (-1.25,1.0){(a)};
    \end{tikzpicture}
%%%%%%%%%%%%%%%%%%%%%%%%%%%%%%%%
\end{subfigure}
%%%%%%%%%%%%%%%%%%%%%%%%%%%%%%%%
%%%%%%%%%%%%%%%%%%%%
\hfill
%%%%%%%%%%%%%%%%%%%%
%%%%%%%%%%%%%%%%%%%%%%%%%%%%%%%%
\begin{subfigure}{0.333\linewidth-9pt}
\centering
%%%%%%%%%%%%%%%%%%%%%%%%%%%%%%%%
    \begin{tikzpicture}
        \node[inner sep=0pt] (BLVC) at (0,0)
        {
            \includegraphics[width=.99\textwidth]{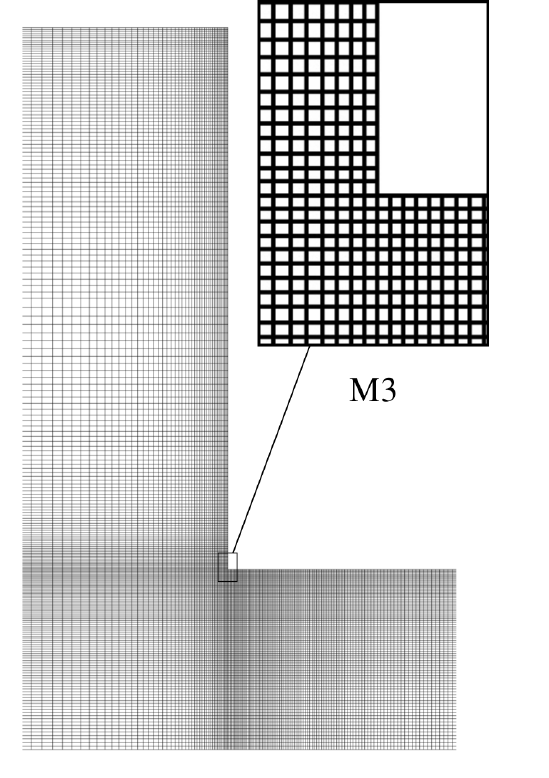}
        };
%        (0,0) at (2.65,-1.0){$\operatorname{Wi} = 0.5$};
    \end{tikzpicture}
%%%%%%%%%%%%%%%%%%%%%%%%%%%%%%%%
\end{subfigure}
%%%%%%%%%%%%%%%%%%%%%%%%%%%%%%%%
%%%%%%%%%%%%%%%%%%%%%%%%%%%%%%%%
%%%%%%%%%%%%%%%%%%%%
\hfill
%%%%%%%%%%%%%%%%%%%%
%%%%%%%%%%%%%%%%%%%%%%%%%%%%%%%%
\begin{subfigure}{0.333\linewidth-9pt}
\centering
%%%%%%%%%%%%%%%%%%%%%%%%%%%%%%%%
    \begin{tikzpicture}
        \node[inner sep=0pt] (BLVC) at (0,0)
        {
            \includegraphics[width=.99\textwidth]{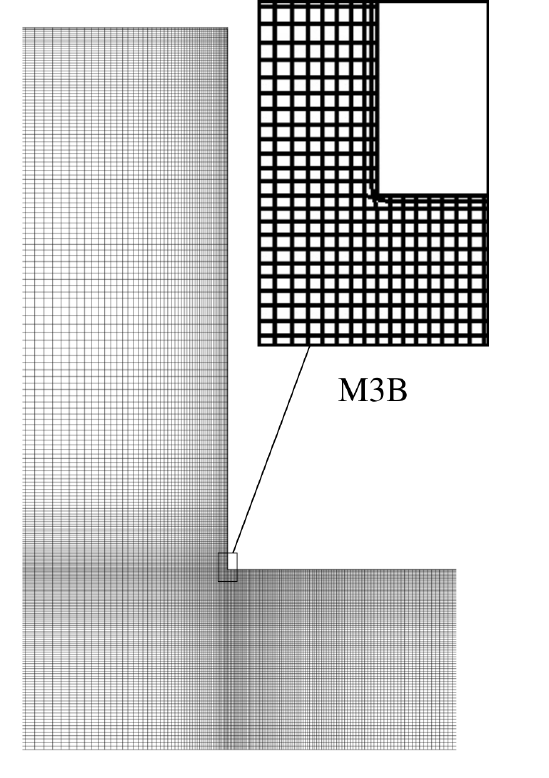}
        };
%        (0,0) at (2.65,-1.0){$\operatorname{Wi} = 0.5$};
    \end{tikzpicture}
%%%%%%%%%%%%%%%%%%%%%%%%%%%%%%%%
\end{subfigure}
%%%%%%%%%%%%%%%%%%%%%%%%%%%%%%%%
%%%%%%%%%%%%%%%%%%%%
%%%%%%%%%%%%%%%%%%%%%%%%%%%%%%%%
    \caption{Geometry and detail of the meshes M3 and M3B.}
    \label{fig:geometryandmesh}
\end{figure}

We consider the entire computational domain of a planar 4:1 contraction without imposing any symmetry conditions. This is due to the unsteady nature of the flow at higher Weissenberg numbers, which causes at least temporarily asymmetric flow fields. 
The total length of the computational domain is at least $154 L$ and the length of the inlet channel is $56 L$; cf. Figure \ref{fig:geometryandmesh}. These dimensions have been considered as adequate for capturing the elastic flow fields and to obtain accurate vortex data in previous studies \cite{Alves2003}.
%%%%%%%%%%%%%%%%%%%%%%%%%%%%%%%%%%%%%%%%%%%%%%%%%%%%%%%%%%%%%%%%
%%%%%%%%%%%%%%%%%%%%%%%%%%%%%%%%%%%%%%%%%%%%%%%%%%%%%%%%%%%%%%%%

%%%%%%%%%%%%%%%%%%%%%%%%%%%%%%%%%%%%%%%%%%%%%%%%%%%%%%%%%%%%%%%%
As described in the FV discretization in section \ref{Numerical Method}, the computational domain is subdivided into a set of convex, non-overlapping CVs, that fill the domain completely. All computational grids are non-uniform with local refinement near the re-entrant corner. A detail of the medium refined mesh M3 near the re-entrant corner is given in Figure \ref{fig:geometryandmesh}. It is part of the series of successively refined orthogonal meshes M1 to M5; the mesh data is given in Table \ref{table:meshdata}. Moreover, we use a non-orthogonal mesh M3B where an additional cell boundary layer is inserted next to the channel walls. Mesh M3B has a smaller cell spacing as M5 next to the wall, but only slightly more cells than M3. On this mesh, we study the impact of non-orthogonal CVs on the result and the effect if only the spatial resolution in the wall boundary layer is increased, but not in the entire domain.
%%%%%%%%%%%%%%%%%%%%%%%%%%%%%%%%%%%%%%%%%%%%%%%%%%%%%%%%%%%%%%%%
\begin{table}[htbp]
\begin{center}
\begin{tabular}{@{}lrrr@{}}
\toprule
Mesh \quad & \quad CVs  \quad & \quad DoF  \quad &  \quad $\vert \vec{s}_{f,\: \operatorname{min}} \vert/L^2 \times 10^5$ \\ 
\midrule 
M1 & $29316$ & 175896 & 14.3178 \\ 
M2 & $52224$ & 313344 & 7.8395 \\ 
M3 & $118740$ & 712440 & 3.6057 \\ 
M4 & $210224$ & 1261344 & 2.0403 \\ 
M5 & $396550$ & 2379300 & 1.3001 \\ 
M3B & $131376$ & 788256 & 1.0738 \\
\bottomrule
\end{tabular}
\end{center}
\caption{Mesh characteristics: Number of CVs, degrees of freedom (DoF), minimum face area $\vert \vec{s}_{f,\: \operatorname{min}} \vert/L^2 \times 10^5$.}
\label{table:meshdata}
\end{table}

We prescribe no-slip boundary conditions on the channel walls and a constant velocity at the inlet. On the outlet, we prescribe a mixed boundary condition which is locally of Neumann-type if the outflux through a cell-face is positive, or of Dirichlet-type with a zero velocity vector if the cell-face flux is directed into the domain. We specify a constant pressure at the outlet and zero normal pressure gradients at all the other boundaries. Since the domain length of the inflow channel is considered long enough to obtain fully developed profiles prior to the entry flow region, it is sufficient to impose a zero polymer stress tensor on the inlet. Therefore, we impose a constant Diriclet boundary condition for the polymer stress on the inlet and Neumann boundary conditions on the remaining boundaries. We emphasize that in the change-of-variable representations the constitutive working variable is not the stress, but the generic tensor $\GT$. Thus, it is important to specify boundary conditions for $\GT$, too. Since $\stressP$ and $\GT$ are directly related to each other, we use similar boundary conditions for both fields. On all Dirichlet boundaries, we transform the profiles of $\stressP$ to $\GT$ and vice versa, to ensure that the boundary profiles are always consistent to each other.

The discretized systems of linear equations are solved by indirect iterative methods. A conjugate gradient (CG) method with algebraic multigrid (AMG) preconditioning is used for the pressure. A bi-conjugate gradient stabilized (BiCGstab) method with incomplete lower-upper (ILU) preconditioning is used for the stress.

%%%%%%%%%%%%%%%%%%%%%%%%%%%%%%%%%%%%%%%%%%%%%%%%%%%%%%%%%%%%%%%%
%%%%%%%%%%%%%%%%%%%%%%%%%%%%%%%%%%%%%%%%%%%%%%%%%%%%%%%%%%%%%%%%
\subsection{Results for small Weissenberg numbers}
\begin{figure}[h!]
\centering
%%%%%%%%%%%%%%%%%%%%%%%%%%%%%%%%
\begin{subfigure}{202pt}
\centering
%%%%%%%%%%%%%%%%%%%%%%%%%%%%%%%%
    \begin{tikzpicture}
        \node[inner sep=0pt] (BLVC) at (0,0)
        {
            \includegraphics[width=.96\textwidth]{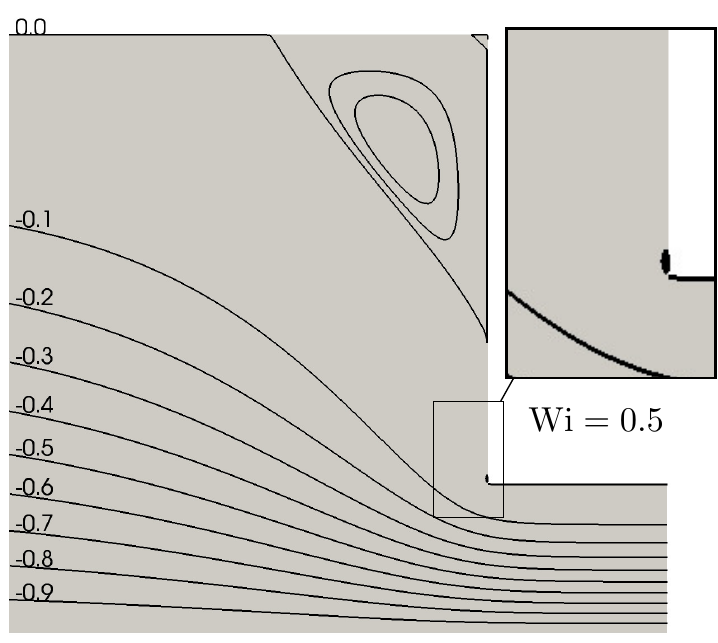}
        };
    \end{tikzpicture}
%%%%%%%%%%%%%%%%%%%%%%%%%%%%%%%%
\end{subfigure}
%%%%%%%%%%%%%%%%%%%%%%%%%%%%%%%%
%%%%%%%%%%%%%%%%%%%%
\quad
%%%%%%%%%%%%%%%%%%%%
%%%%%%%%%%%%%%%%%%%%%%%%%%%%%%%%
\begin{subfigure}{202pt}
\centering
%%%%%%%%%%%%%%%%%%%%%%%%%%%%%%%%
    \begin{tikzpicture}
        \node[inner sep=0pt] (BLVC) at (0,0)
        {
            \includegraphics[width=.96\textwidth]{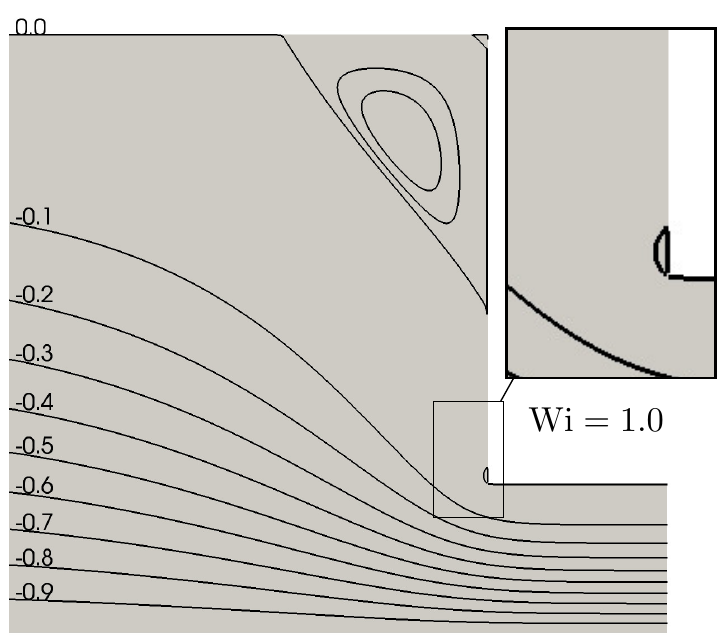}
        };
    \end{tikzpicture}
%%%%%%%%%%%%%%%%%%%%%%%%%%%%%%%%
\end{subfigure}
%%%%%%%%%%%%%%%%%%%%%%%%%%%%%%%%
%%%%%%%%%%%%%%%%%%%%
\\
%%%%%%%%%%%%%%%%%%%%
%%%%%%%%%%%%%%%%%%%%%%%%%%%%%%%%
\begin{subfigure}{202pt}
\centering
%%%%%%%%%%%%%%%%%%%%%%%%%%%%%%%%
    \begin{tikzpicture}
        \node[inner sep=0pt] (BLVC) at (0,0)
        {
            \includegraphics[width=.96\textwidth]{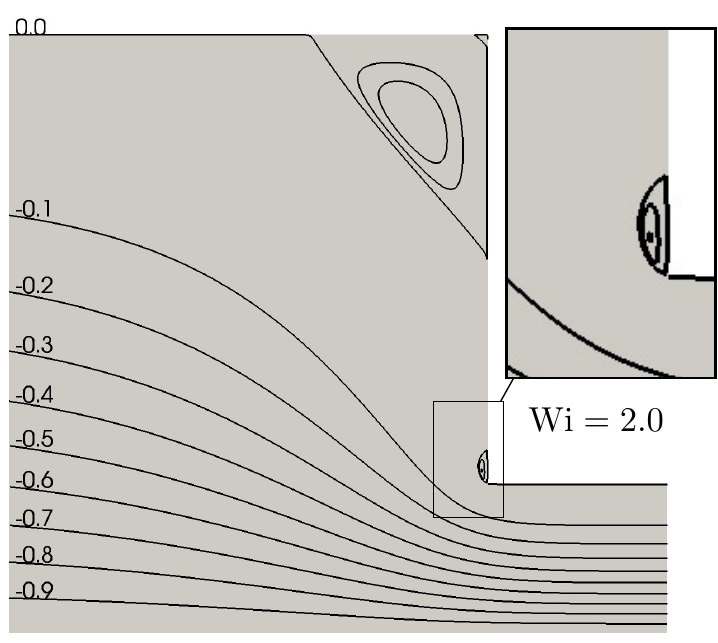}
        };
    \end{tikzpicture}
%%%%%%%%%%%%%%%%%%%%%%%%%%%%%%%%
\end{subfigure}
%%%%%%%%%%%%%%%%%%%%%%%%%%%%%%%%
%%%%%%%%%%%%%%%%%%%%
\quad
%%%%%%%%%%%%%%%%%%%%
%%%%%%%%%%%%%%%%%%%%%%%%%%%%%%%%
\begin{subfigure}{202pt}
\centering
%%%%%%%%%%%%%%%%%%%%%%%%%%%%%%%%
    \begin{tikzpicture}
        \node[inner sep=0pt] (BLVC) at (0,0)
        {
            \includegraphics[width=.96\textwidth]{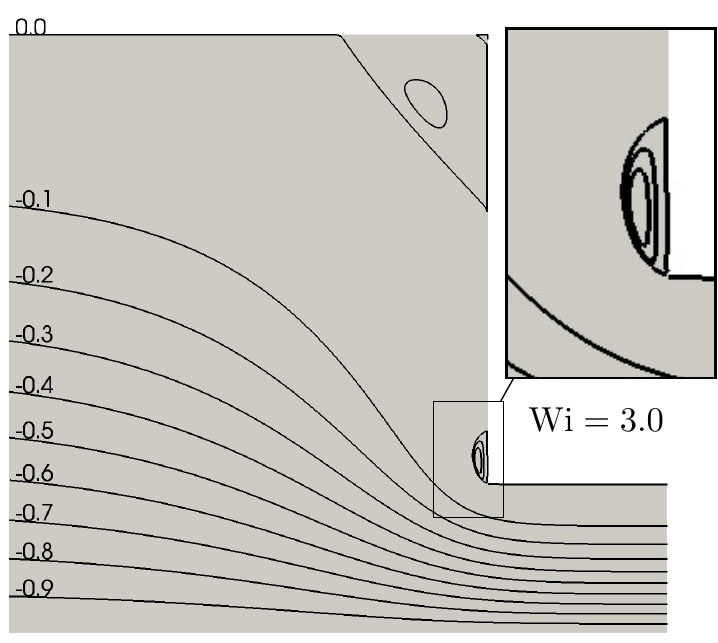}
        };
    \end{tikzpicture}
%%%%%%%%%%%%%%%%%%%%%%%%%%%%%%%%
\end{subfigure}
%%%%%%%%%%%%%%%%%%%%%%%%%%%%%%%%
    \caption{Flow patterns of an Oldroyd-B fluid in a planar 4:1 contraction up to $\operatorname{Wi} = 3$, carried out with the natural logarithm conformation representation. The streamlines are normalized, such that the zero line represents the separatrix between the recirculation vortices and the internal flow field.}
    \label{fig:contoursWi3}
\end{figure}

%%%%%%%%%%%%%%%%%%%%%%%%%%%%%%%%%%%%%%%%%%%%%%%%%%%%%%%%%%%%%%%%
%%%%%%%%%%%%%%%%%%%%%%%%%%%%%%%%%%%%%%%%%%%%%%%%%%%%%%%%%%%%%%%%
In Figure \ref{fig:contoursWi3}, we first present the streamlines to qualitatively illustrate the flow structures of an Oldroyd-B fluid for Weissenberg numbers from $0.5$ to $3$. A recirculating secondary flow develops at the outer corner of the large channel. The streamlines are normalized, such that the zero level represents the separatrix between the corner vortex and the internal flow field. The size of this corner vortex is characterized by a dimensionless reattachment length $\operatorname{L}_c = l_c/{L}$, where $l_c$ is the upstream horizontal extent of the corner vortex. Additionally, a smaller vortex is formed upstream of the re-entrant corner, where the fluid enters into the smaller cross-section. The appearance of this lip vortex is a characteristic flow phenomenon of elastic fluids and cannot be observed when the flow is completely dominated by viscous effects at Weissenberg numbers tending to zero. As the elasticity is increased, the elastic lip vortex expands in upstream direction while the corner vortex decreases in size. The lip vortex is spatially fully separated from the corner vortex, which indicates that the flow field is well-resolved. The prediction of the two vortices is highly sensitive to both, the numerical discretization scheme used for the advection term in the constitutive equation and the spatial resolution of the computational grid. These two influencing parameters need to be evaluated in order to obtain reliable results, because if one or both of them are chosen improperly, the separatrix might not converge entirely to the vertical channel wall such that the two vortices might erroneously be connected with each other along the near-wall region. We have observed such inaccuracies in the flow structure, e.g.\ when a first-order upwind scheme was used. Prior to the benchmark simulations, we have evaluated several high-resolution schemes for the advection term, namely, the Gamma scheme, the CUBISTA scheme and the MINMOD limiter. Best accuracy and convergence was obtained with the second-order accurate Gamma scheme, using the blending constant $\beta_m = 1/10$. The Gamma scheme is used throughout all simulations presented in this work.
%%%%%%%%%%%%%%%%%%%%%%%%%%%%%%%%%%%%%%%%%%%%%%%%%%%%%%%%%%%%%%%%
%%%%%%%%%%%%%%%%%%%%%%%%%%%%%%%%%%%%%%%%%%%%%%%%%%%%%%%%%%%%%%%%

\begin{figure}[h!]
\centering
    \begin{tikzpicture}[]
        \node[inner sep=0pt] (BLVC) at (0,0)
        {
            \includegraphics[width=265pt]{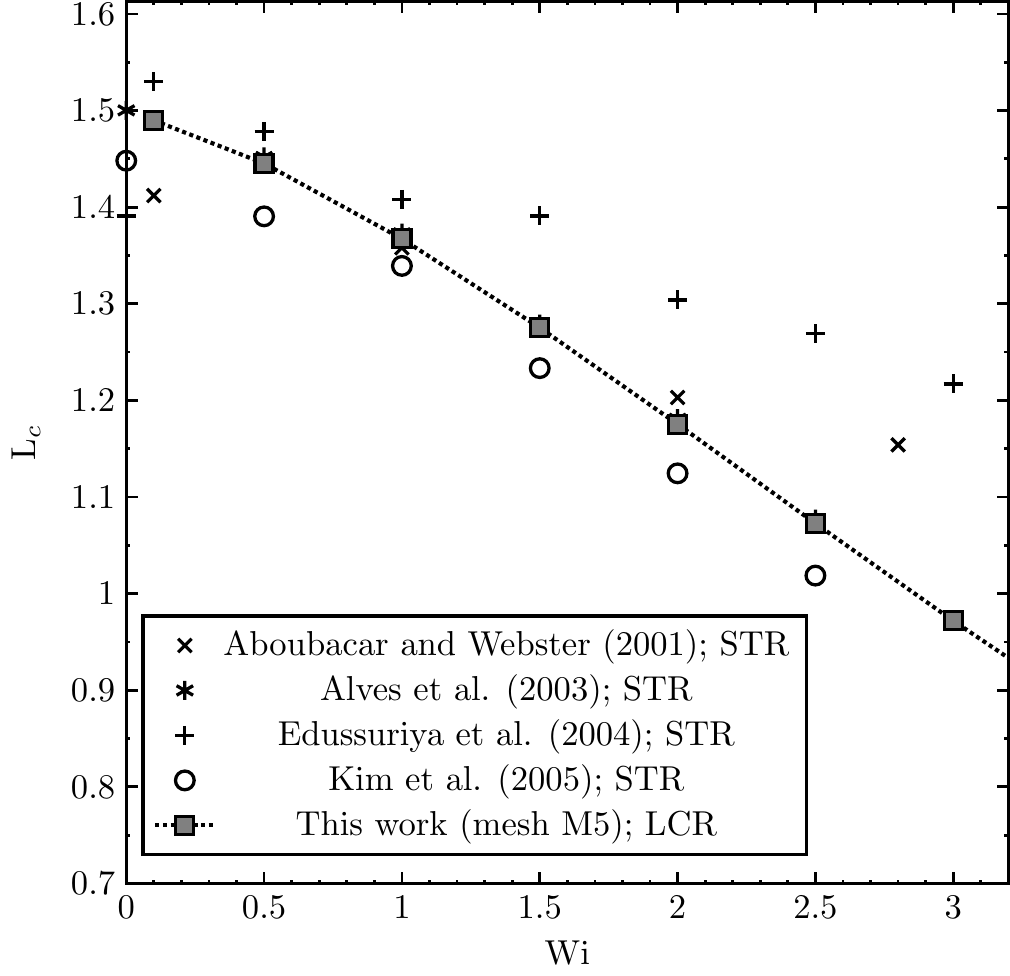}
        };
    \end{tikzpicture}
%%%%%%%%%%%%%%%%%%%%%%%%%%%%%%%%
%%%%%%%%%%%%%%%%%%%%%%%%%%%%%%%%
    \caption{Corner vortex size $\operatorname{L}_c$ over the Weissenberg number $\operatorname{Wi}$ of an Oldroyd-B fluid in a planar 4:1 contraction up to $\operatorname{Wi} = 3$. Comparison between published values and the benchmark results from this work, carried out with the natural logarithm conformation representation (LCRN) on the finest mesh M5.}
    \label{fig:cvcomparison}
\end{figure}

To quantitatively compare some numerical data from literature with the results of the present work, we plot in Figure \ref{fig:cvcomparison} the dimensionless corner vortex length $\operatorname{L}_c$ over $\operatorname{Wi}$. The results from the finest mesh M5, carried out with the natural logarithm conformation representation (LCRN), are chosen as a reference of this work and are plotted as a dashed curve with diamond symbols. The literature data is taken from both, the stress tensor representation (STR) and the LCRN. The corner vortex lengths from this work show closest agreement with the benchmark results in Alves et al.\ \cite{Alves2003}. It is important to note that similar to Alves et al. \cite{Alves2003}
our results converge towards $\operatorname{L}_c \approx 1.5$ for the limiting case of vanishing Weissenberg numbers. In this respect, the results in Aboubacar \& Webster \cite{Aboubacar2001}, Edussuriya et al.\ \cite{Edussuriya2004} and Kim et al.\ \cite{Kim2005} prove to be less reliable, since they deviate considerably from the Newtonian limit. Note that in Figure \ref{fig:cvcomparison} some of the results from literature are obtained from numerical simulations in which the advection term in the momentum balance was neglected. This is strictly speaking ony valid if both $\operatorname{Re}$ and $\operatorname{Wi}$ tend to zero. In the present work, the advection term is always considered and the Reynolds number is set to 0.01.
%%%%%%%%%%%%%%%%%%%%%%%%%%%%%%%%%%%%%%%%%%%%%%%%%%%%%%%%%%%%%%%%
%%%%%%%%%%%%%%%%%%%%%%%%%%%%%%%%%%%%%%%%%%%%%%%%%%%%%%%%%%%%%%%%
%%%%%%%%%%%%%%%%%%%%%%%%%%%%%%%%%%%%%%%%%%%%%%%%%%%%%%%%%%%%%%%%
%%%%%%%%%%%%%%%%%%%%%%%%%%%%%%%%%%%%%%%%%%%%%%%%%%%%%%%%%%%%%%%%

\begin{figure}[h!]
\centering
%%%%%%%%%%%%%%%%%%%%%%%%%%%%%%%%
\begin{subfigure}{0.5\linewidth-9pt}
\centering
%%%%%%%%%%%%%%%%%%%%%%%%%%%%%%%%
    \begin{tikzpicture}
        \node[inner sep=0pt] (BLVC) at (0,0)
        {
            \includegraphics[width=.99\textwidth]{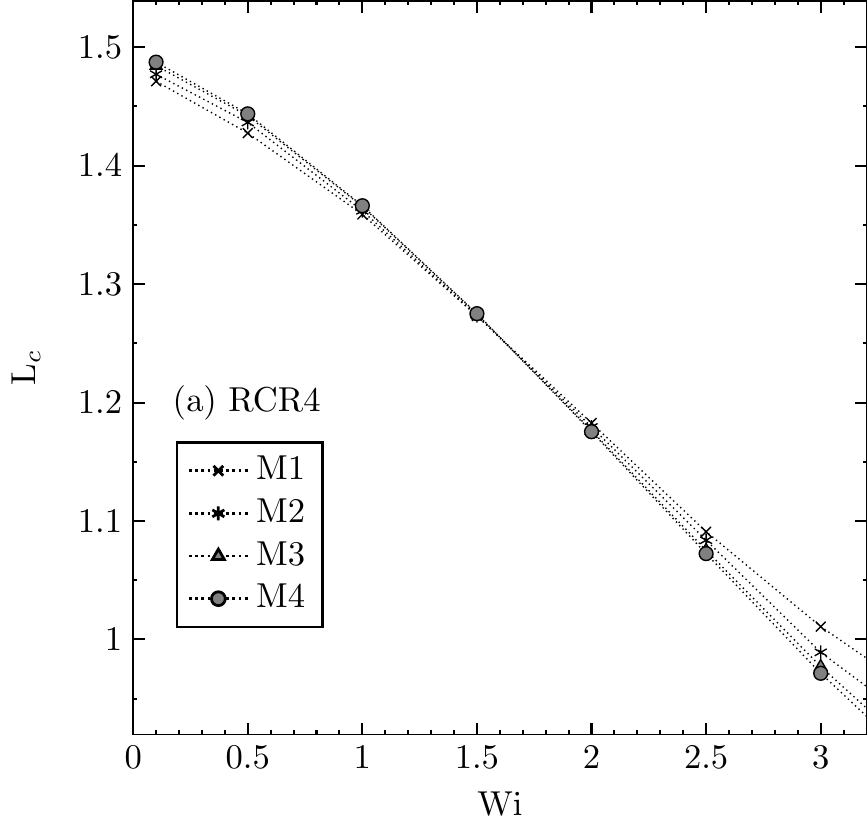}
        };
        \node[]
        (0,0) at (0.8,2.4){\small{RCR4}};
    \end{tikzpicture}
%%%%%%%%%%%%%%%%%%%%%%%%%%%%%%%%
\end{subfigure}
%%%%%%%%%%%%%%%%%%%%%%%%%%%%%%%%
%%%%%%%%%%%%%%%%%%%%
\hfill
%%%%%%%%%%%%%%%%%%%%
%%%%%%%%%%%%%%%%%%%%%%%%%%%%%%%%
\begin{subfigure}{0.5\linewidth-9pt}
\centering
%%%%%%%%%%%%%%%%%%%%%%%%%%%%%%%%
    \begin{tikzpicture}
        \node[inner sep=0pt] (BLVC) at (0,0)
        {
            \includegraphics[width=.99\textwidth]{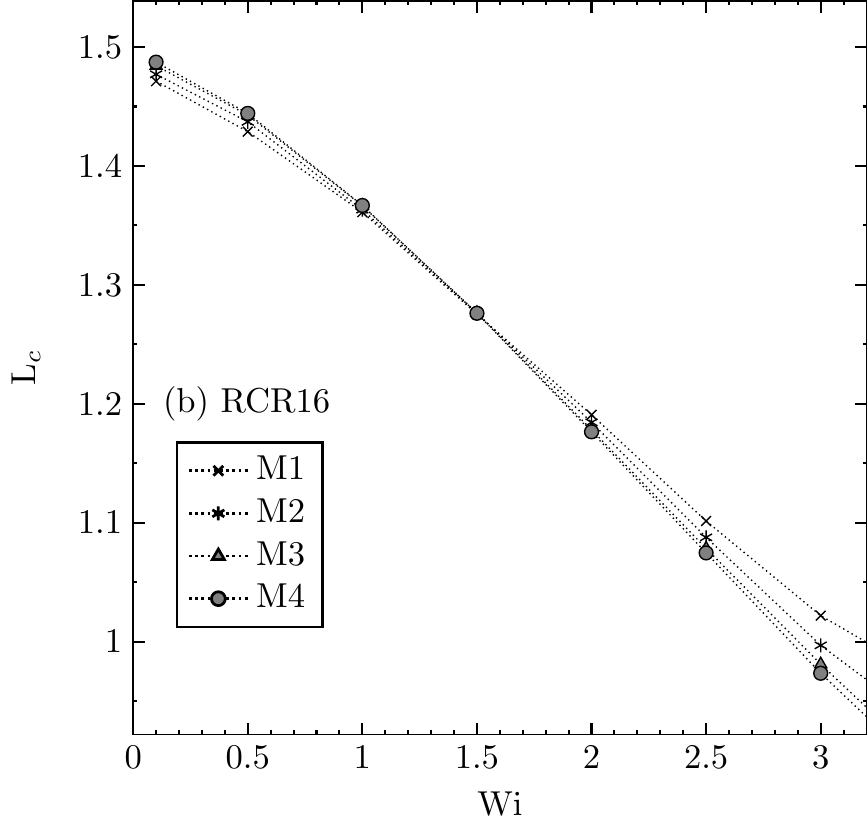}
        };
        \node[]
        (0,0) at (0.8,2.4){\small{RCR16}};
    \end{tikzpicture}
%%%%%%%%%%%%%%%%%%%%%%%%%%%%%%%%
\end{subfigure}
%%%%%%%%%%%%%%%%%%%%%%%%%%%%%%%%
%%%%%%%%%%%%%%%%%%%%
\\
%%%%%%%%%%%%%%%%%%%%
%%%%%%%%%%%%%%%%%%%%%%%%%%%%%%%%
\begin{subfigure}{0.5\linewidth-9pt}
\centering
%%%%%%%%%%%%%%%%%%%%%%%%%%%%%%%%
    \begin{tikzpicture}
        \node[inner sep=0pt] (BLVC) at (0,0)
        {
            \includegraphics[width=.99\textwidth]{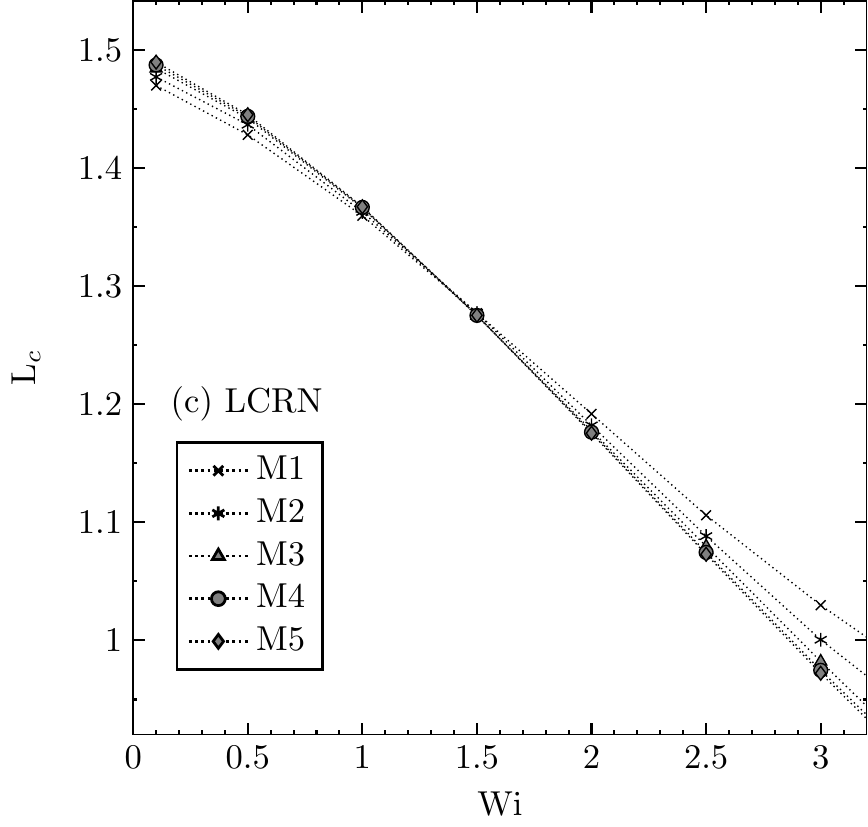}
        };
        \node[]
        (0,0) at (0.8,2.4){\small{LCRN}};
    \end{tikzpicture}
%%%%%%%%%%%%%%%%%%%%%%%%%%%%%%%%
\end{subfigure}
%%%%%%%%%%%%%%%%%%%%%%%%%%%%%%%%
%%%%%%%%%%%%%%%%%%%%
\hfill
%%%%%%%%%%%%%%%%%%%%
%%%%%%%%%%%%%%%%%%%%%%%%%%%%%%%%
\begin{subfigure}{0.5\linewidth-9pt}
\centering
%%%%%%%%%%%%%%%%%%%%%%%%%%%%%%%%
    \begin{tikzpicture}
        \node[inner sep=0pt] (BLVC) at (0,0)
        {
            \includegraphics[width=.99\textwidth]{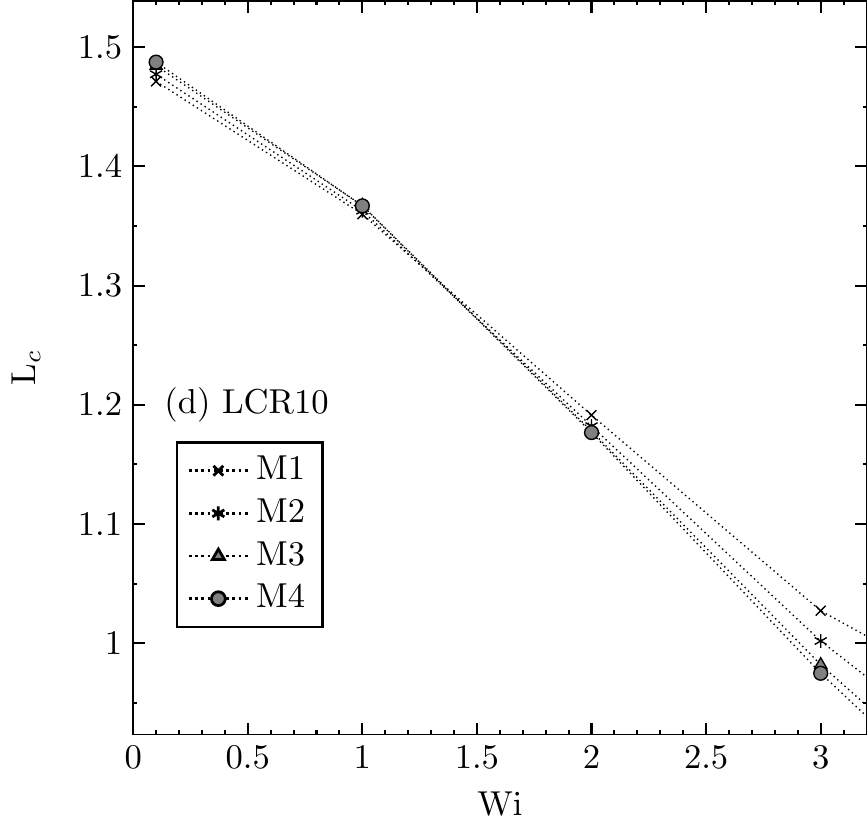}
        };
        \node[]
        (0,0) at (0.8,2.4){\small{LCR10}};
    \end{tikzpicture}
%%%%%%%%%%%%%%%%%%%%%%%%%%%%%%%%
\end{subfigure}
%%%%%%%%%%%%%%%%%%%%%%%%%%%%%%%%
    \label{fig:contourSmallWi}
    \caption{Corner vortex size $\operatorname{L}_c$ over the Weissenberg number $\operatorname{Wi}$ of an Oldroyd-B fluid in a planar 4:1 contraction for $\operatorname{Wi} \leq 3$. Comparison of the results from different change-of-variable representations on the meshes M1 to M5: (a) RCR4, (b) RCR16, (c) LCRN, (d) LCR10.}
    \label{fig:mcsWi3}
\end{figure}
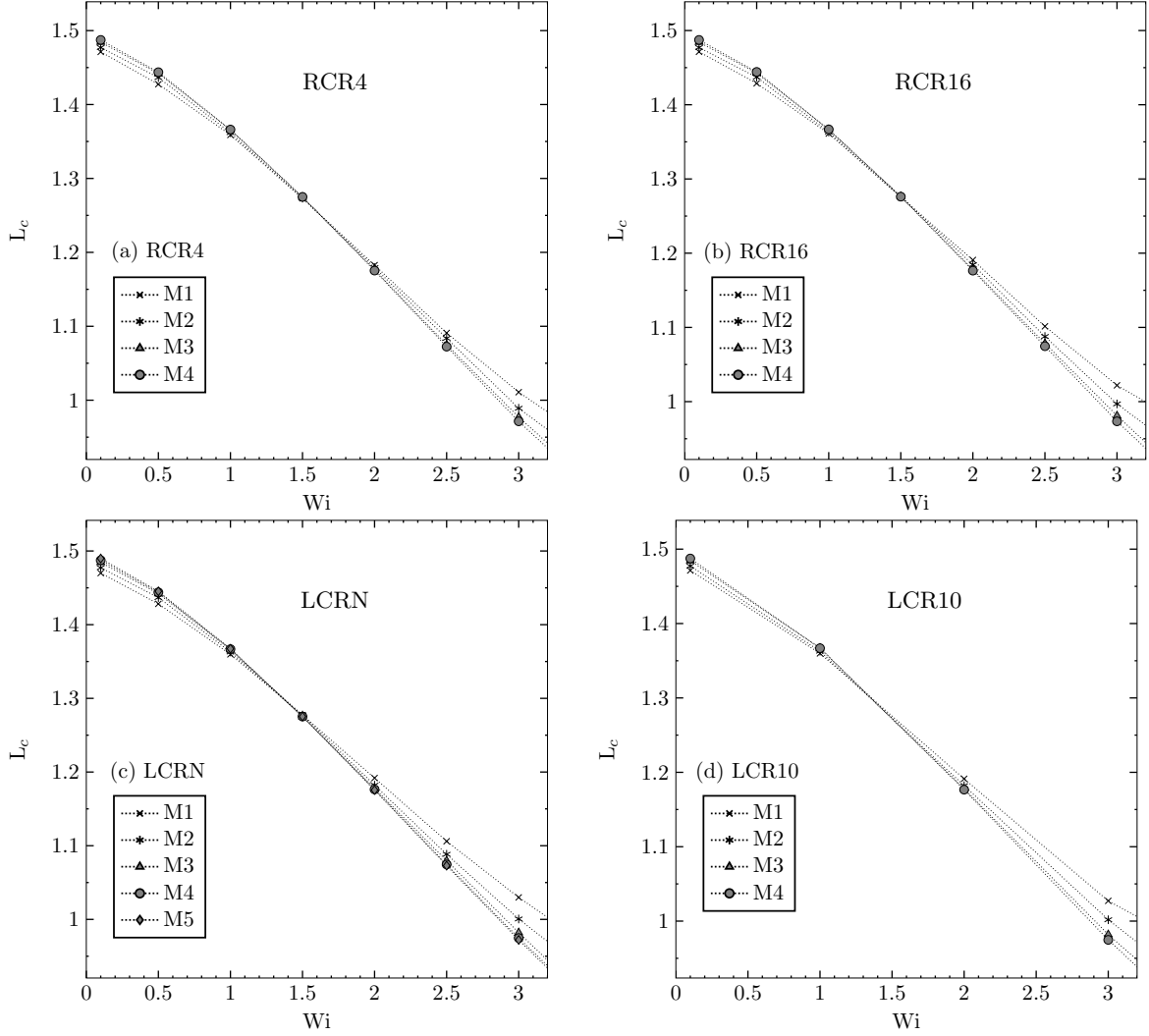

To compare the convergence, the corner vortex length obtained on the meshes M1 to M5 is plotted in Figure \ref{fig:mcsWi3} for different logarithm and root conformation representations. The data for the LCRN is given in Table \ref{table:LCRdata}. There are only minor variations between the results obtained from different change-of-variable representations. This should generally be the case when the elasticity is rather small, i.e.\ $\operatorname{Wi} < 1$. As the elasticity is increased, the influence of the constitutive equation representation on the result becomes more pronounced. For $\operatorname{Wi}$ up to three there is a close consistency of the results from different representations on the medium and fine meshes M3 to M5, which indicates that the methods converge to the same solution and have been implemented properly. Slight variations in the results become more pronounced, the less spatial mesh resolution is provided. For ascending Weissenberg numbers starting from $\operatorname{Wi} = 1.5$, the absolute range of variation between the result of the coarsest mesh M1 and the finest mesh M5 increases. This implies that the mesh-sensitivity of the result increases considerably as $\operatorname{Wi}$ is increased. The largest mesh-sensitivity is found at $\operatorname{Wi} = 3$, where we observe a significant increase of the variation in the corner vortex size between the meshes M1 to M5. Moreover, the mesh-sensitivity depends on the type of representation, e.g.\ for $\operatorname{Wi} = 3$ the variation in $\operatorname{L}_c$ between the meshes M1 to M5 is smaller for the RCR4 than for the LCRN.
%%%%%%%%%%%%%%%%%%%%%%%%%%%%%%%%%%%%%%%%%%%%%%%%%%%%%%%%%%%%%%%%
%%%%%%%%%%%%%%%%%%%%%%%%%%%%%%%%%%%%%%%%%%%%%%%%%%%%%%%%%%%%%%%%
%%%%%%%%%%%%%%%%%%%%%%%%%%%%%%%%%%%%%%%%%%%%%%%%%%%%%%%%%%%%%%%%
%%%%%%%%%%%%%%%%%%%%%%%%%%%%%%%%%%%%%%%%%%%%%%%%%%%%%%%%%%%%%%%%

\begin{figure}[h!]
\centering
%%%%%%%%%%%%%%%%%%%%%%%%%%%%%%%%
\begin{subfigure}{0.5\linewidth-9pt}
\centering
%%%%%%%%%%%%%%%%%%%%%%%%%%%%%%%%
    \begin{tikzpicture}
        \node[inner sep=0pt] (BLVC) at (0,0)
        {
            \includegraphics[width=.99\textwidth]{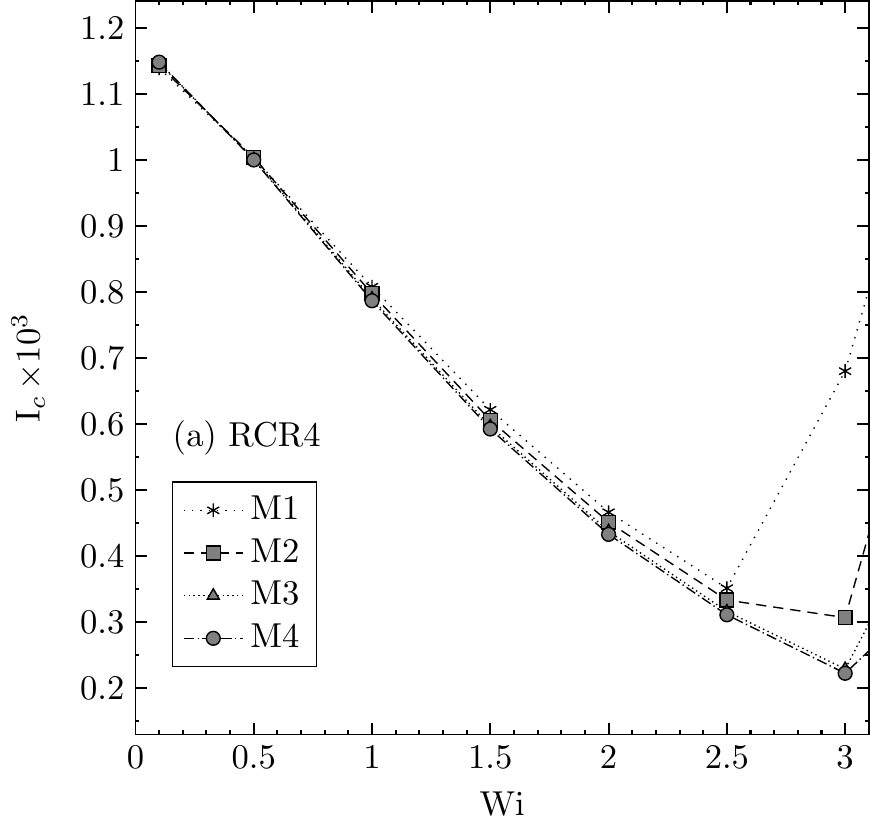}
        };
%        (0,0) at (-1.25,1.0){(a)};
    \end{tikzpicture}
%%%%%%%%%%%%%%%%%%%%%%%%%%%%%%%%
\end{subfigure}
%%%%%%%%%%%%%%%%%%%%%%%%%%%%%%%%
%%%%%%%%%%%%%%%%%%%%
\hfill
%%%%%%%%%%%%%%%%%%%%
%%%%%%%%%%%%%%%%%%%%%%%%%%%%%%%%
\begin{subfigure}{0.5\linewidth-9pt}
\centering
%%%%%%%%%%%%%%%%%%%%%%%%%%%%%%%%
    \begin{tikzpicture}
        \node[inner sep=0pt] (BLVC) at (0,0)
        {
            \includegraphics[width=.99\textwidth]{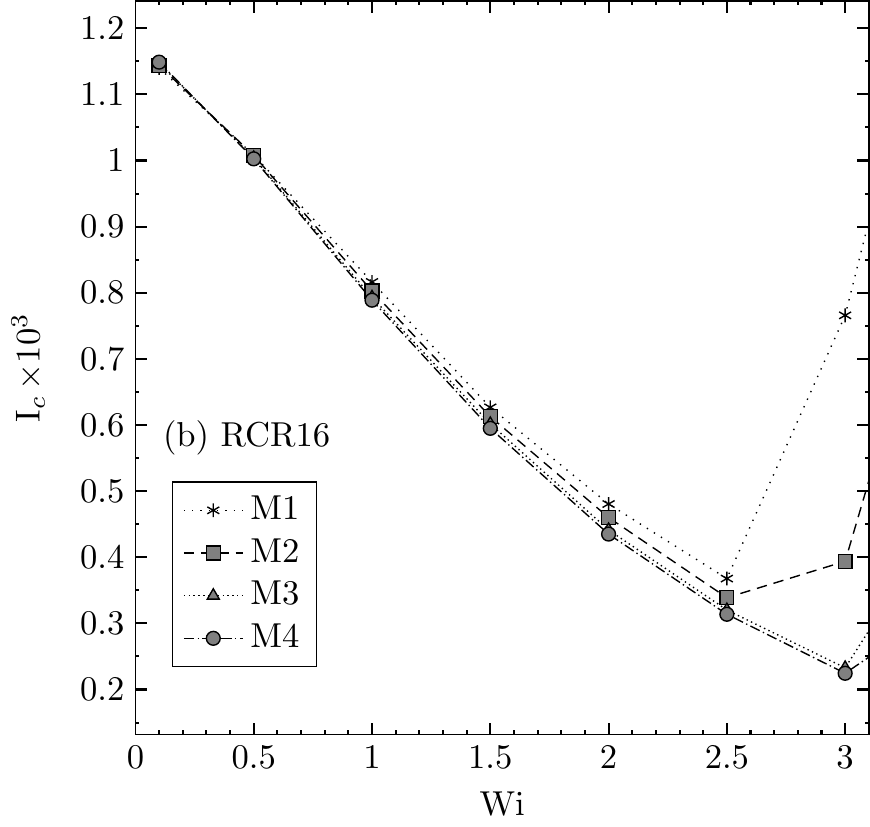}
        };
%        (0,0) at (2.65,-1.0){$\operatorname{Wi} = 0.5$};
    \end{tikzpicture}
%%%%%%%%%%%%%%%%%%%%%%%%%%%%%%%%
\end{subfigure}
%%%%%%%%%%%%%%%%%%%%%%%%%%%%%%%%
%%%%%%%%%%%%%%%%%%%%
\\
%%%%%%%%%%%%%%%%%%%%
%%%%%%%%%%%%%%%%%%%%%%%%%%%%%%%%
\begin{subfigure}{0.5\linewidth-9pt}
\centering
%%%%%%%%%%%%%%%%%%%%%%%%%%%%%%%%
    \begin{tikzpicture}
        \node[inner sep=0pt] (BLVC) at (0,0)
        {
            \includegraphics[width=.99\textwidth]{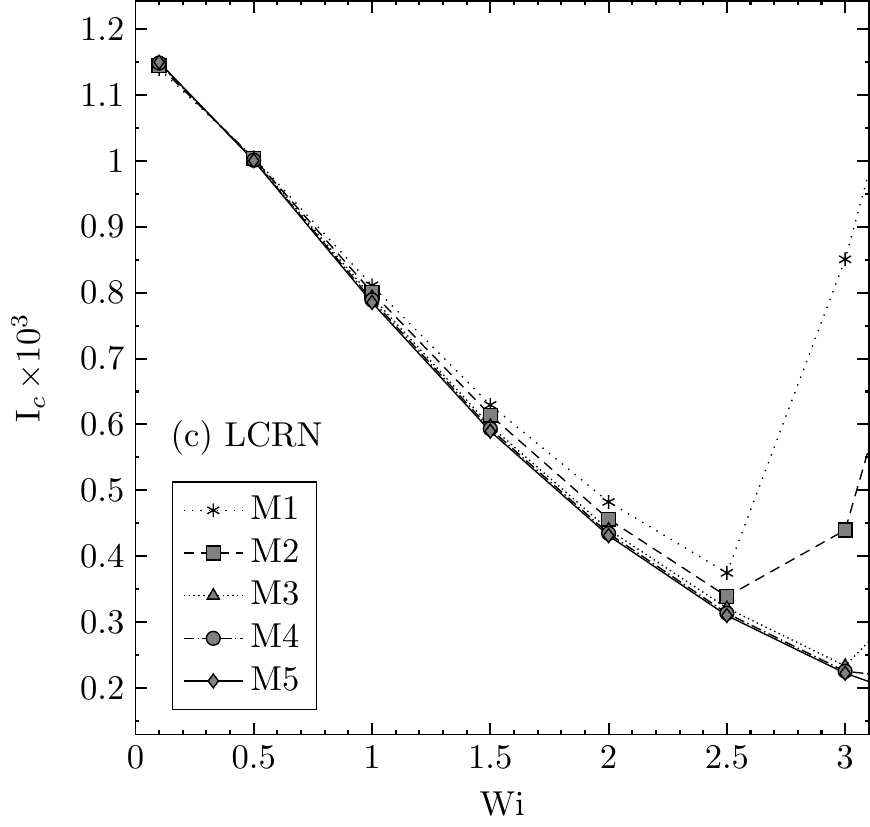}
        };
%        (0,0) at (2.65,-1.0){$\operatorname{Wi} = 0.5$};
    \end{tikzpicture}
%%%%%%%%%%%%%%%%%%%%%%%%%%%%%%%%
\end{subfigure}
%%%%%%%%%%%%%%%%%%%%%%%%%%%%%%%%
%%%%%%%%%%%%%%%%%%%%
\hfill
%%%%%%%%%%%%%%%%%%%%
%%%%%%%%%%%%%%%%%%%%%%%%%%%%%%%%
\begin{subfigure}{0.5\linewidth-9pt}
\centering
%%%%%%%%%%%%%%%%%%%%%%%%%%%%%%%%
    \begin{tikzpicture}
        \node[inner sep=0pt] (BLVC) at (0,0)
        {
            \includegraphics[width=.99\textwidth]{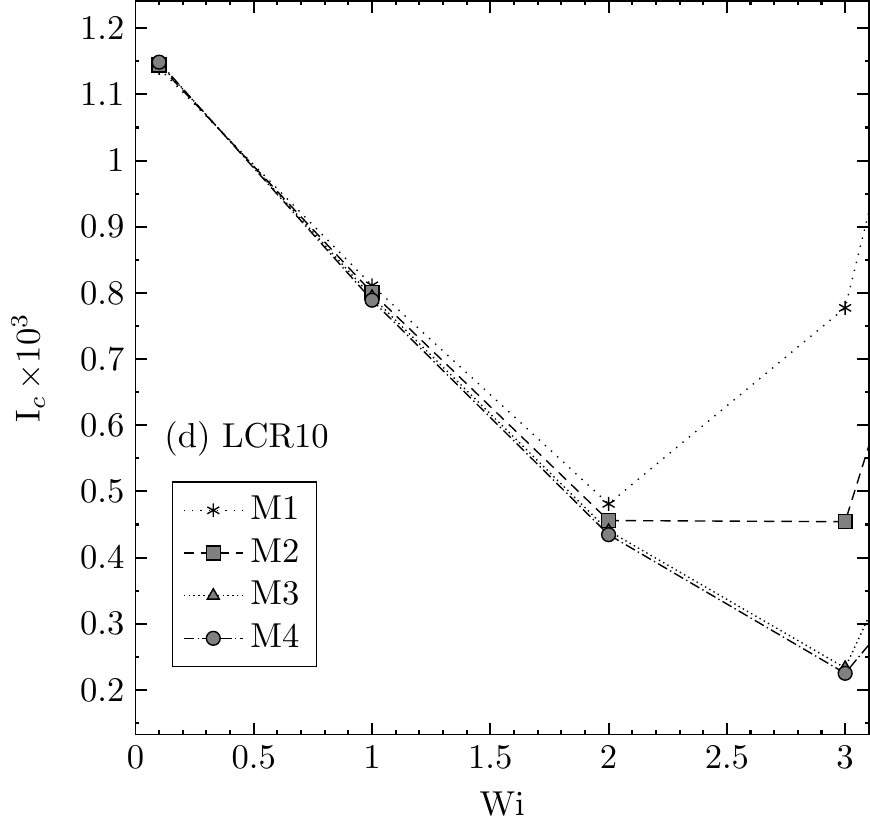}
        };
%        (0,0) at (2.65,-1.0){$\operatorname{Wi} = 0.5$};
    \end{tikzpicture}
%%%%%%%%%%%%%%%%%%%%%%%%%%%%%%%%
\end{subfigure}
%%%%%%%%%%%%%%%%%%%%%%%%%%%%%%%%
    \caption{Corner vortex intensity $\operatorname{I}_c$ over the Weissenberg number $\operatorname{Wi}$ for $\operatorname{Wi} \leq 3$. Comparison of the results from different change-of-variable representations on the meshes M1 to M5: (a) RCR4, (b) RCR16, (c) LCRN, (d) LCR10.}
    \label{fig:intensityWi3}
\end{figure}

The findings manifest itself in the corner vortex intensity plots shown in Figure \ref{fig:intensityWi3}: The results from different change-of-variable representations converge to the same solution as the spatial mesh resolution is increased. An increasing mesh sensitivity at larger Weissenberg numbers is observed in all representations as well. A remarkable increase in the variation of the result from the coarsest and the finest mesh is detected at $\operatorname{Wi} = 3$.

%%%%%%%%%%%%%%%%%%%%%%%%%%%%%%%%%%%%%%%%%%%%%%%%%%%%%%%%%%%%%%%%
%%%%%%%%%%%%%%%%%%%%%%%%%%%%%%%%%%%%%%%%%%%%%%%%%%%%%%%%%%%%%%%%
%%%%%%%%%%%%%%%%%%%%%%%%%%%%%%%%%%%%%%%%%%%%%%%%%%%%%%%%%%%%%%%%
%%%%%%%%%%%%%%%%%%%%%%%%%%%%%%%%%%%%%%%%%%%%%%%%%%%%%%%%%%%%%%%%

%%%%%%%%%%%%%%%%%%%%%%%%%%%%%%%%%%%%%%%%%%%%%%%%%%%%%%%%%%%%%%%%
%%%%%%%%%%%%%%%%%%%%%%%%%%%%%%%%%%%%%%%%%%%%%%%%%%%%%%%%%%%%%%%%
\subsection{Results for high Weissenberg numbers}
%%%%%%%%%%%%%%%%%%%%%%%%%%%%%%%%%%%%%%%%%%%%%%%%%%%%%%%%%%%%%%%%
%%%%%%%%%%%%%%%%%%%%%%%%%%%%%%%%%%%%%%%%%%%%%%%%%%%%%%%%%%%%%%%%
%Stable simulations at Weissenberg numbers larger than three
To the best of our knowledge \cite{Afonso2011, Castillo2014, Comminal2016, Pimenta2017} are so far the only references providing results far beyond the critical Weissenberg number limit in the Oldroyd-B fluid planar 4:1 contraction benchmark. In most publications \cite{Afonso2011, Comminal2016, Pimenta2017} the natural logarithm conformation representation was used to carry out stable numerical simulations. The stability of the LCRN was demonstrated for Weissenberg numbers up to one hundred \cite{Afonso2011}, however without verifying the mesh-convergence of the results. The more recent work \cite{Pimenta2017} goes back to Weissenberg numbers smaller than $12$ to study the convergence of the LCRN. The results from the present work confirm that this makes sense, because the sensitivity to the spatial resolution increases dramatically for $\operatorname{Wi} \geq 3$ for all change-of-variable representations, causing severe uncertainties on under-resolved meshes. Moreover, the mesh-sensitivity is not independent from the change-of-variable representation. Some root conformation representations have better convergence properties than the LCRN, meaning that the choice of the tranformation function is of considerable importance.
Let us note that for all $\operatorname{Wi} > 2$ we observe a superposition of the flow with very small fluctuations, inducing small-scale oscillations in the corner vortex size. These oscillations are getting more intense as the Weissenberg number is increased. Our results suggest that starting from Weissenberg numbers between $2 < \operatorname{Wi} < 2.5$, there is a continuous transition regime where the unsteadiness of flow is growing in magnitude with increasing $\operatorname{Wi}$. While for $\operatorname{Wi} < 3$ the impact of such unsteady effects on $\operatorname{L}_c$ is negligibly small, we observe a considerable extent at higher Weissenberg numbers. For the purpose of quantitative comparison, we evaluate only the mean transient effects on the flow by computing time-averaged $\operatorname{L}_c$ in all simulations for $\operatorname{Wi} > 2$ and no flow symmetry is used in the mesh.

\begin{figure}[h!]
\centering
    \begin{tikzpicture}[]
        \node[inner sep=0pt] (BLVC) at (0,0)
        {
            \includegraphics[width=265pt]{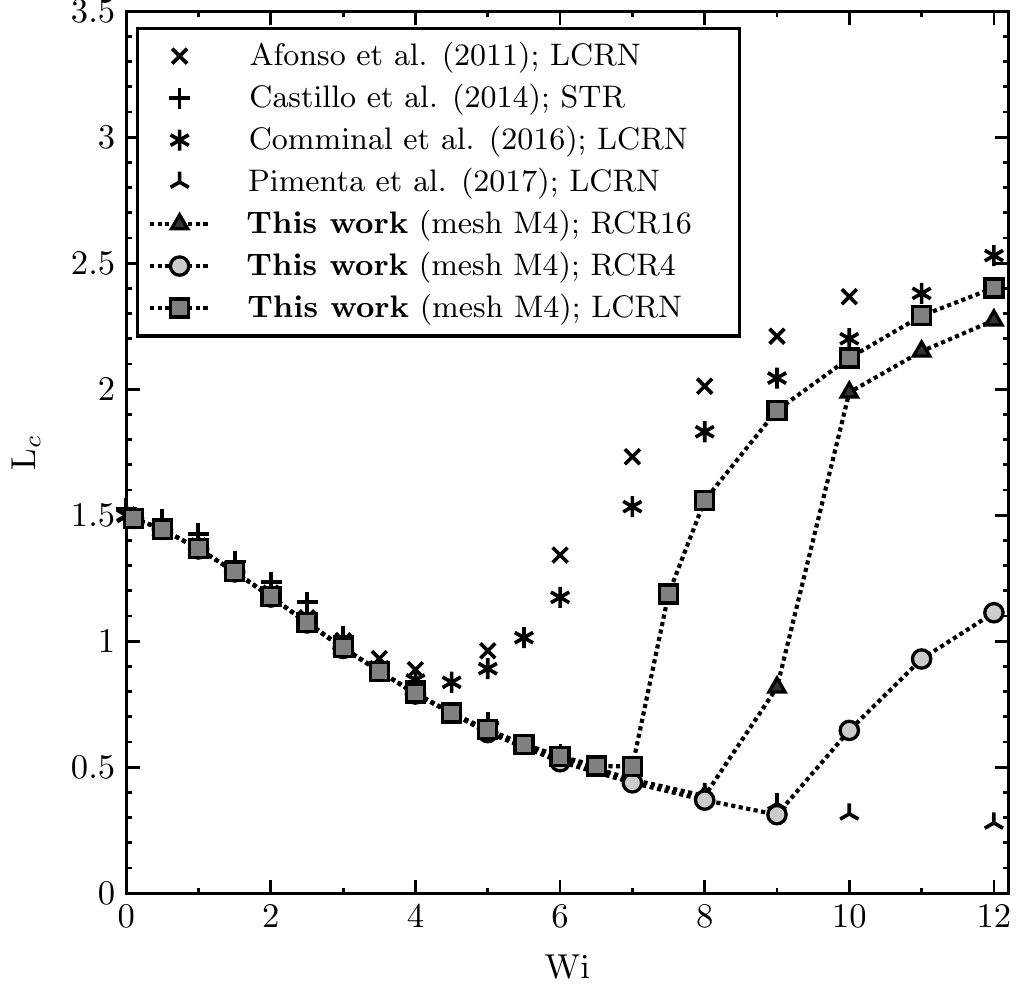}
        };
    \end{tikzpicture}
%%%%%%%%%%%%%%%%%%%%%%%%%%%%%%%%
%%%%%%%%%%%%%%%%%%%%%%%%%%%%%%%%
    \caption{Corner vortex size over $\operatorname{Wi}$ of an Oldroyd-B fluid in a planar 4:1 contraction. Comparison between published values and this work (mesh M4) up to $\operatorname{Wi} = 12$. The results obtained from the natural logarithm conformation representation (LCRN), the fourth root conformation representation (RCR4) and the sixteenth root conformation representation (RCR16) are plotted.}
    \label{fig:cvcomparisonHigh}
\end{figure}

A quantitative comparison of the dimensionless corner vortex length $\operatorname{L}_c$ over $\operatorname{Wi}$ between the numerical data in \cite{Afonso2011, Castillo2014, Comminal2016, Pimenta2017} and the results from the present work on the mesh M4 is presented in Figure \ref{fig:cvcomparisonHigh} for $\operatorname{Wi}$ up to twelve. Three $\operatorname{L}_c$-curves are plotted representing the results of the natural logarithm conformation representation, the fourth root conformation representation and the sixteenth root conformation representation, respectively. Above the critical Weissenberg number, we obtain significant deviations in $\operatorname{L}_c$ from the references \cite{Afonso2011, Comminal2016}. These deviations result from an increase of mesh-sensitivity with increasing Weissenberg numbers. The spatial resolution in the computational mesh M4 is to a considerable degree higher than the ones used in \cite{Afonso2011, Comminal2016}. While for small $\operatorname{Wi} < 3$ the resolution in \cite{Afonso2011, Comminal2016} is still satisfactory, it is not for the critical Weissenberg number and above. For $\operatorname{Wi} > 3$ the results of the present work show a better agreement with \cite{Castillo2014, Pimenta2017}. It was already demonstrated in Figure \ref{fig:cvcomparison} that for $\operatorname{Wi} \geq 2.5$ the spatial resolution in our two coarsest meshes M1 and M2 is insufficient to obtain relative errors $< 1 \%$ for $\operatorname{L}_c$, compared to the results from our finest mesh M5; the values are given in Table \ref{table:LCRdata}. With increasing $\operatorname{Wi}$ the relative errors in these under-resolved meshes increase monotonically, which is shown later in detail up to $\operatorname{Wi} = 6$. For $\operatorname{Wi} \geq 6$ even the spatial resolution on mesh M4 is found to be insufficient: For different representations $\operatorname{L}_c$ does not converge to the same value. From this we conclude that on the one hand within the range of $\operatorname{Wi}$ up to twelve the change-of-variable representations alleviate the HWNP; no stability issues were found for all RCR and LCR used in this work. But on the other hand, even though being stable our simulations are subject to further limitations regarding the required spatial resolution to converge to a mesh-independent solution. In this regard, it is interesting that the relation between the quality of the results and the spatial resolution depends on the specific type of constitutive equation representation, which can be seen from the variation in $\operatorname{L}_c$ between the results from the RCR4 and the LCRN. Our results on the mesh M4 suggest that the spatial resolution must be further increased to obtain reliable results for $\operatorname{Wi} > 6$. However, considering the large number of different simulations carried out for this work, the spatial resolution could not be further increased with the available computational resources. At $\operatorname{Wi} > 6$ we expect a dramatical increase of computational costs, not only because of the finer required spatial resolution, but also because of the higher physical time-intervals needed to compute the time-averaged $\operatorname{L}_c$.
Consequently, proper quantitative benchmark results are given, based on the meshes M4 and M5, up to twice the critical Weissenberg number, where the spatial resolution proves to be satisfactory. The quality of the results for $\operatorname{Wi} \leq 6$ is quantified in a detailed study below, demonstrating the high accuracy of the benchmark results beyond the critical Weissenberg number.
%%%%%%%%%%%%%%%%%%%%%%%%%%%%%%%%%%%%%%%%%%%%%%%%%%%%%%%%%%%%%%%%
%%%%%%%%%%%%%%%%%%%%%%%%%%%%%%%%%%%%%%%%%%%%%%%%%%%%%%%%%%%%%%%%
%%%%%%%%%%%%%%%%%%%%%%%%%%%%%%%%%%%%%%%%%%%%%%%%%%%%%%%%%%%%%%%%
%%%%%%%%%%%%%%%%%%%%%%%%%%%%%%%%%%%%%%%%%%%%%%%%%%%%%%%%%%%%%%%%

\begin{figure}[h!]
\centering
    \begin{tikzpicture}[]
        \node[inner sep=0pt] (BLVC) at (0,0)
        {
            \includegraphics[width=265pt]{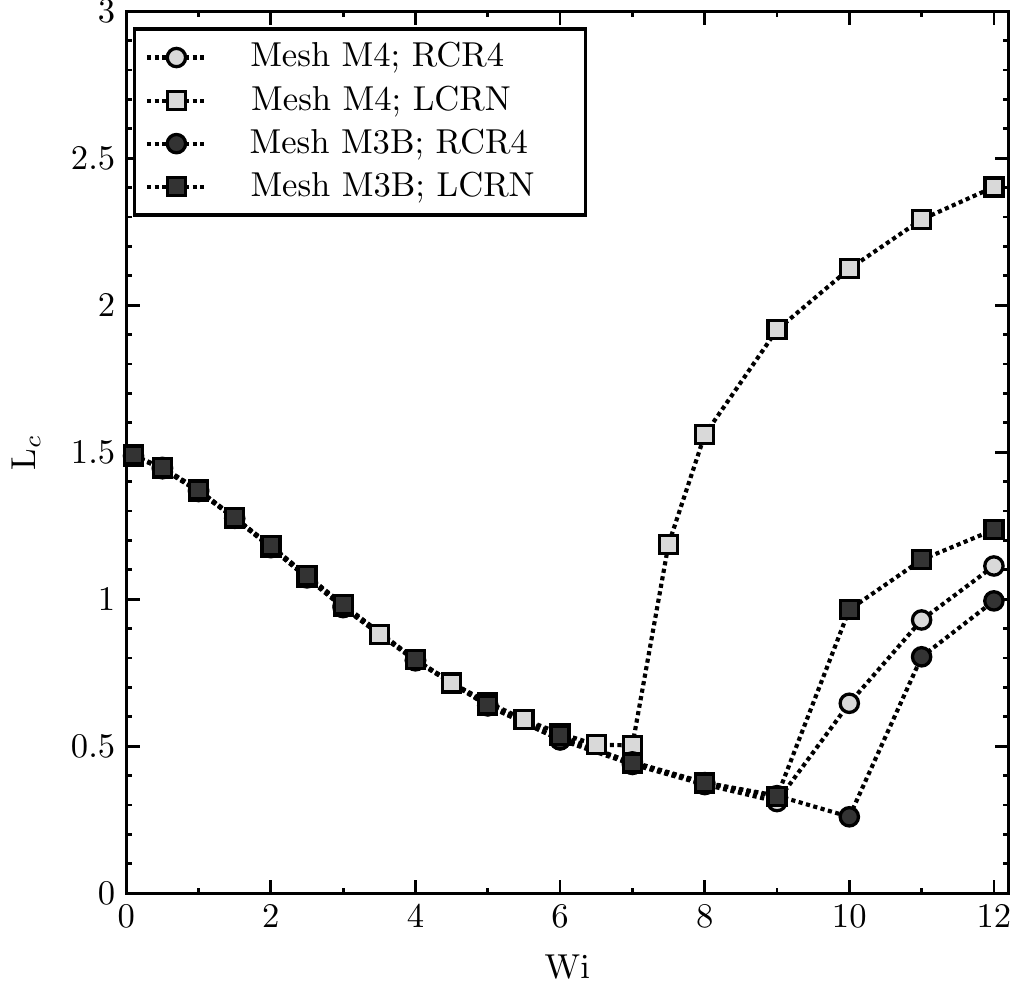}
        };
    \end{tikzpicture}
%%%%%%%%%%%%%%%%%%%%%%%%%%%%%%%%
%%%%%%%%%%%%%%%%%%%%%%%%%%%%%%%%
    \caption{Corner vortex size over $\operatorname{Wi}$ of an Oldroyd-B fluid in a planar 4:1 contraction up to $\operatorname{Wi} = 12$. Comparison between the results from mesh M4 and from the modified mesh M3B. In the latter, an additional boundary layer of cells is added at the channel walls. Mesh M3B has a finer spatial resolution in the wall region, but is coarser in the bulk than mesh M4. The results obtained from the natural logarithm conformation representation (LCRN) and the fourth root conformation representation (RCR4) are shown.}
    \label{fig:cvcomparisonHighBL}
\end{figure}

\begin{figure}[h!]
\centering
%%%%%%%%%%%%%%%%%%%%%%%%%%%%%%%%
\begin{subfigure}{206pt}
\centering
%%%%%%%%%%%%%%%%%%%%%%%%%%%%%%%%
    \begin{tikzpicture}
        \node[inner sep=0pt] (BLVC) at (0,0)
        {
            \includegraphics[width=.96\textwidth]{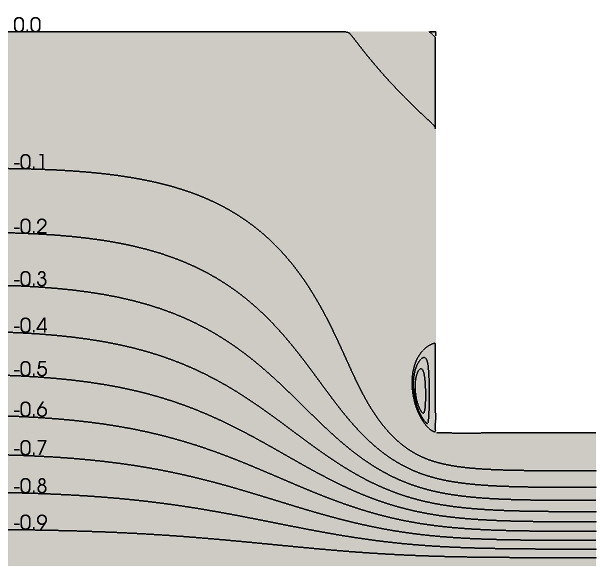}
        };
        \node[]
        (0,0) at (2.65,-1.0){$\operatorname{Wi} = 5$};
    \end{tikzpicture}
%%%%%%%%%%%%%%%%%%%%%%%%%%%%%%%%
\end{subfigure}
%%%%%%%%%%%%%%%%%%%%%%%%%%%%%%%%
%%%%%%%%%%%%%%%%%%%%
\quad
%%%%%%%%%%%%%%%%%%%%
%%%%%%%%%%%%%%%%%%%%%%%%%%%%%%%%
\begin{subfigure}{206pt}
\centering
%%%%%%%%%%%%%%%%%%%%%%%%%%%%%%%%
    \begin{tikzpicture}
        \node[inner sep=0pt] (BLVC) at (0,0)
        {
            \includegraphics[width=.96\textwidth]{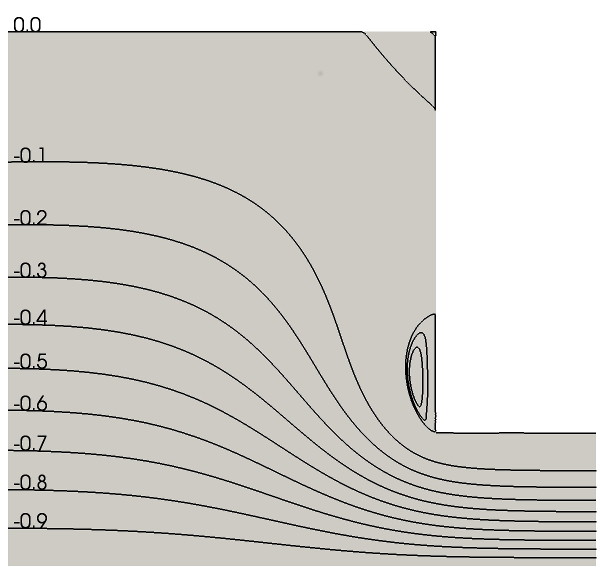}
        };
        \node[]
        (0,0) at (2.65,-1.0){$\operatorname{Wi} = 6$};
    \end{tikzpicture}
%%%%%%%%%%%%%%%%%%%%%%%%%%%%%%%%
\end{subfigure}
%%%%%%%%%%%%%%%%%%%%%%%%%%%%%%%%
%%%%%%%%%%%%%%%%%%%%
\\
%%%%%%%%%%%%%%%%%%%%
%%%%%%%%%%%%%%%%%%%%%%%%%%%%%%%%
\begin{subfigure}{206pt}
\centering
%%%%%%%%%%%%%%%%%%%%%%%%%%%%%%%%
    \begin{tikzpicture}
        \node[inner sep=0pt] (BLVC) at (0,0)
        {
            \includegraphics[width=.96\textwidth]{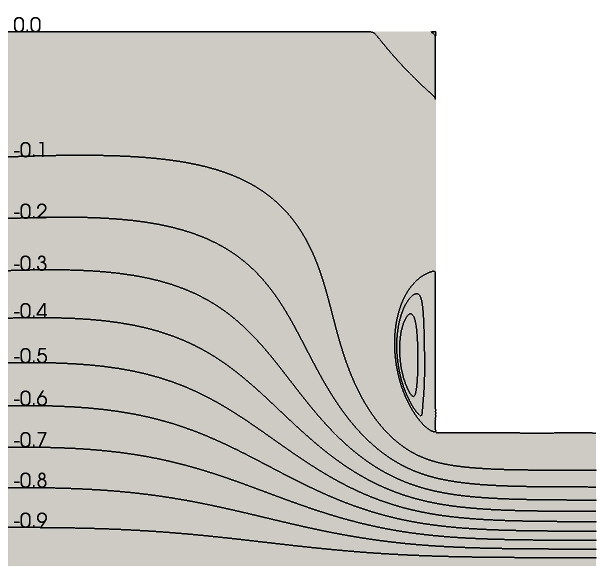}
        };
        \node[]
        (0,0) at (2.65,-1.0){$\operatorname{Wi} = 7$};
    \end{tikzpicture}
%%%%%%%%%%%%%%%%%%%%%%%%%%%%%%%%
\end{subfigure}
%%%%%%%%%%%%%%%%%%%%%%%%%%%%%%%%
%%%%%%%%%%%%%%%%%%%%
\quad
%%%%%%%%%%%%%%%%%%%%
%%%%%%%%%%%%%%%%%%%%%%%%%%%%%%%%
\begin{subfigure}{206pt}
\centering
%%%%%%%%%%%%%%%%%%%%%%%%%%%%%%%%
    \begin{tikzpicture}
        \node[inner sep=0pt] (BLVC) at (0,0)
        {
            \includegraphics[width=.96\textwidth]{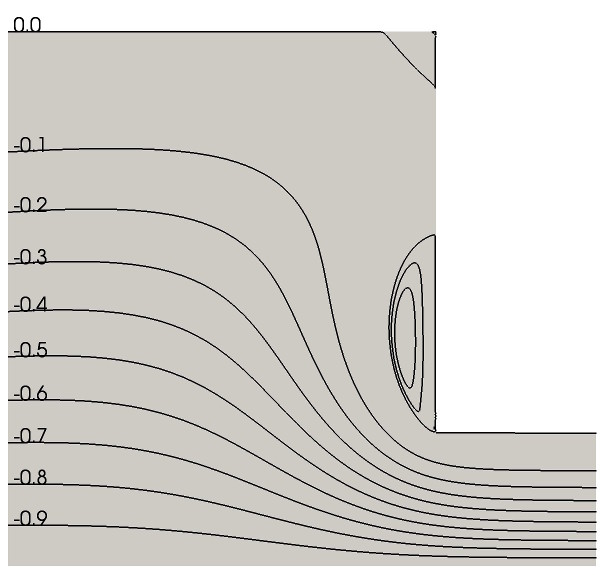}
        };
        \node[]
        (0,0) at (2.65,-1.0){$\operatorname{Wi} = 8$};
    \end{tikzpicture}
%%%%%%%%%%%%%%%%%%%%%%%%%%%%%%%%
\end{subfigure}
%%%%%%%%%%%%%%%%%%%%%%%%%%%%%%%%
\\
%%%%%%%%%%%%%%%%%%%%
%%%%%%%%%%%%%%%%%%%%%%%%%%%%%%%%
\begin{subfigure}{206pt}
\centering
%%%%%%%%%%%%%%%%%%%%%%%%%%%%%%%%
    \begin{tikzpicture}
        \node[inner sep=0pt] (BLVC) at (0,0)
        {
            \includegraphics[width=.96\textwidth]{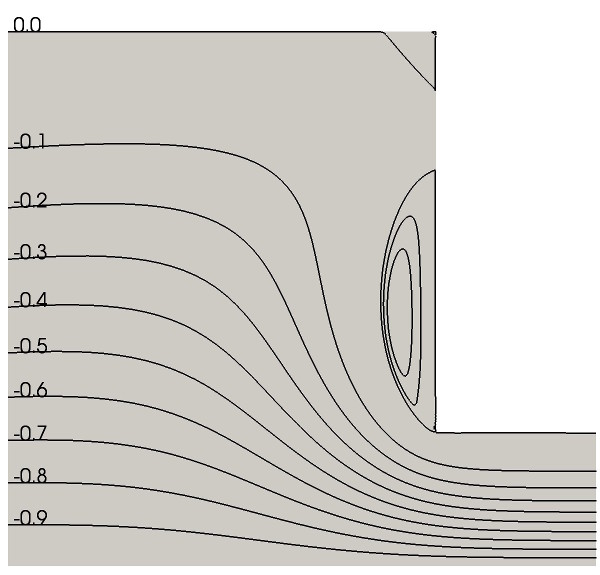}
        };
        \node[]
        (0,0) at (2.65,-1.0){$\operatorname{Wi} = 9$};
    \end{tikzpicture}
%%%%%%%%%%%%%%%%%%%%%%%%%%%%%%%%
\end{subfigure}
%%%%%%%%%%%%%%%%%%%%%%%%%%%%%%%%
%%%%%%%%%%%%%%%%%%%%
\quad
%%%%%%%%%%%%%%%%%%%%
%%%%%%%%%%%%%%%%%%%%%%%%%%%%%%%%
\begin{subfigure}{206pt}
\centering
%%%%%%%%%%%%%%%%%%%%%%%%%%%%%%%%
    \begin{tikzpicture}
        \node[inner sep=0pt] (BLVC) at (0,0)
        {
            \includegraphics[width=.96\textwidth]{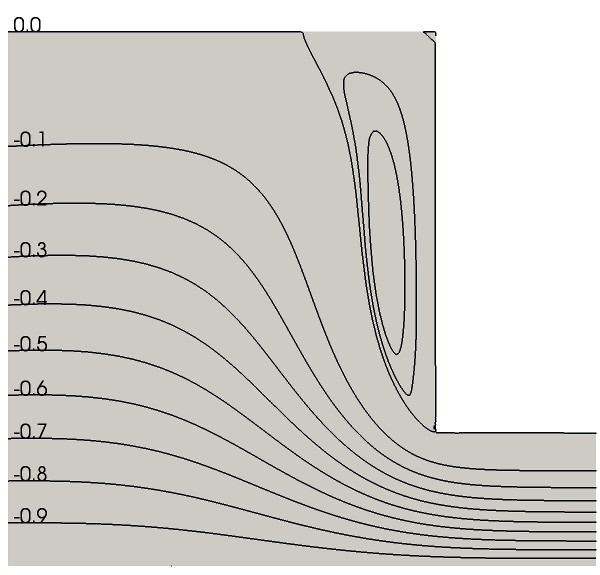}
        };
        \node[]
        (0,0) at (2.65,-1.0){$\operatorname{Wi} = 10$};
    \end{tikzpicture}
%%%%%%%%%%%%%%%%%%%%%%%%%%%%%%%%
\end{subfigure}
%%%%%%%%%%%%%%%%%%%%%%%%%%%%%%%%
    \caption{Qualitative comparison of the flow patterns above the critical Weissenberg number limit for the range $5 \leq \operatorname{Wi} \leq 10$, carried out with the natural logarithm conformation representation on mesh M3B. }
    \label{fig:contoursWiHigh}
\end{figure}
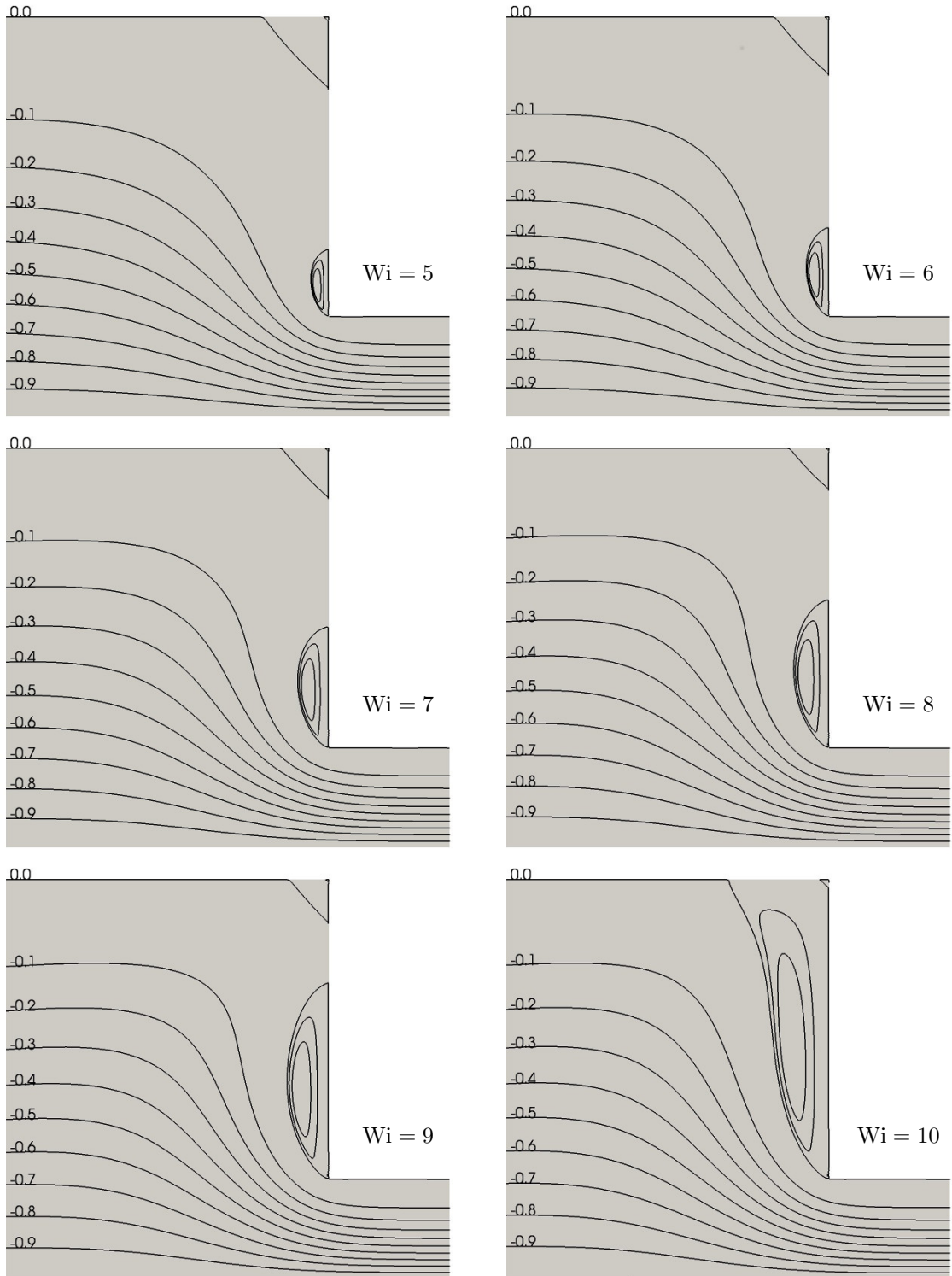

Keeping in mind that the results in Figure \ref{fig:cvcomparisonHigh} for $\operatorname{Wi} > 6$ are of qualitative nature, we identify two flow regimes in the range $0 < \operatorname{Wi} < 12$: Starting from small Weissenberg numbers $\operatorname{L}_c$ decreases in magnitude as $\operatorname{Wi}$ is increased. This regime of corner vortex shrinkage ends abruptly at a certain limit of $\operatorname{Wi}$ and is then followed by a second regime of vortex growth when $\operatorname{Wi}$ is further increased. The exact turning point between the two regimes cannot be quantified accurately based on the present data, i.e.\ on mesh M4 it appears for the LCRN at smaller $\operatorname{Wi}$ as for the RCR4. However, Figure \ref{fig:cvcomparisonHighBL} indicates that the conversion occurs at $\operatorname{Wi} > 9$. Moreover, we observe that there happens to be a sudden change of $\operatorname{L}_c$ between the two flow regimes in all change-of-variable representations. As can be seen e.g.\ from the streamlines in Figure \ref{fig:contoursWiHigh}, the reason for the second vortex growth regime is related to the independent elastic lip vortex which appears first at $\operatorname{Wi} \approx 0.5$ at the re-entrant corner of the smaller channel. While the corner vortex decreases in size, the elastic lip vortex grows towards the outer wall of the channel as the Weissenberg number is increased. Eventually, the elastic lip vortex reaches the outer wall and coincides with the smaller corner vortex such that there is only one large elastic vortex left in the domain. This change in the flow structure is observed in $\operatorname{L}_c$ as the sudden turn between the two flow regimes.

Regarding the limitations in computational costs when the entire domain is successively refined, a question of remarkable relevance is whether the accuracy can be increased by generating some local refinement in the regions of steep stress profiles at the channel walls only. We study the effect when one additional cell boundary layer is added into the medium-refined mesh M3 by cutting the first cell layer next to the channel walls into two parts. A detail of this mesh M3B is shown above in Figure \ref{fig:geometryandmesh}. Because of the additional cell boundary layer in mesh M3B, the spacing between the first two computational nodes in normal direction to the wall is significantly reduced; it is slightly smaller than in the finest mesh M5. On the other hand, the total number of cells in mesh M3B is by factor 3 smaller than in mesh M5, which implies less computational costs. The evaluation of the corner vortex size is shown in Figure \ref{fig:cvcomparisonHighBL} for the natural logarithm conformation representation and the fourth root conformation representation. The comparison between the results from mesh M3B and M4 reveals that the accuracy increases considerably if mesh M3B is used. The deviation in $\operatorname{L}_c$ between the two change-of-variable representations is clearly smaller for mesh M3B, in particular at higher Weissenberg numbers. For the logarithm conformation representation, the difference between the results on mesh M3B and M4 is significantly larger than for the fourth root conformation representation, which indicates that the latter representation is more accurate. However, the accuracy is still not good enough to exactly determine the Weissenberg number of the sudden transition between the vortex shrinkage and the vortex growth regime. We conclude that more accurate $\operatorname{L}_c$ predictions are obtained when only the resolution of the wall boundary layer is increased. This finding might help to reduce the limitations in computational costs because it allows to keep the number of CVs inside the domain at a moderate level.
%%%%%%%%%%%%%%%%%%%%%%%%%%%%%%%%%%%%%%%%%%%%%%%%%%%%%%%%%%%%%%%%
%%%%%%%%%%%%%%%%%%%%%%%%%%%%%%%%%%%%%%%%%%%%%%%%%%%%%%%%%%%%%%%%
%%%%%%%%%%%%%%%%%%%%%%%%%%%%%%%%%%%%%%%%%%%%%%%%%%%%%%%%%%%%%%%%
%%%%%%%%%%%%%%%%%%%%%%%%%%%%%%%%%%%%%%%%%%%%%%%%%%%%%%%%%%%%%%%%

%%%%%%%%%%%%%%%%%%%%%%%%%%%%%%%%%%%%%%%%%%%%%%%%%%%%%%%%%%%%%%%%
%%%%%%%%%%%%%%%%%%%%%%%%%%%%%%%%%%%%%%%%%%%%%%%%%%%%%%%%%%%%%%%%

%%%%%%%%%%%%%%%%%%%%%%%%%%%%%%%%%%%%%%%%%%%%%%%%%%%%%%%%%%%%%%%%
%%%%%%%%%%%%%%%%%%%%%%%%%%%%%%%%%%%%%%%%%%%%%%%%%%%%%%%%%%%%%%%%
\subsection{Accuracy for Weissenberg numbers $\leq 6$}
\begin{figure*}[h!]
\centering
%%%%%%%%%%%%%%%%%%%%%%%%%%%%%%%%
\begin{subfigure}{0.5\linewidth-9pt}
\centering
%%%%%%%%%%%%%%%%%%%%%%%%%%%%%%%%
    \begin{tikzpicture}
        \node[inner sep=0pt] (BLVC) at (0,0)
        {
            \includegraphics[width=.99\textwidth]{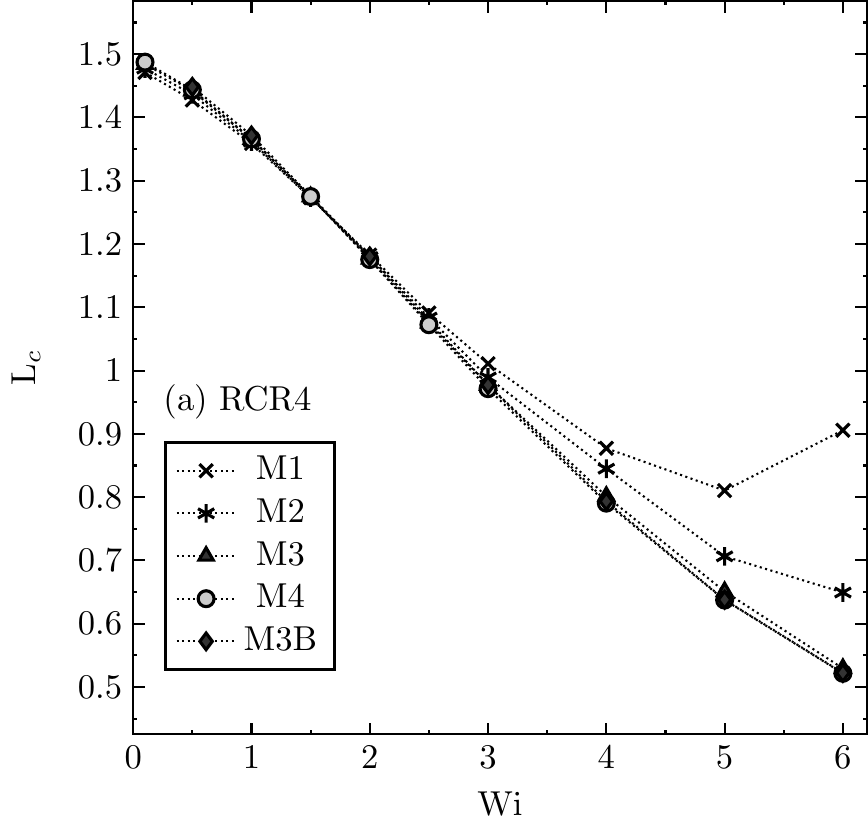}
        };
%        \node[]
%        (0,0) at (0.8,2.4){\small{RCR4}};
    \end{tikzpicture}
%%%%%%%%%%%%%%%%%%%%%%%%%%%%%%%%
\end{subfigure}
%%%%%%%%%%%%%%%%%%%%%%%%%%%%%%%%
%%%%%%%%%%%%%%%%%%%%
\hfill
%%%%%%%%%%%%%%%%%%%%
%%%%%%%%%%%%%%%%%%%%%%%%%%%%%%%%
\begin{subfigure}{0.5\linewidth-9pt}
\centering
%%%%%%%%%%%%%%%%%%%%%%%%%%%%%%%%
    \begin{tikzpicture}
        \node[inner sep=0pt] (BLVC) at (0,0)
        {
            \includegraphics[width=.99\textwidth]{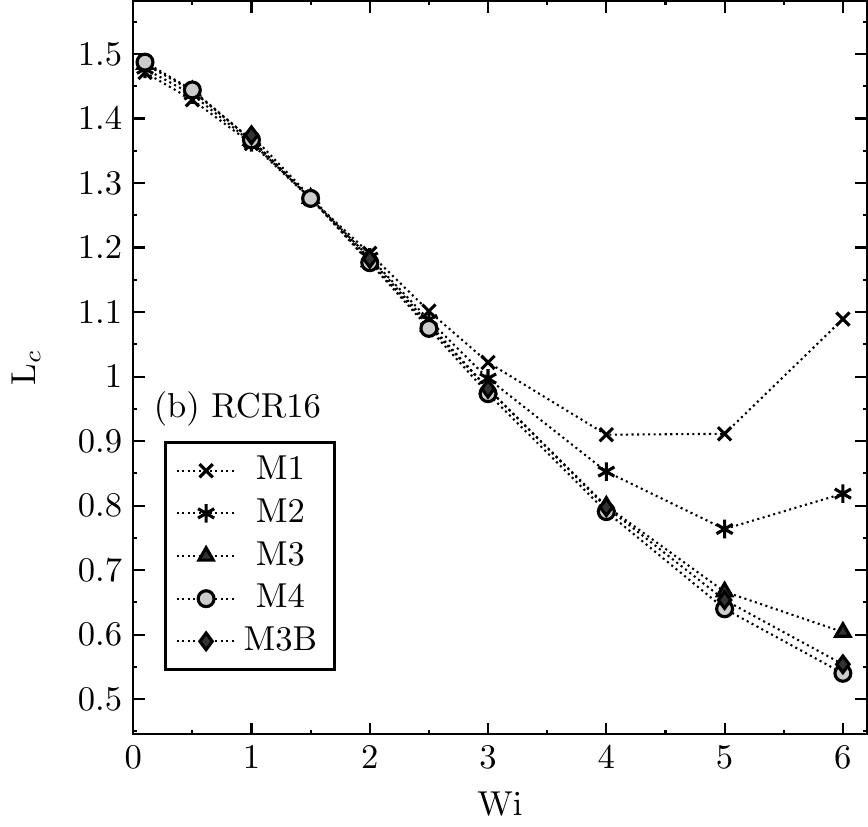}
        };
%        \node[]
%        (0,0) at (0.8,2.4){\small{RCR16}};
    \end{tikzpicture}
%%%%%%%%%%%%%%%%%%%%%%%%%%%%%%%%
\end{subfigure}
%%%%%%%%%%%%%%%%%%%%%%%%%%%%%%%%
%%%%%%%%%%%%%%%%%%%%
\\
%%%%%%%%%%%%%%%%%%%%
%%%%%%%%%%%%%%%%%%%%%%%%%%%%%%%%
\begin{subfigure}{0.5\linewidth-9pt}
\centering
%%%%%%%%%%%%%%%%%%%%%%%%%%%%%%%%
    \begin{tikzpicture}
        \node[inner sep=0pt] (BLVC) at (0,0)
        {
            \includegraphics[width=.99\textwidth]{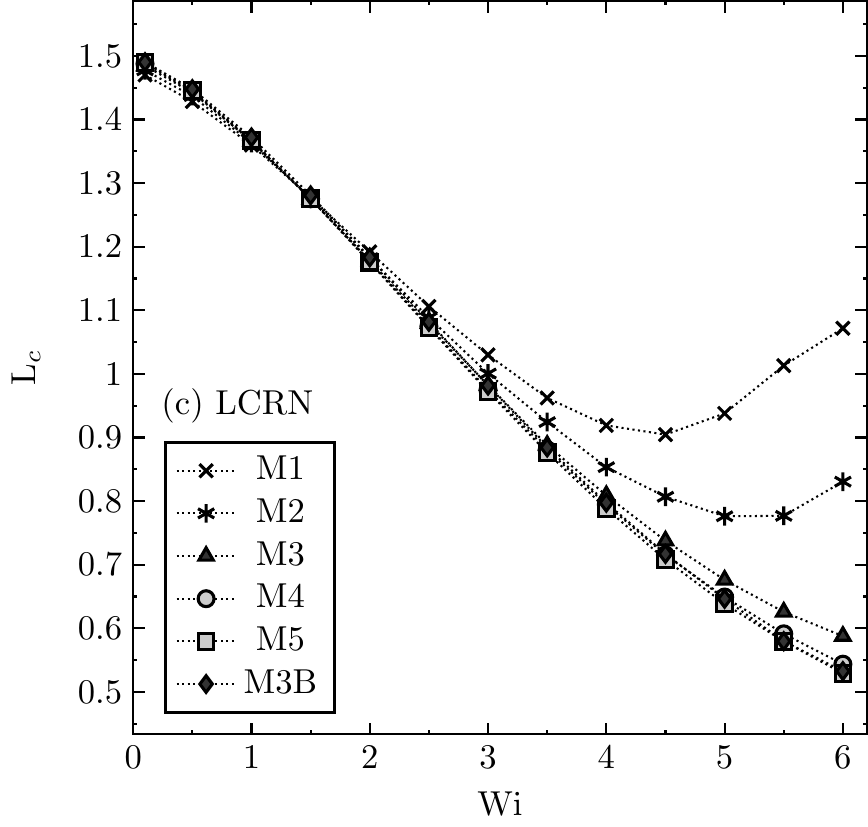}
        };
%        \node[]
%        (0,0) at (0.8,2.4){\small{LCRN}};
    \end{tikzpicture}
%%%%%%%%%%%%%%%%%%%%%%%%%%%%%%%%
\end{subfigure}
%%%%%%%%%%%%%%%%%%%%%%%%%%%%%%%%
%%%%%%%%%%%%%%%%%%%%
\hfill
%%%%%%%%%%%%%%%%%%%%
%%%%%%%%%%%%%%%%%%%%%%%%%%%%%%%%
\begin{subfigure}{0.5\linewidth-9pt}
\centering
%%%%%%%%%%%%%%%%%%%%%%%%%%%%%%%%
    \begin{tikzpicture}
        \node[inner sep=0pt] (BLVC) at (0,0)
        {
            \includegraphics[width=.99\textwidth]{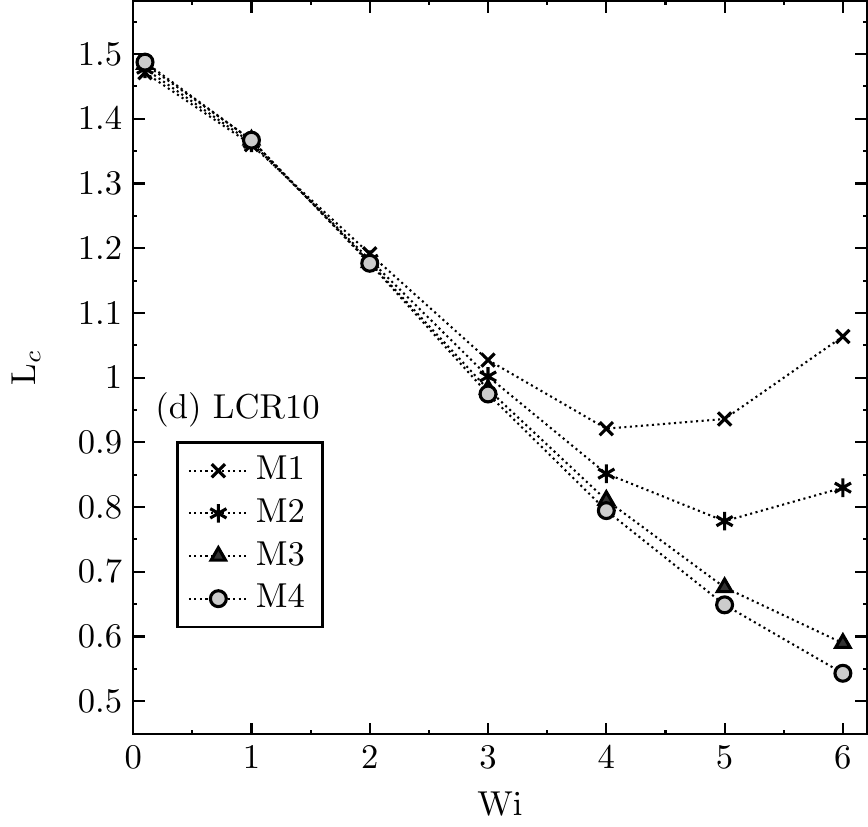}
        };
%        \node[]
%        (0,0) at (0.8,2.4){\small{LCR10}};
    \end{tikzpicture}
%%%%%%%%%%%%%%%%%%%%%%%%%%%%%%%%
\end{subfigure}
%%%%%%%%%%%%%%%%%%%%%%%%%%%%%%%%
    \caption{Corner vortex size $\operatorname{L}_c$ over the Weissenberg number $\operatorname{Wi}$ of an Oldroyd-B fluid in a planar 4:1 contraction for $\operatorname{Wi} \leq 6$. Comparison of the results from different change-of-variable representations on the meshes M1 to M5: (a) RCR4, (b) RCR16, (c) LCRN, (d) LCR10.
}
    \label{fig:contourWiSix}
\end{figure*}

%%%%%%%%%%%%%%%%%%%%%%%%%%%%%%%%%%%%%%%%%%%%%%%%%%%%%%%%%%%%%%%%
%%%%%%%%%%%%%%%%%%%%%%%%%%%%%%%%%%%%%%%%%%%%%%%%%%%%%%%%%%%%%%%%
The corner vortex size, the corner vortex intensity and the normal stress differences are studied more rigorously for $\operatorname{Wi} \leq 6$. Figure \ref{fig:contourWiSix} shows $\operatorname{L}_c$ over $\operatorname{Wi}$ for all meshes M1 to M5 and for the natural logarithm conformation representation, the common logarithm conformation representation, the fourth root conformation representation and the sixteenth root conformation representation, respectively. The corresponding data is provided in Table \ref{table:LCRdata}. The variation in the results between different change-of-variable representations decreases monotonically in magnitude when the spatial resolution is increased, which demonstrates the consistence of the results, irrespective of the particular change in the constitutive variables. All change-of-variable representations converge towards a homogeneous solution as the mesh is successively refined. Moreover, we observe considerable variation in the results from different representations on the coarser meshes, i.e.\ for the under-resolved meshes M1 and M2 the results from the RCR4 are more accurate than those from the logarithm conformation representations. This shows that the mesh-sensitivity depends on the respective constitutive equation representation. The discrepancy in $\operatorname{L}_c$ between the finest and the coarsest mesh increases in magnitude for all representations when the Weissenberg number is increased, which indicates that on a particular mesh the quality of the solution is getting worse when $\operatorname{Wi}$ is increased.
%%%%%%%%%%%%%%%%%%%%%%%%%%%%%%%%%%%%%%%%%%%%%%%%%%%%%%%%%%%%%%%%
%%%%%%%%%%%%%%%%%%%%%%%%%%%%%%%%%%%%%%%%%%%%%%%%%%%%%%%%%%%%%%%%
%%%%%%%%%%%%%%%%%%%%%%%%%%%%%%%%%%%%%%%%%%%%%%%%%%%%%%%%%%%%%%%%
%%%%%%%%%%%%%%%%%%%%%%%%%%%%%%%%%%%%%%%%%%%%%%%%%%%%%%%%%%%%%%%%

\begin{figure*}[h!]
\centering
%%%%%%%%%%%%%%%%%%%%%%%%%%%%%%%%
\begin{subfigure}{0.5\linewidth-9pt}
\centering
%%%%%%%%%%%%%%%%%%%%%%%%%%%%%%%%
    \begin{tikzpicture}
        \node[inner sep=0pt] (BLVC) at (0,0)
        {
            \includegraphics[width=.99\textwidth]{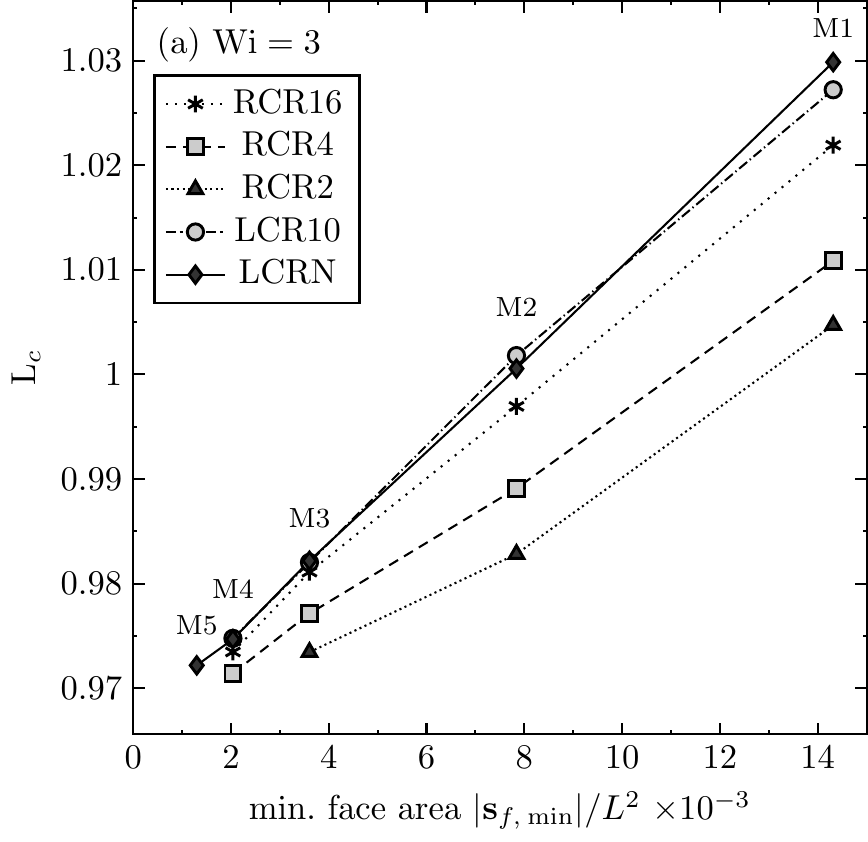}
        };
%        (0,0) at (-1.25,1.0){(a)};
    \end{tikzpicture}
%%%%%%%%%%%%%%%%%%%%%%%%%%%%%%%%
\end{subfigure}
%%%%%%%%%%%%%%%%%%%%%%%%%%%%%%%%
%%%%%%%%%%%%%%%%%%%%
\hfill
%%%%%%%%%%%%%%%%%%%%
%%%%%%%%%%%%%%%%%%%%%%%%%%%%%%%%
\begin{subfigure}{0.5\linewidth-9pt}
\centering
%%%%%%%%%%%%%%%%%%%%%%%%%%%%%%%%
    \begin{tikzpicture}
        \node[inner sep=0pt] (BLVC) at (0,0)
        {
            \includegraphics[width=.99\textwidth]{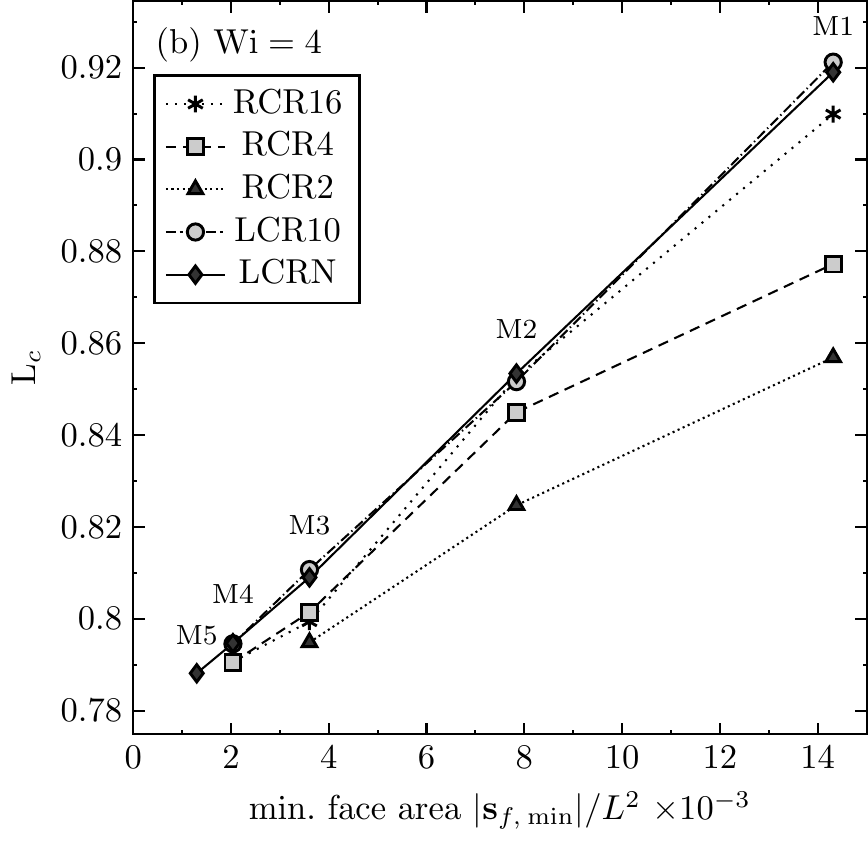}
        };
%        (0,0) at (2.65,-1.0){$\operatorname{Wi} = 0.5$};
    \end{tikzpicture}
%%%%%%%%%%%%%%%%%%%%%%%%%%%%%%%%
\end{subfigure}
%%%%%%%%%%%%%%%%%%%%%%%%%%%%%%%%
%%%%%%%%%%%%%%%%%%%%
\\
%%%%%%%%%%%%%%%%%%%%
%%%%%%%%%%%%%%%%%%%%%%%%%%%%%%%%
\begin{subfigure}{0.5\linewidth-9pt}
\centering
%%%%%%%%%%%%%%%%%%%%%%%%%%%%%%%%
    \begin{tikzpicture}
        \node[inner sep=0pt] (BLVC) at (0,0)
        {
            \includegraphics[width=.99\textwidth]{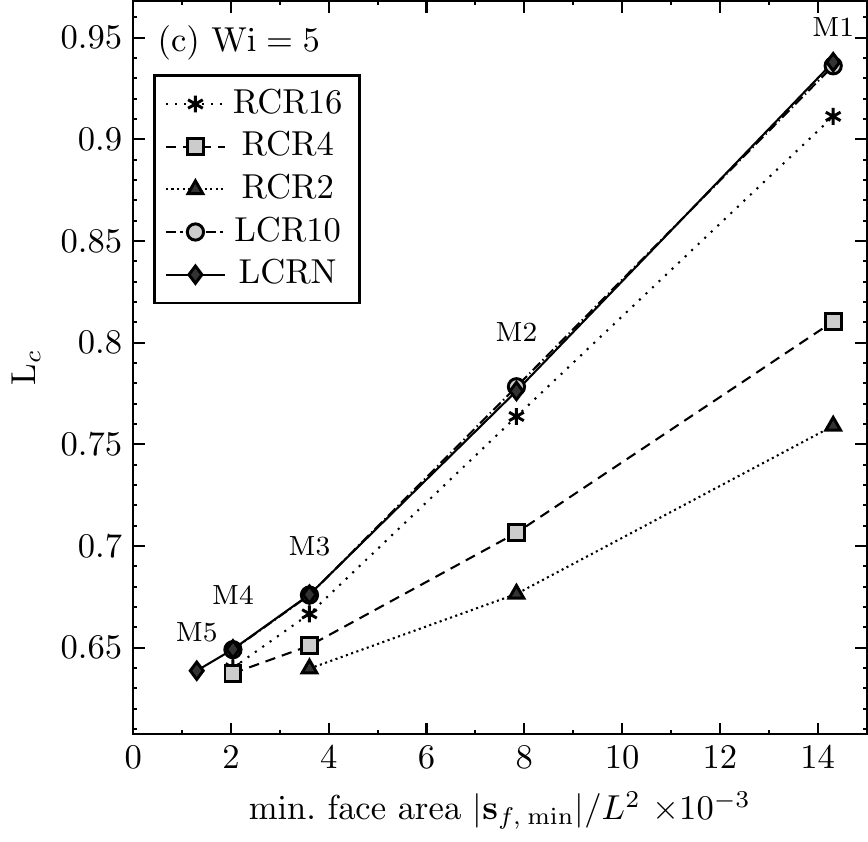}
        };
%        (0,0) at (2.65,-1.0){$\operatorname{Wi} = 0.5$};
    \end{tikzpicture}
%%%%%%%%%%%%%%%%%%%%%%%%%%%%%%%%
\end{subfigure}
%%%%%%%%%%%%%%%%%%%%%%%%%%%%%%%%
%%%%%%%%%%%%%%%%%%%%
\hfill
%%%%%%%%%%%%%%%%%%%%
%%%%%%%%%%%%%%%%%%%%%%%%%%%%%%%%
\begin{subfigure}{0.5\linewidth-9pt}
\centering
%%%%%%%%%%%%%%%%%%%%%%%%%%%%%%%%
    \begin{tikzpicture}
        \node[inner sep=0pt] (BLVC) at (0,0)
        {
            \includegraphics[width=.99\textwidth]{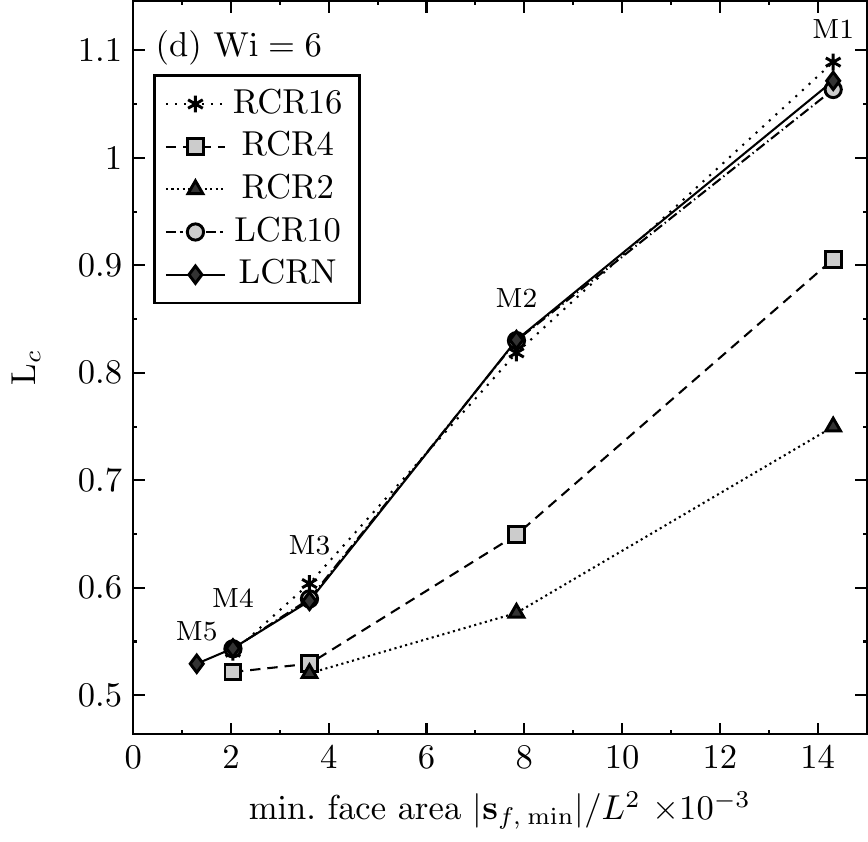}
        };
%        (0,0) at (2.65,-1.0){$\operatorname{Wi} = 0.5$};
    \end{tikzpicture}
%%%%%%%%%%%%%%%%%%%%%%%%%%%%%%%%
\end{subfigure}
%%%%%%%%%%%%%%%%%%%%%%%%%%%%%%%%
    \caption{Mesh convergence of the corner vortex size as a function of the normalized minimum cell-face area of the meshes $\vert \vec{s}_{f,\:\operatorname{min}} \vert /L^2$ for \mbox{(a) $\operatorname{Wi} = 3$}, \mbox{(b) $\operatorname{Wi} = 4$}, \mbox{(c) $\operatorname{Wi} = 5$}, \mbox{(d) $\operatorname{Wi} = 6$}.}
    \label{fig:convergenceWiSix}
\end{figure*}

The mesh-convergence of each representation becomes more evident when $\operatorname{L}_c$ is plotted as a function of the normalized minimum cell-face area of the meshes $\vert \vec{s}_{f,\:\operatorname{min}} \vert /L^2$, which is shown in Figure \ref{fig:convergenceWiSix}. On all meshes we obtain a similar corner vortex size when different logarithm transformations are applied, i.e.\ the natural logarithm and the common logarithm result in nearly identical $\operatorname{L}_c$, provided that the same mesh is used for both representations. From this we conclude that changing the logarithm base does not affect the quality of the simulation results, regardless of the spatial resolution of the mesh. But contrary to this, the root of the conformation tensor does indeed influence the quality of the results considerably. In all simulations, small root functions provide more accurate predictions of the corner vortex size than the logarithm conformation representation, in particular when the mesh is under-resolved. With the second root or the fourth root we obtain on each particular mesh quite similar accuracy than with the logarithm conformation representation used on a one-level of refinement finer mesh, respectively. For greater root functions the accuracy becomes worse, i.e.\ with the sixteenth root conformation representation the results are approximately comparable to the respective values from the logarithm conformation representations. For all Weissenberg numbers the absolute variation of the corner vortex size between all different representations decreases in magnitude when the spatial resolution of the computational domain is increased. On the mesh M4 this range of variation in $\operatorname{L}_c$ is so small that it falls below a tolerance of $2 \%$ for the maximum variation in $\operatorname{L}_c$ when different representations are used. This confirms that the results from various change-of-variable representations converge to a unique solution when the spatial resolution is successively increased, and furthermore that the numerical framework in this work is consistent. 
%%%%%%%%%%%%%%%%%%%%%%%%%%%%%%%%%%%%%%%%%%%%%%%%%%%%%%%%%%%%%%%%
%%%%%%%%%%%%%%%%%%%%%%%%%%%%%%%%%%%%%%%%%%%%%%%%%%%%%%%%%%%%%%%%
%%%%%%%%%%%%%%%%%%%%%%%%%%%%%%%%%%%%%%%%%%%%%%%%%%%%%%%%%%%%%%%%
%%%%%%%%%%%%%%%%%%%%%%%%%%%%%%%%%%%%%%%%%%%%%%%%%%%%%%%%%%%%%%%%

\begin{figure*}[h!]
\centering
%%%%%%%%%%%%%%%%%%%%%%%%%%%%%%%%
\begin{subfigure}{0.5\linewidth-9pt}
\centering
%%%%%%%%%%%%%%%%%%%%%%%%%%%%%%%%
    \begin{tikzpicture}
        \node[inner sep=0pt] (BLVC) at (0,0)
        {
            \includegraphics[width=.99\textwidth]{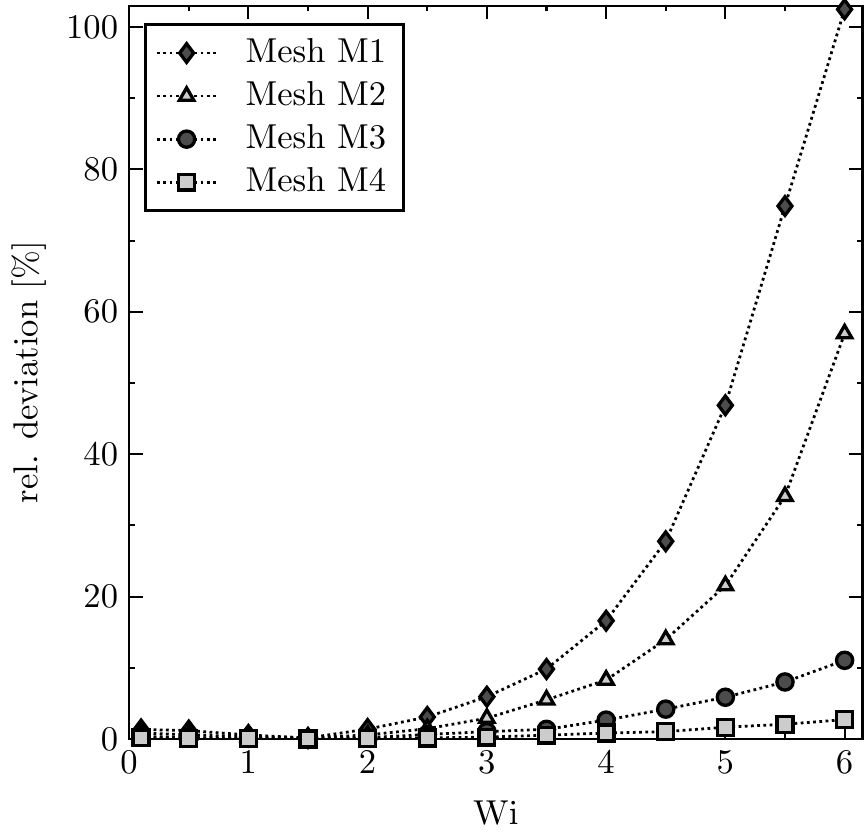}
        };
%        (0,0) at (-1.25,1.0){(a)};
    \end{tikzpicture}
%%%%%%%%%%%%%%%%%%%%%%%%%%%%%%%%
\end{subfigure}
%%%%%%%%%%%%%%%%%%%%%%%%%%%%%%%%
%%%%%%%%%%%%%%%%%%%%
\hfill
%%%%%%%%%%%%%%%%%%%%
%%%%%%%%%%%%%%%%%%%%%%%%%%%%%%%%
\begin{subfigure}{0.5\linewidth-9pt}
\centering
%%%%%%%%%%%%%%%%%%%%%%%%%%%%%%%%
    \begin{tikzpicture}
        \node[inner sep=0pt] (BLVC) at (0,0)
        {
            \includegraphics[width=.99\textwidth]{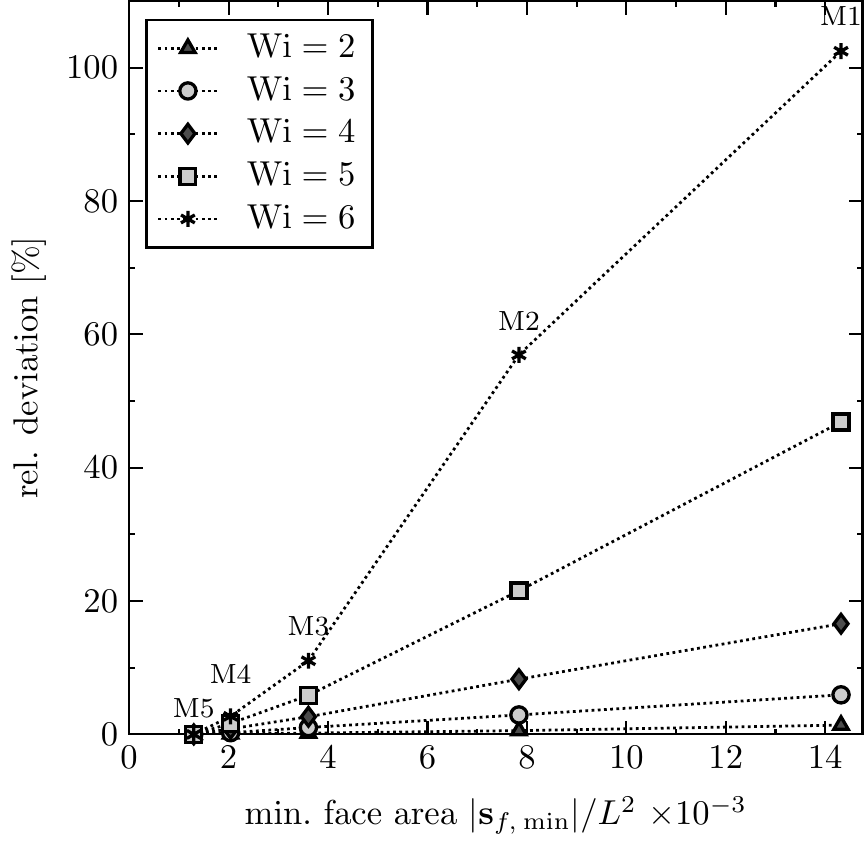}
        };
%        (0,0) at (2.65,-1.0){$\operatorname{Wi} = 0.5$};
    \end{tikzpicture}
%%%%%%%%%%%%%%%%%%%%%%%%%%%%%%%%
\end{subfigure}
%%%%%%%%%%%%%%%%%%%%%%%%%%%%%%%%
%%%%%%%%%%%%%%%%%%%%
%%%%%%%%%%%%%%%%%%%%%%%%%%%%%%%%
%%%%%%%%%%%%%%%%%%%%%%%%%%%%%%%%
    \caption{Relative deviation in the corner vortex length $\operatorname{L}_c$ from the reference solution, obtained on the finest mesh M5 with the LCRN, in $\%$. Left: Relative deviation in $\operatorname{L}_c$ over the Weissenberg number $\operatorname{Wi}$ for the meshes M1 to M4. Right: Relative deviation in $\operatorname{L}_c$ as a function of the normalized minimum cell-face area of the meshes $\vert \vec{s}_{f,\:\operatorname{min}} \vert /L^2$ for Weissenberg numbers $2$ to $6$.}
    \label{fig:relErrorsLc}
\end{figure*}

The relative deviation in the corner vortex size between the reference mesh M5 and the coarser meshes M1-M4 is shown in Figure \ref{fig:relErrorsLc}. All simulations have been carried out with the natural logarithm conformation representation. On the left-hand side, the relative deviation in the corner vortex size is plotted over $\operatorname{Wi}$. On the right hand side, the relative deviation is shown as a function of the normalized minimum cell-face area of the meshes $\vert \vec{s}_{f,\:\operatorname{min}} \vert /L^2$. The corresponding data is provided in Table \ref{table:LCRdata}. Assuming that the result from the finest mesh M5 approximately equals the mesh-independent solution, the relative deviation quantifies the error in $\operatorname{L}_c$ when coarser meshes are used. It is an interesting result that the relative deviation increases nearly exponentially for increasing $\operatorname{Wi}$ when a constant spatial resolution is used. This relation between relative deviation and the Weissenberg number is of considerable importance regarding the requirements of spatial resolution inside the domain to obtain mesh-convergence: It is demanding that the spatial resolution must be successively increased when $\operatorname{Wi}$ is increased to achieve equivalent accuracy. From the plot of the relative deviation over $\vert \vec{s}_{f,\:\operatorname{min}} \vert /L^2$ we get an idea on the required grid spacing to obtain a certain degree of accuracy. It is notable that while this dependence is approximately linear for smaller $\operatorname{Wi}$, it is rather parabolic for higher $\operatorname{Wi}$. With these considerations, it is not surprising that we run into limitations regarding the computational costs, preventing us from computing mesh-independent solutions at high Weissenberg numbers, even though we have overcome the HWNP by the change-of-variable representations.
%%%%%%%%%%%%%%%%%%%%%%%%%%%%%%%%%%%%%%%%%%%%%%%%%%%%%%%%%%%%%%%%
%%%%%%%%%%%%%%%%%%%%%%%%%%%%%%%%%%%%%%%%%%%%%%%%%%%%%%%%%%%%%%%%
%%%%%%%%%%%%%%%%%%%%%%%%%%%%%%%%%%%%%%%%%%%%%%%%%%%%%%%%%%%%%%%%

%%%%%%%%%%%%%%%%%%%%%%%%%%%%%%%%%%%%%%%%%%%%%%%%%%%%%%%%%%%%%%%%
%%%%%%%%%%%%%%%%%%%%%%%%%%%%%%%%%%%%%%%%%%%%%%%%%%%%%%%%%%%%%%%%
\begin{table}[h!]
\begin{center}
\begin{tabular}{@{}lrrrrrrrrrr@{}}
\toprule
\quad &
\multicolumn{3}{c}{vortex size $\operatorname{L}_c$} &
\multicolumn{4}{c}{relative deviation in $\operatorname{L}_c$ $[\%]$} &
\multicolumn{3}{c}{vortex intensity $\operatorname{I}_c$} \\
%%%%%%%%%%%%%%%%%%%%%%%%%%%%%%%%%%%%%%%%%%%%%%%%%%%%%%%%%%%%%%%%
$\operatorname{Wi}$ \quad &
\multicolumn{1}{c}{M3} & \multicolumn{1}{c}{M4} & \multicolumn{1}{c}{M5} &
\multicolumn{1}{c}{M4} & \multicolumn{1}{c}{M3} & \multicolumn{1}{c}{M2} & \multicolumn{1}{c}{M1} &
\multicolumn{1}{c}{M3} & \multicolumn{1}{c}{M4} & \multicolumn{1}{c}{M5} \\
\midrule
$0.1$ & $1.484$ & $1.487$ & $1.490$ & $0.16$ & $0.36$ & $0.84$ & $1.33$ & $1.147$ & $1.149$ & $1.150$ \\
$0.5$ & $1.442$ & $1.444$ & $1.445$ & $0.10$ & $0.21$ & $0.59$ & $1.18$ & $1.003$ & $1.001$ & $1.000$ \\
$1.0$ & $1.366$ & $1.367$ & $1.367$ & $0.04$ & $0.08$ & $0.32$ & $0.55$ & $0.793$ & $0.789$ & $0.785$ \\
$1.5$ & $1.275$ & $1.275$ & $1.275$ & $0.01$ & $0.03$ & $0.15$ & $0.14$ & $0.598$ & $0.593$ & $0.590$ \\
$2.0$ & $1.178$ & $1.176$ & $1.175$ & $0.08$ & $0.19$ & $0.56$ & $1.39$ & $0.440$ & $0.435$ & $0.431$ \\
$2.5$ & $1.080$ & $1.074$ & $1.073$ & $0.15$ & $0.63$ & $1.44$ & $3.08$ & $0.321$ & $0.313$ & $0.310$ \\
$3.0$ & $0.982$ & $0.975$ & $0.972$ & $0.26$ & $1.03$ & $2.92$ & $5.93$ & $0.233$ & $0.225$ & $0.222$ \\
$3.5$ & $0.888$ & $0.880$ & $0.876$ & $0.54$ & $1.35$ & $5.50$ & $9.82$ & $0.419$ & $0.206$ & $0.160$ \\
$4.0$ & $0.809$ & $0.795$ & $0.788$ & $0.83$ & $2.65$ & $8.28$ & $16.60$ & $0.978$ & $0.652$ & $0.336$ \\
$4.5$ & $0.738$ & $0.715$ & $0.708$ & $1.03$ & $4.19$ & $13.96$ & $27.76$ & $1.438$ & $1.034$ & $0.730$ \\
$5.0$ & $0.676$ & $0.649$ & $0.639$ & $1.65$ & $5.86$ & $21.54$ & $46.86$ & $1.866$ & $1.456$ & $1.155$ \\
$5.5$ & $0.626$ & $0.591$ & $0.579$ & $2.07$ & $8.00$ & $34.08$ & $74.85$ & $2.366$ & $1.846$ & $1.550$ \\
$6.0$ & $0.588$ & $0.544$ & $0.529$ & $2.71$ & $11.04$ & $56.92$ & $102.47$ & $2.926$ & $2.348$ & $2.007$ \\
\bottomrule
\end{tabular}
\end{center}
\caption{Benchmark results for the Oldroyd-B fluid, carried out with the natural logarithm conformation representation. The corner vortex size $\operatorname{L}_c$ and the corner vortex intensity $\operatorname{I}_c$ are listed for the meshes M3, M4 and M5. The results from mesh M5 represent the most accurate and should be used as reference to this work. The relative deviation in $\operatorname{L}_c$ from the reference results on the mesh M5 is given for all coarser meshes M1 to M4.}
\label{table:LCRdata}
\end{table}
%%%%%%%%%%%%%%%%%%%%%%%%%%%%%%%%%%%%%%%%%%%%%%%%%%%%%%%%%%%%%%%%
%%%%%%%%%%%%%%%%%%%%%%%%%%%%%%%%%%%%%%%%%%%%%%%%%%%%%%%%%%%%%%%%

\begin{figure*}[h!]
\centering
%%%%%%%%%%%%%%%%%%%%%%%%%%%%%%%%
\begin{subfigure}{0.5\linewidth-9pt}
\centering
%%%%%%%%%%%%%%%%%%%%%%%%%%%%%%%%
    \begin{tikzpicture}
        \node[inner sep=0pt] (BLVC) at (0,0)
        {
            \includegraphics[width=.99\textwidth]{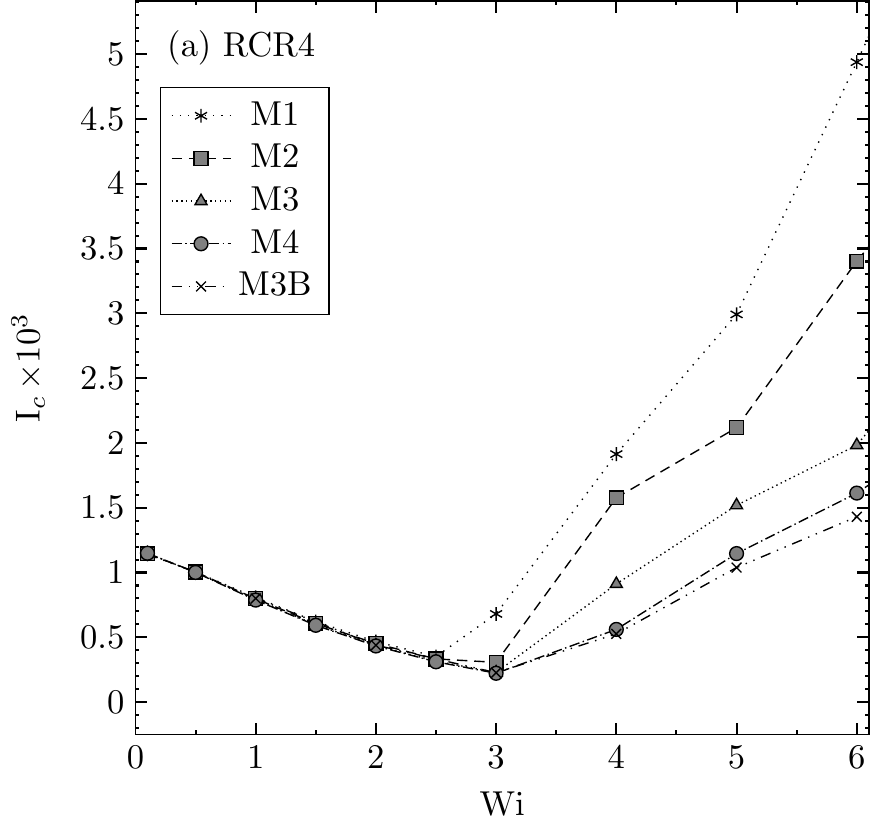}
        };
%        (0,0) at (-1.25,1.0){(a)};
    \end{tikzpicture}
%%%%%%%%%%%%%%%%%%%%%%%%%%%%%%%%
\end{subfigure}
%%%%%%%%%%%%%%%%%%%%%%%%%%%%%%%%
%%%%%%%%%%%%%%%%%%%%
\hfill
%%%%%%%%%%%%%%%%%%%%
%%%%%%%%%%%%%%%%%%%%%%%%%%%%%%%%
\begin{subfigure}{0.5\linewidth-9pt}
\centering
%%%%%%%%%%%%%%%%%%%%%%%%%%%%%%%%
    \begin{tikzpicture}
        \node[inner sep=0pt] (BLVC) at (0,0)
        {
            \includegraphics[width=.99\textwidth]{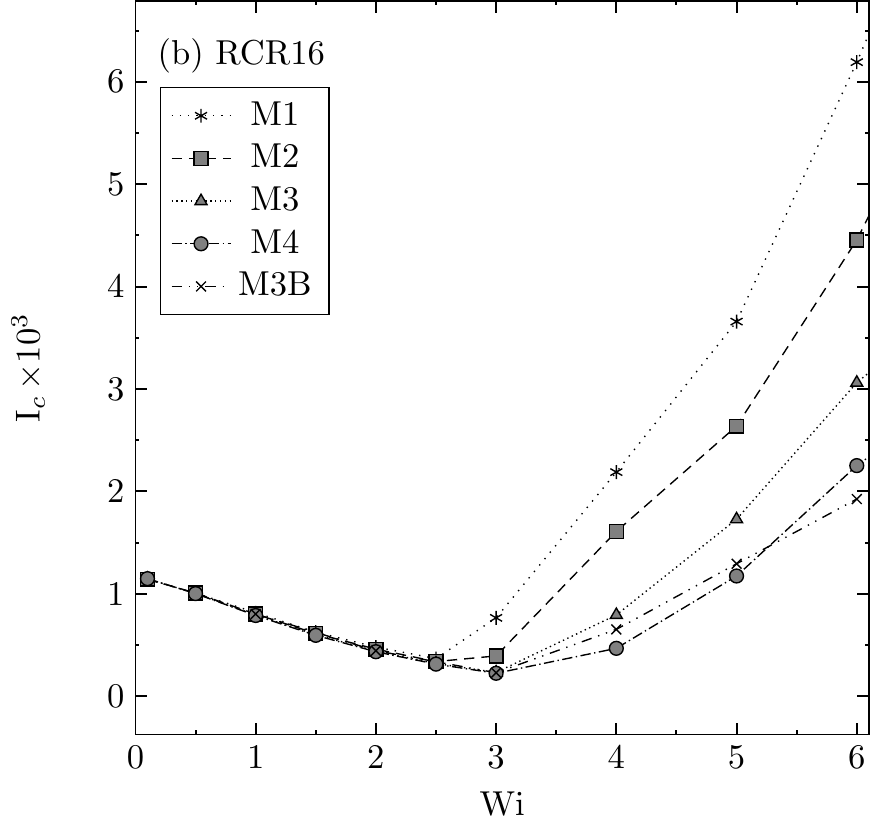}
        };
%        (0,0) at (2.65,-1.0){$\operatorname{Wi} = 0.5$};
    \end{tikzpicture}
%%%%%%%%%%%%%%%%%%%%%%%%%%%%%%%%
\end{subfigure}
%%%%%%%%%%%%%%%%%%%%%%%%%%%%%%%%
%%%%%%%%%%%%%%%%%%%%
\\
%%%%%%%%%%%%%%%%%%%%
%%%%%%%%%%%%%%%%%%%%%%%%%%%%%%%%
\begin{subfigure}{0.5\linewidth-9pt}
\centering
%%%%%%%%%%%%%%%%%%%%%%%%%%%%%%%%
    \begin{tikzpicture}
        \node[inner sep=0pt] (BLVC) at (0,0)
        {
            \includegraphics[width=.99\textwidth]{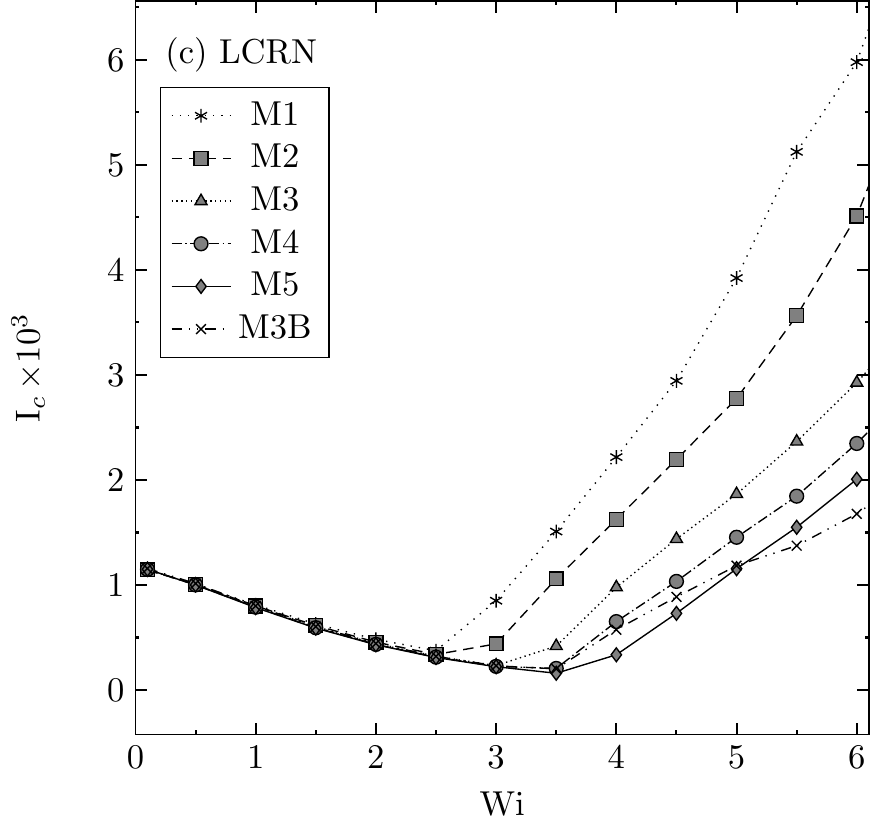}
        };
%        (0,0) at (2.65,-1.0){$\operatorname{Wi} = 0.5$};
    \end{tikzpicture}
%%%%%%%%%%%%%%%%%%%%%%%%%%%%%%%%
\end{subfigure}
%%%%%%%%%%%%%%%%%%%%%%%%%%%%%%%%
%%%%%%%%%%%%%%%%%%%%
\hfill
%%%%%%%%%%%%%%%%%%%%
%%%%%%%%%%%%%%%%%%%%%%%%%%%%%%%%
\begin{subfigure}{0.5\linewidth-9pt}
\centering
%%%%%%%%%%%%%%%%%%%%%%%%%%%%%%%%
    \begin{tikzpicture}
        \node[inner sep=0pt] (BLVC) at (0,0)
        {
            \includegraphics[width=.99\textwidth]{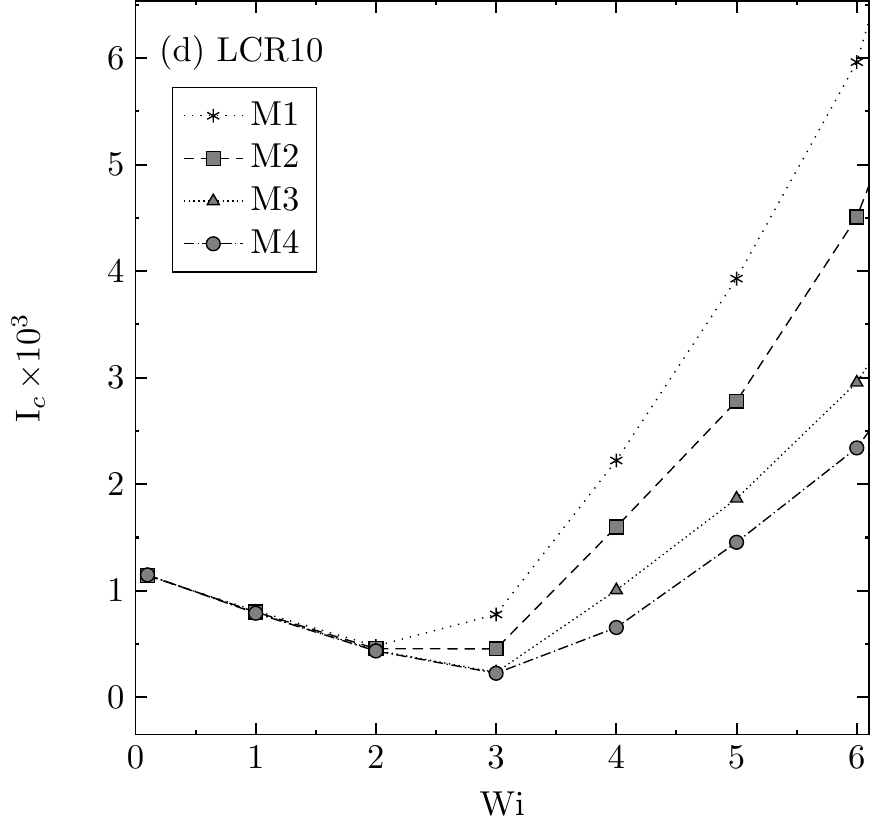}
        };
%        (0,0) at (2.65,-1.0){$\operatorname{Wi} = 0.5$};
    \end{tikzpicture}
%%%%%%%%%%%%%%%%%%%%%%%%%%%%%%%%
\end{subfigure}
%%%%%%%%%%%%%%%%%%%%%%%%%%%%%%%%
    \caption{Corner vortex intensity $\operatorname{I}_c$ over the Weissenberg number $\operatorname{Wi}$ for $\operatorname{Wi} \leq 6$. Comparison of the results from different change-of-variable representations on the meshes M1 to M5: (a) RCR4, (b) RCR16, (c) LCRN, (d) LCR10.}
    \label{fig:intensityWiSix}
\end{figure*}

Figure \ref{fig:intensityWiSix} shows the corner vortex intensity $\operatorname{I}_c$ as a function of the Weissenberg number for the meshes M1 to M5 and for the four change-of-variable representations: LCRN, LCR10, RCR4 and RCR16, respectively. The vortex intensity decreases in magnitude for increasing $\operatorname{Wi}$ up to $\operatorname{Wi} \approx 3.5$. From this point on, the vortex intensity increases as $\operatorname{Wi}$ is further increased. The change between the two regimes is located at smaller Weissenberg numbers when a coarser mesh is used. We observe that $\operatorname{I}_c$ is overpredicted for all representations on the coarse meshes. The variation in $\operatorname{I}_c$ between the finest and the coarsest mesh is to a considerable degree higher in the second regime, which shows that the mesh-sensitivity in $\operatorname{I}_c$ escalates above the critical Weissenberg number. For $\operatorname{Wi} > 3.5$, the corner vortex intensity has significantly poorer convergence properties as the corner vortex size, i.e.\ even on the finest mesh M5 the accuracy of $\operatorname{I}_c$ is not yet satisfactory. Therefore, $\operatorname{I}_c$ poses even stricter conditions to the mesh-convergence as $\operatorname{L}_c$. For the corner vortex intensity, we have not achieved mesh-independent solutions for $\operatorname{Wi} > 3.5$.
%%%%%%%%%%%%%%%%%%%%%%%%%%%%%%%%%%%%%%%%%%%%%%%%%%%%%%%%%%%%%%%%
%%%%%%%%%%%%%%%%%%%%%%%%%%%%%%%%%%%%%%%%%%%%%%%%%%%%%%%%%%%%%%%%
%%%%%%%%%%%%%%%%%%%%%%%%%%%%%%%%%%%%%%%%%%%%%%%%%%%%%%%%%%%%%%%%
%%%%%%%%%%%%%%%%%%%%%%%%%%%%%%%%%%%%%%%%%%%%%%%%%%%%%%%%%%%%%%%%

\begin{figure*}[h!]
\centering
%%%%%%%%%%%%%%%%%%%%%%%%%%%%%%%%
\begin{subfigure}{0.5\linewidth-9pt}
\centering
%%%%%%%%%%%%%%%%%%%%%%%%%%%%%%%%
    \begin{tikzpicture}
        \node[inner sep=0pt] (BLVC) at (0,0)
        {
            \includegraphics[width=.99\textwidth]{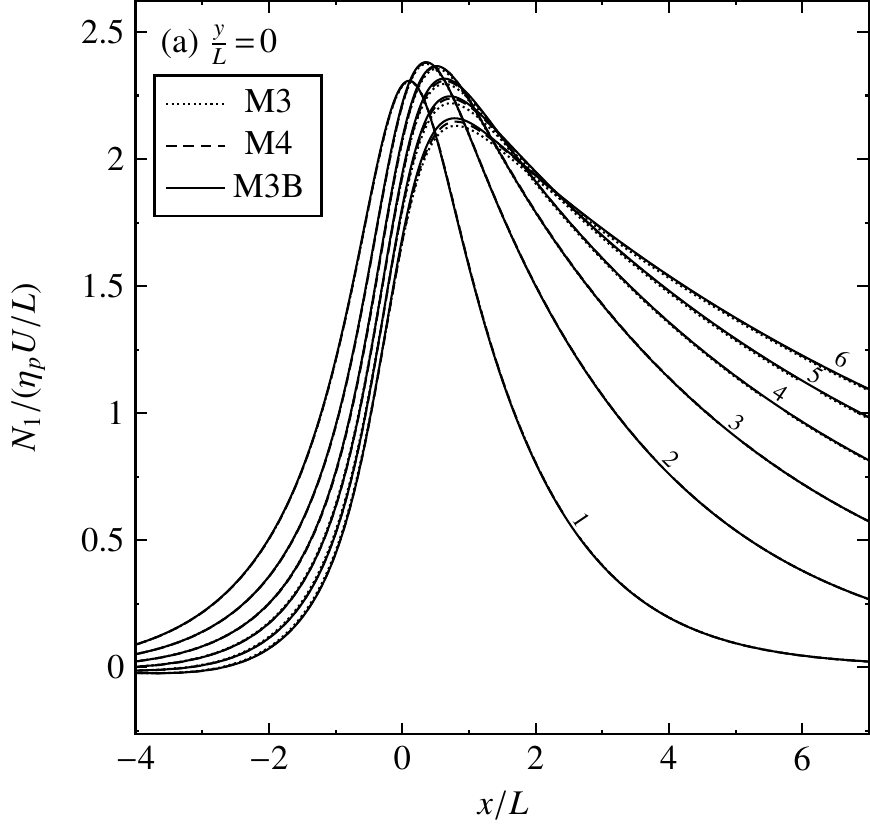}
        };
%        (0,0) at (-1.25,1.0){(a)};
    \end{tikzpicture}
%%%%%%%%%%%%%%%%%%%%%%%%%%%%%%%%
\end{subfigure}
%%%%%%%%%%%%%%%%%%%%%%%%%%%%%%%%
%%%%%%%%%%%%%%%%%%%%
\hfill
%%%%%%%%%%%%%%%%%%%%
%%%%%%%%%%%%%%%%%%%%%%%%%%%%%%%%
\begin{subfigure}{0.5\linewidth-9pt}
\centering
%%%%%%%%%%%%%%%%%%%%%%%%%%%%%%%%
    \begin{tikzpicture}
        \node[inner sep=0pt] (BLVC) at (0,0)
        {
            \includegraphics[width=.99\textwidth]{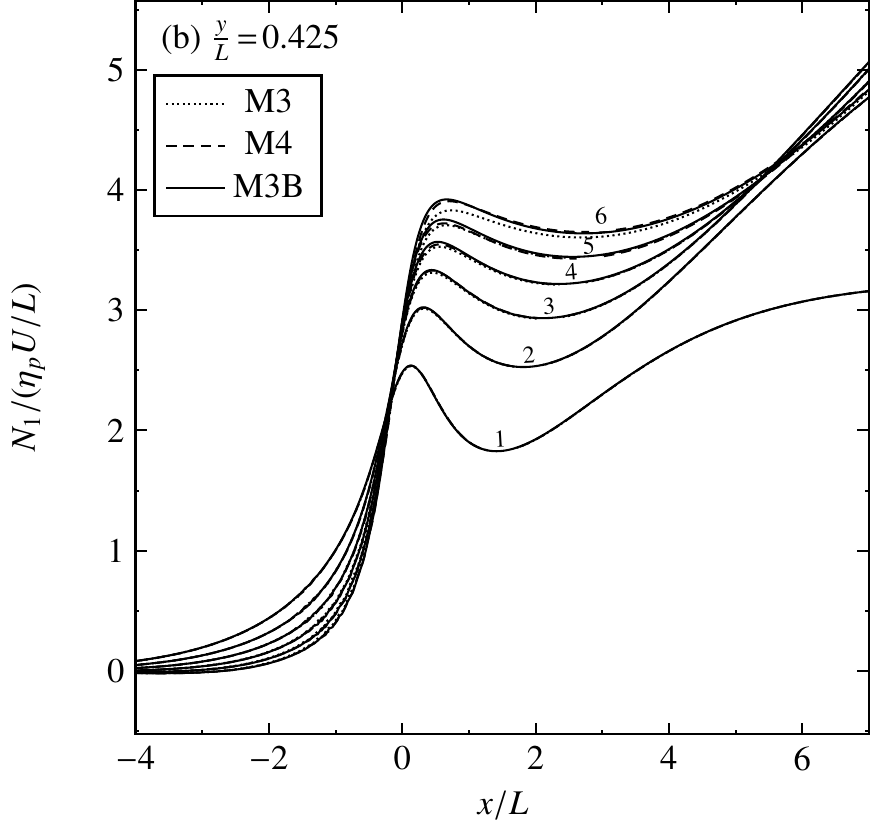}
        };
%        (0,0) at (2.65,-1.0){$\operatorname{Wi} = 0.5$};
    \end{tikzpicture}
%%%%%%%%%%%%%%%%%%%%%%%%%%%%%%%%
\end{subfigure}
%%%%%%%%%%%%%%%%%%%%%%%%%%%%%%%%
%%%%%%%%%%%%%%%%%%%%
\\
%%%%%%%%%%%%%%%%%%%%
%%%%%%%%%%%%%%%%%%%%%%%%%%%%%%%%
\begin{subfigure}{0.5\linewidth-9pt}
\centering
%%%%%%%%%%%%%%%%%%%%%%%%%%%%%%%%
    \begin{tikzpicture}
        \node[inner sep=0pt] (BLVC) at (0,0)
        {
            \includegraphics[width=.99\textwidth]{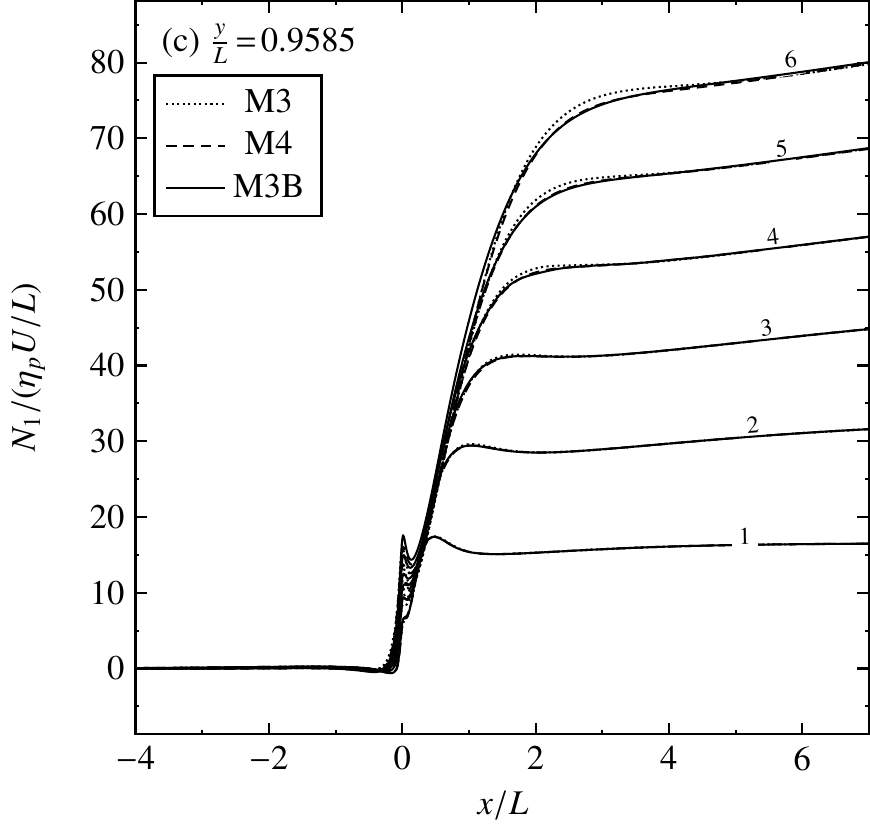}
        };
%        (0,0) at (2.65,-1.0){$\operatorname{Wi} = 0.5$};
    \end{tikzpicture}
%%%%%%%%%%%%%%%%%%%%%%%%%%%%%%%%
\end{subfigure}
%%%%%%%%%%%%%%%%%%%%%%%%%%%%%%%%
%%%%%%%%%%%%%%%%%%%%
\hfill
%%%%%%%%%%%%%%%%%%%%
%%%%%%%%%%%%%%%%%%%%%%%%%%%%%%%%
\begin{subfigure}{0.5\linewidth-9pt}
\centering
%%%%%%%%%%%%%%%%%%%%%%%%%%%%%%%%
    \begin{tikzpicture}
        \node[inner sep=0pt] (BLVC) at (0,0)
        {
            \includegraphics[width=.99\textwidth]{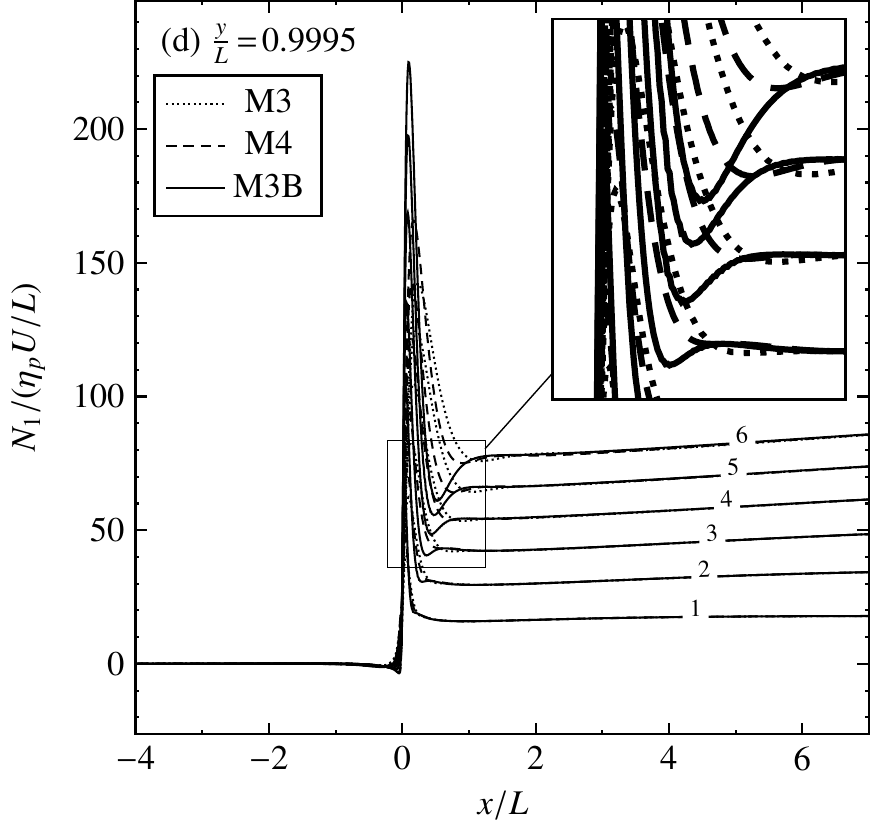}
        };
%        (0,0) at (2.65,-1.0){$\operatorname{Wi} = 0.5$};
    \end{tikzpicture}
%%%%%%%%%%%%%%%%%%%%%%%%%%%%%%%%
\end{subfigure}
%%%%%%%%%%%%%%%%%%%%%%%%%%%%%%%%
    \caption{Distribution of the RCR4 axial first normal stress differences for $1 \leq \operatorname{Wi} \leq 6$ along: (a) center line, (b) $y/L = 0.425$, (c) y/L = 0.9585, (d) y/L = 0.9995}
    \label{fig:normalstressdiffRCR4}
\end{figure*}

\begin{figure*}[h!]
\centering
%%%%%%%%%%%%%%%%%%%%%%%%%%%%%%%%
\begin{subfigure}{0.5\linewidth-9pt}
\centering
%%%%%%%%%%%%%%%%%%%%%%%%%%%%%%%%
    \begin{tikzpicture}
        \node[inner sep=0pt] (BLVC) at (0,0)
        {
            \includegraphics[width=.99\textwidth]{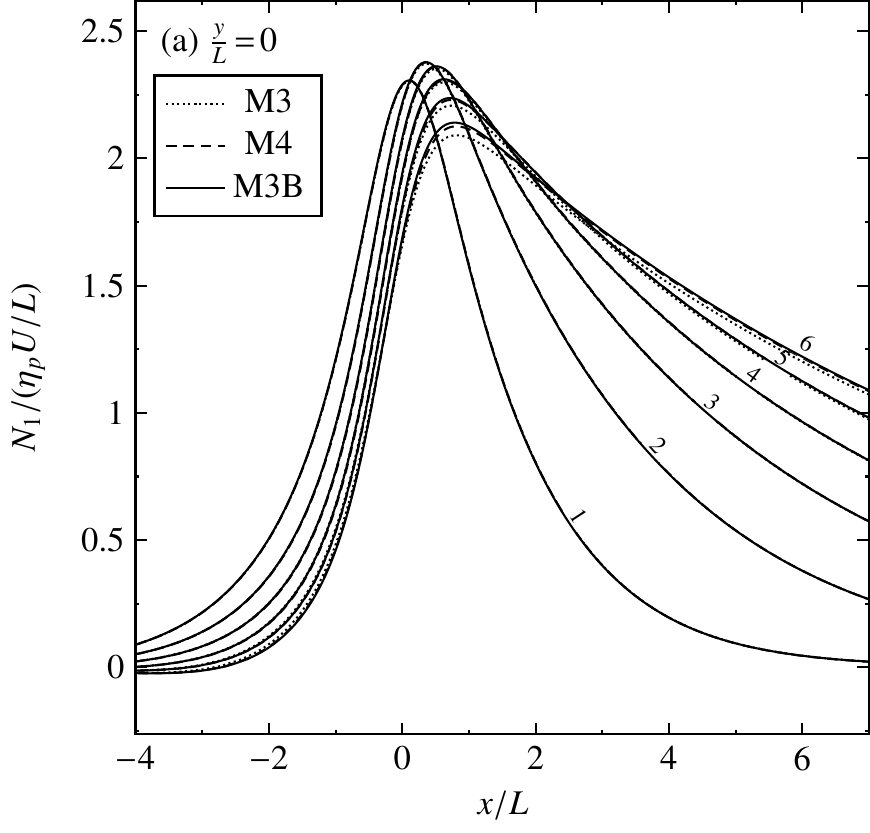}
        };
%        (0,0) at (-1.25,1.0){(a)};
    \end{tikzpicture}
%%%%%%%%%%%%%%%%%%%%%%%%%%%%%%%%
\end{subfigure}
%%%%%%%%%%%%%%%%%%%%%%%%%%%%%%%%
%%%%%%%%%%%%%%%%%%%%
\hfill
%%%%%%%%%%%%%%%%%%%%
%%%%%%%%%%%%%%%%%%%%%%%%%%%%%%%%
\begin{subfigure}{0.5\linewidth-9pt}
\centering
%%%%%%%%%%%%%%%%%%%%%%%%%%%%%%%%
    \begin{tikzpicture}
        \node[inner sep=0pt] (BLVC) at (0,0)
        {
            \includegraphics[width=.99\textwidth]{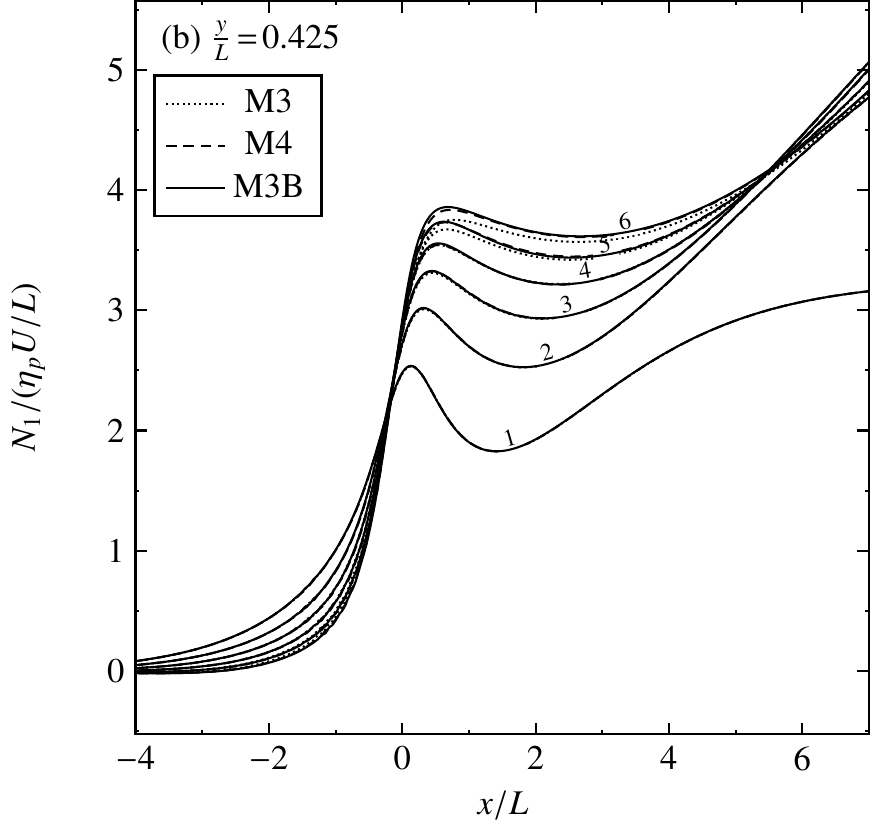}
        };
%        (0,0) at (2.65,-1.0){$\operatorname{Wi} = 0.5$};
    \end{tikzpicture}
%%%%%%%%%%%%%%%%%%%%%%%%%%%%%%%%
\end{subfigure}
%%%%%%%%%%%%%%%%%%%%%%%%%%%%%%%%
%%%%%%%%%%%%%%%%%%%%
\\
%%%%%%%%%%%%%%%%%%%%
%%%%%%%%%%%%%%%%%%%%%%%%%%%%%%%%
\begin{subfigure}{0.5\linewidth-9pt}
\centering
%%%%%%%%%%%%%%%%%%%%%%%%%%%%%%%%
    \begin{tikzpicture}
        \node[inner sep=0pt] (BLVC) at (0,0)
        {
            \includegraphics[width=.99\textwidth]{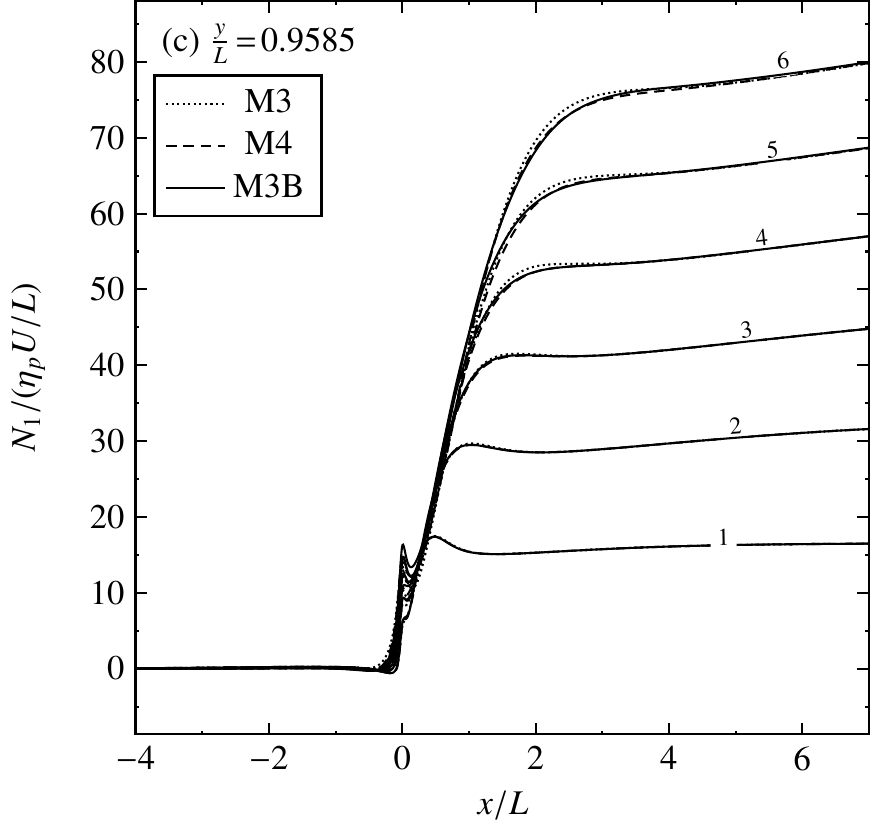}
        };
%        (0,0) at (2.65,-1.0){$\operatorname{Wi} = 0.5$};
    \end{tikzpicture}
%%%%%%%%%%%%%%%%%%%%%%%%%%%%%%%%
\end{subfigure}
%%%%%%%%%%%%%%%%%%%%%%%%%%%%%%%%
%%%%%%%%%%%%%%%%%%%%
\hfill
%%%%%%%%%%%%%%%%%%%%
%%%%%%%%%%%%%%%%%%%%%%%%%%%%%%%%
\begin{subfigure}{0.5\linewidth-9pt}
\centering
%%%%%%%%%%%%%%%%%%%%%%%%%%%%%%%%
    \begin{tikzpicture}
        \node[inner sep=0pt] (BLVC) at (0,0)
        {
            \includegraphics[width=.99\textwidth]{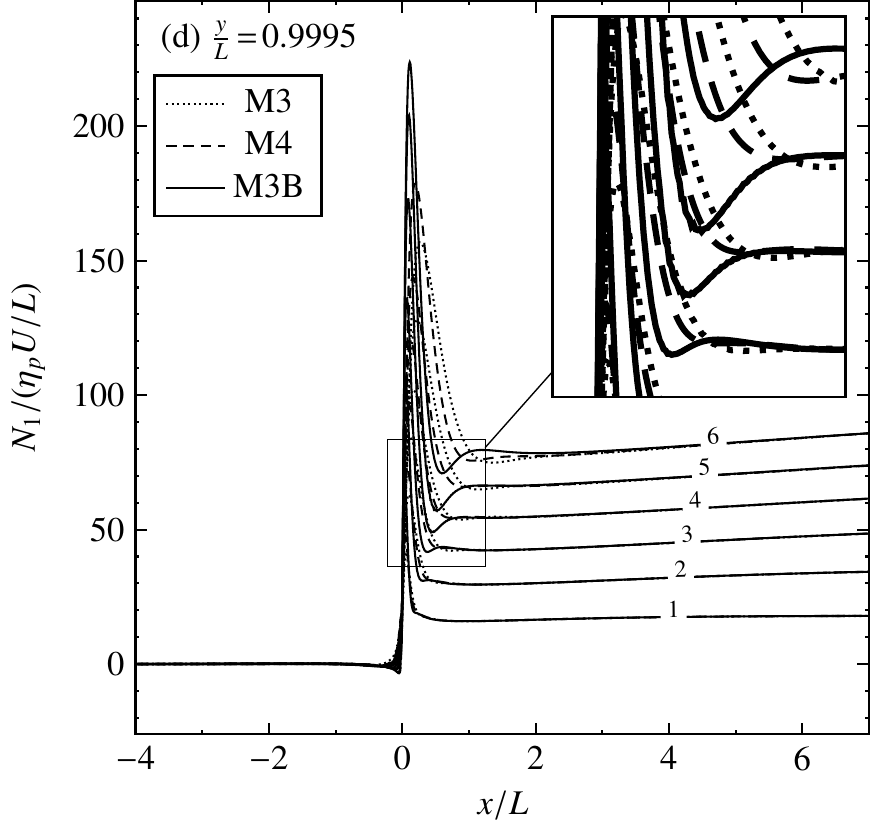}
        };
%        (0,0) at (2.65,-1.0){$\operatorname{Wi} = 0.5$};
    \end{tikzpicture}
%%%%%%%%%%%%%%%%%%%%%%%%%%%%%%%%
\end{subfigure}
%%%%%%%%%%%%%%%%%%%%%%%%%%%%%%%%
    \caption{Distribution of the RCR16 axial first normal stress differences for $1 \leq \operatorname{Wi} \leq 6$ along: (a) center line, (b) $y/L = 0.425$, (c) y/L = 0.9585, (d) y/L = 0.9995}
    \label{fig:normalstressdiffRCR16}
\end{figure*}

\begin{figure*}[h!]
\centering
%%%%%%%%%%%%%%%%%%%%%%%%%%%%%%%%
\begin{subfigure}{0.5\linewidth-9pt}
\centering
%%%%%%%%%%%%%%%%%%%%%%%%%%%%%%%%
    \begin{tikzpicture}
        \node[inner sep=0pt] (BLVC) at (0,0)
        {
            \includegraphics[width=.99\textwidth]{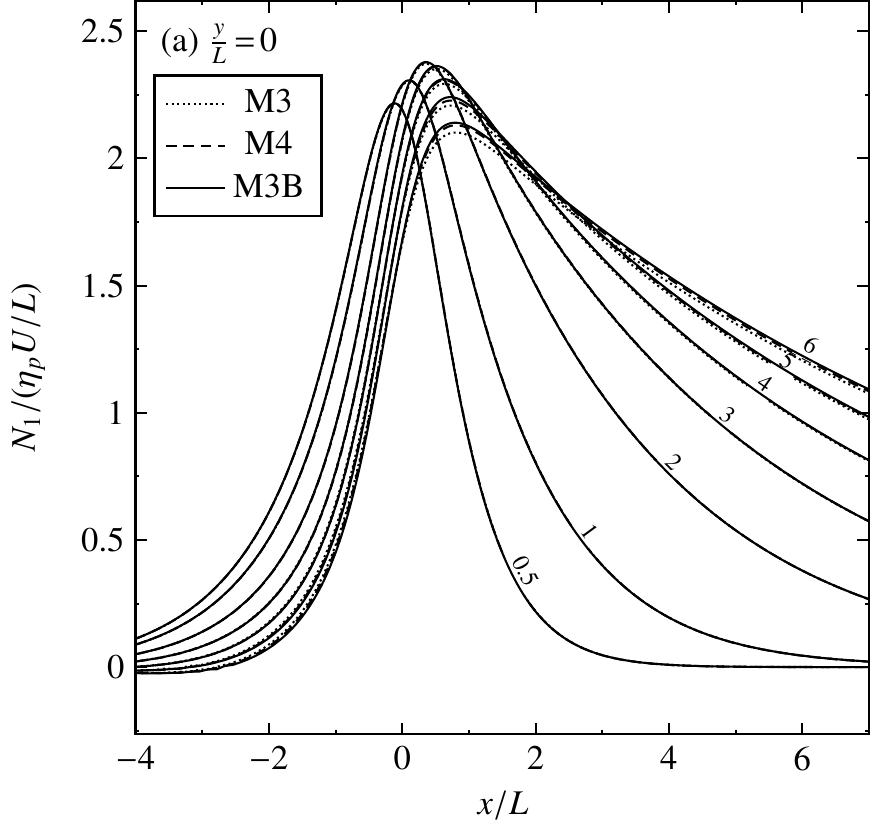}
        };
%        (0,0) at (-1.25,1.0){(a)};
    \end{tikzpicture}
%%%%%%%%%%%%%%%%%%%%%%%%%%%%%%%%
\end{subfigure}
%%%%%%%%%%%%%%%%%%%%%%%%%%%%%%%%
%%%%%%%%%%%%%%%%%%%%
\hfill
%%%%%%%%%%%%%%%%%%%%
%%%%%%%%%%%%%%%%%%%%%%%%%%%%%%%%
\begin{subfigure}{0.5\linewidth-9pt}
\centering
%%%%%%%%%%%%%%%%%%%%%%%%%%%%%%%%
    \begin{tikzpicture}
        \node[inner sep=0pt] (BLVC) at (0,0)
        {
            \includegraphics[width=.99\textwidth]{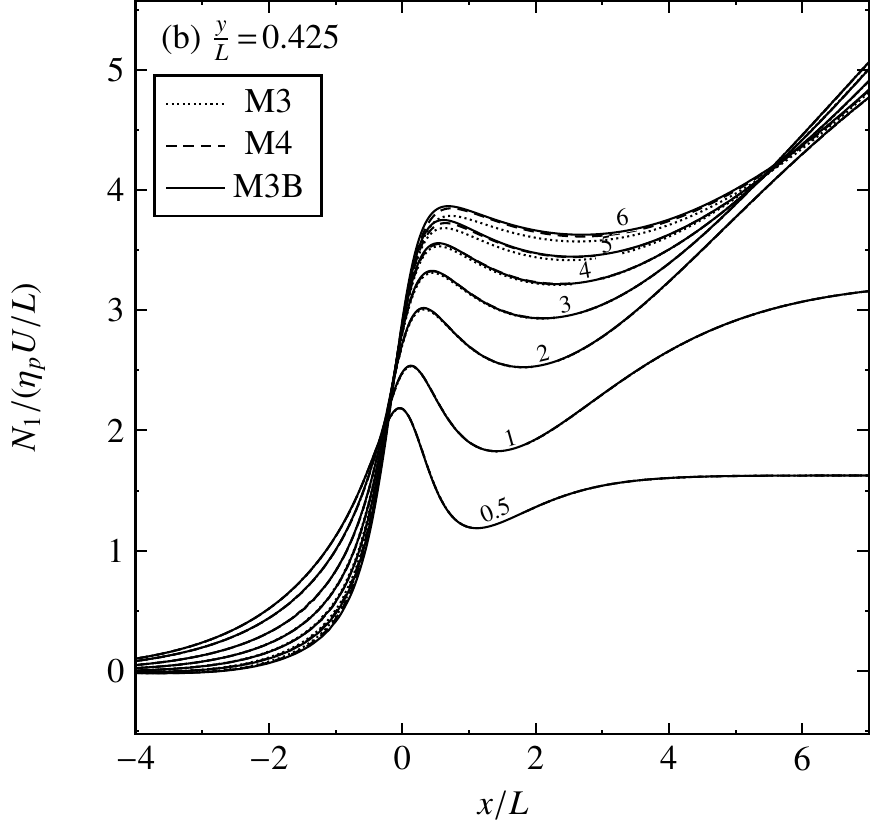}
        };
%        (0,0) at (2.65,-1.0){$\operatorname{Wi} = 0.5$};
    \end{tikzpicture}
%%%%%%%%%%%%%%%%%%%%%%%%%%%%%%%%
\end{subfigure}
%%%%%%%%%%%%%%%%%%%%%%%%%%%%%%%%
%%%%%%%%%%%%%%%%%%%%
\\
%%%%%%%%%%%%%%%%%%%%
%%%%%%%%%%%%%%%%%%%%%%%%%%%%%%%%
\begin{subfigure}{0.5\linewidth-9pt}
\centering
%%%%%%%%%%%%%%%%%%%%%%%%%%%%%%%%
    \begin{tikzpicture}
        \node[inner sep=0pt] (BLVC) at (0,0)
        {
            \includegraphics[width=.99\textwidth]{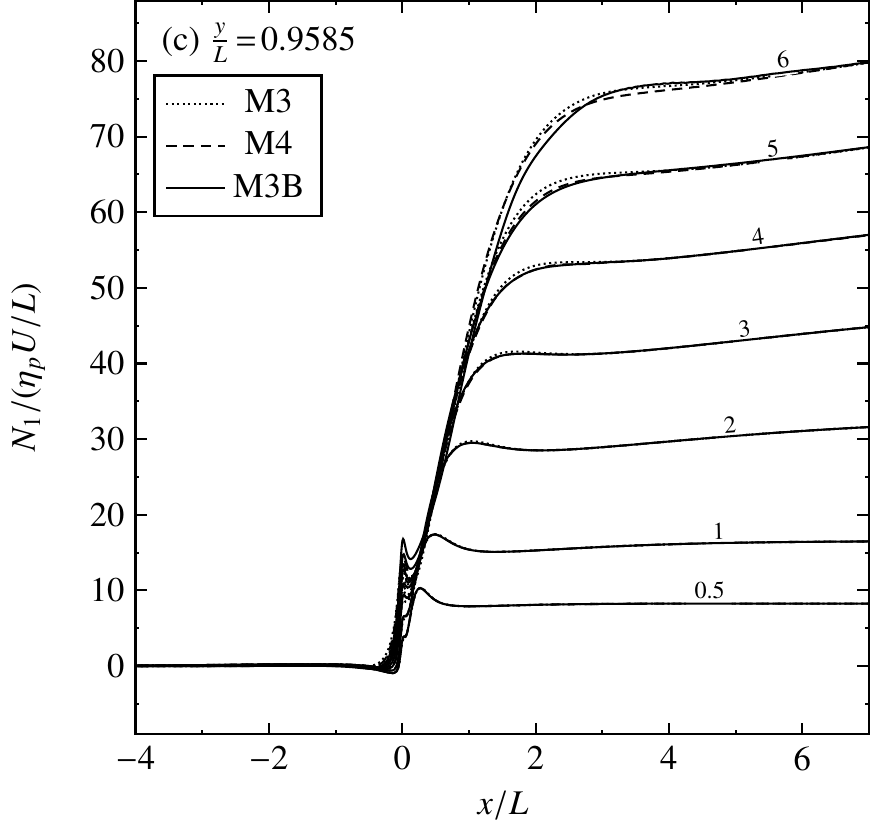}
        };
%        (0,0) at (2.65,-1.0){$\operatorname{Wi} = 0.5$};
    \end{tikzpicture}
%%%%%%%%%%%%%%%%%%%%%%%%%%%%%%%%
\end{subfigure}
%%%%%%%%%%%%%%%%%%%%%%%%%%%%%%%%
%%%%%%%%%%%%%%%%%%%%
\hfill
%%%%%%%%%%%%%%%%%%%%
%%%%%%%%%%%%%%%%%%%%%%%%%%%%%%%%
\begin{subfigure}{0.5\linewidth-9pt}
\centering
%%%%%%%%%%%%%%%%%%%%%%%%%%%%%%%%
    \begin{tikzpicture}
        \node[inner sep=0pt] (BLVC) at (0,0)
        {
            \includegraphics[width=.99\textwidth]{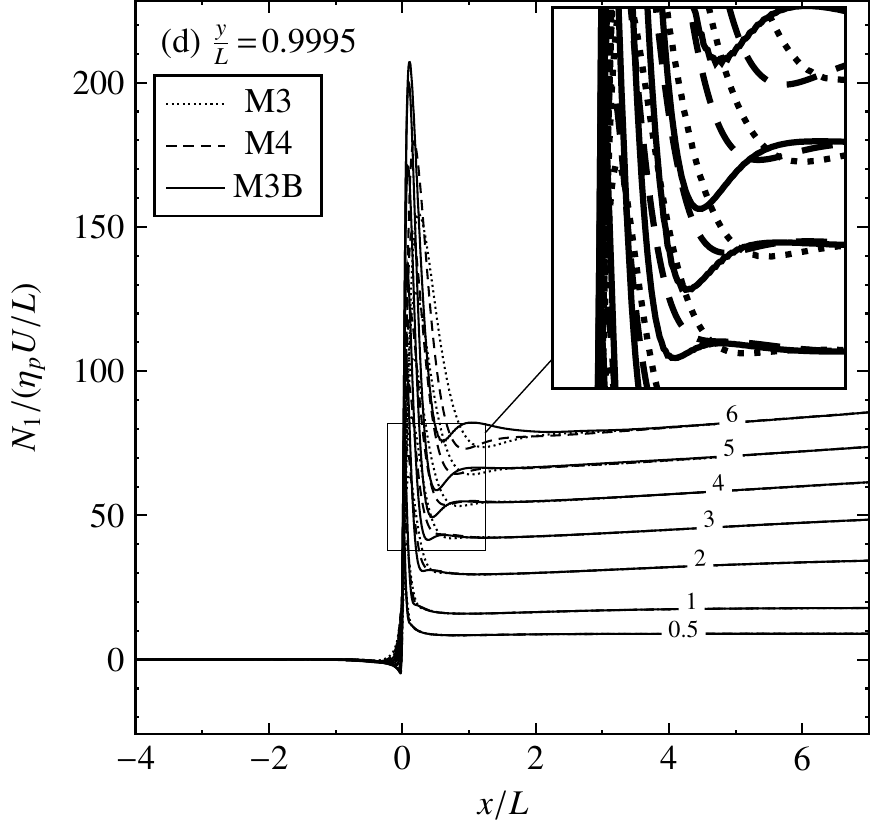}
        };
%        (0,0) at (2.65,-1.0){$\operatorname{Wi} = 0.5$};
    \end{tikzpicture}
%%%%%%%%%%%%%%%%%%%%%%%%%%%%%%%%
\end{subfigure}
%%%%%%%%%%%%%%%%%%%%%%%%%%%%%%%%
    \caption{Distribution of the LCRN axial first normal stress differences for $0.5 \leq \operatorname{Wi} \leq 6$ along: (a) center line, (b) $y/L = 0.425$, (c) y/L = 0.9585, (d) y/L = 0.9995}
    \label{fig:normalstressdiffLCRN}
\end{figure*}

The axial profiles of the normalized first normal stress differences are shown for the fourth root conformation representation in Figure \ref{fig:normalstressdiffRCR4}, for the sixteenth root conformation representation in Figure \ref{fig:normalstressdiffRCR16} and for the natural logarithm representation in Figure \ref{fig:normalstressdiffLCRN}. The normal stresses are normalized as $N_1/(\eta_p {U}/{L})$. The profiles are evaluated along four different lines inside the domain: The center line, $y/L = 0.425$, y/L = 0.9585 and y/L = 0.9995, respectively. The re-entrant corner of the smaller channel (with $x/L$ = 0) is used as a reference point to scale the $x/L$ axis. Therefore, a negative $x/L$ represents points in the domain which are located upstream to the re-entrant corner in the larger channel. A close agreement of the profiles from the meshes M3, M4 and M3B is observed for the center line and for $y/L = 0.425$ in all representations. At higher Weissenberg numbers, there are some slight variations downstream from the re-entrant corner between the profiles from the coarsest mesh M3 and the finer meshes M3B and M4, indicating that the stress boundary layer on the mesh M3 should be resolved slightly better. These variations increase locally near the re-entrant corner as we move closer to the channel wall, i.e. at $y/L = 0.9585$. In these profiles, we observe a local maximum of the first normal stress differences near the re-entrant corner for all $\operatorname{Wi} > 1$, which increases in size for increasing $\operatorname{Wi}$. Around this local maximum, we have higher variations in the first normal stress differences between the results of the three different meshes, which shows that the mesh-sensitivity is locally increased around the re-entrant corner. However, there is still a good agreement of the results for $x/L \gtrless 1$. Therefore, we may assume that the stress profiles are well-resolved in all simulations, except for a very small region near the re-entrant corner. This is not surprising, because of the geometrical re-entrant corner singularity which cannot be resolved, regardless of the provided spatial resolution in the computational grid. However, we shall distinguish between the issue of the geometrical singularity and the difficulty of resolving the stress boundary layers in the neighborhood of the channel walls. The latter can be achieved with a finite spatial resolution. In the profile along $y/L = 0.9995$, which is located in the direct neighborhood of the smaller channel wall, the local maximum near the re-entrant corner increases intensively in magnitude, as compared to $y/L = 0.9585$. In the region around this local peak, the discrepancy between the normal stress differences on the different meshes is notable in all representations, which manifests the issue that we cannot expect any mesh-convergence locally at the geometrical singularity. We observe that the more spatial resolution is provided in the boundary layer, the higher the peak is in the normal stresses. It is interesting that the profiles from mesh M3B manifest some local details in the normal stresses near the re-entrant corner which are not seen in the meshes M3 and M4 with larger cell-spacing normal to the channel walls: For $\operatorname{Wi} \geq 2$, there is additionally a local minimum, slightly downstream to the re-entrant corner, which increases in magnitude for increasing $\operatorname{Wi}$. This local undershoot in the normal stress-differences at the entrance into the smaller channel is observed only when considerably more spatial resolution is provided to resolve the stress boundary layer, as is done in the mesh M3B.
%%%%%%%%%%%%%%%%%%%%%%%%%%%%%%%%%%%%%%%%%%%%%%%%%%%%%%%%%%%%%%%%
%%%%%%%%%%%%%%%%%%%%%%%%%%%%%%%%%%%%%%%%%%%%%%%%%%%%%%%%%%%%%%%%
%%%%%%%%%%%%%%%%%%%%%%%%%%%%%%%%%%%%%%%%%%%%%%%%%%%%%%%%%%%%%%%%
%%%%%%%%%%%%%%%%%%%%%%%%%%%%%%%%%%%%%%%%%%%%%%%%%%%%%%%%%%%%%%%%

%%%%%%%%%%%%%%%%%%%%%%%%%%%%%%%%%%%%%%%%%%%%%%%%%%%%%%%%%%%%%%%%
%%%%%%%%%%%%%%%%%%%%%%%%%%%%%%%%%%%%%%%%%%%%%%%%%%%%%%%%%%%%%%%%
%%%%%%%%%%%%%%%%%%%%%%%%%%%%%%%%%%%%%%%%%%%%%%%%%%%%%%%%%%%%%%%%
%%%%%%%%%%%%%%%%%%%%%%%%%%%%%%%%%%%%%%%%%%%%%%%%%%%%%%%%%%%%%%%%

%%%%%%%%%%%%%%%%%%%%%%%%%%%%%%%%%%%%%%%%%%%%%%%%%%%%%%%%%%%%%%%%
%%%%%%%%%%%%%%%%%%%%%%%%%%%%%%%%%%%%%%%%%%%%%%%%%%%%%%%%%%%%%%%%
%%%%%%%%%%%%%%%%%%%%%%%%%%%%%%%%%%%%%%%%%%%%%%%%%%%%%%%%%%%%%%%%
%%%%%%%%%%%%%%%%%%%%%%%%%%%%%%%%%%%%%%%%%%%%%%%%%%%%%%%%%%%%%%%%

\subsection{Results for the exponential PTT fluid}
\begin{figure}[h!]
\centering
    \begin{tikzpicture}[]
        \node[inner sep=0pt] (BLVC) at (0,0)
        {
            \includegraphics[width=265pt]{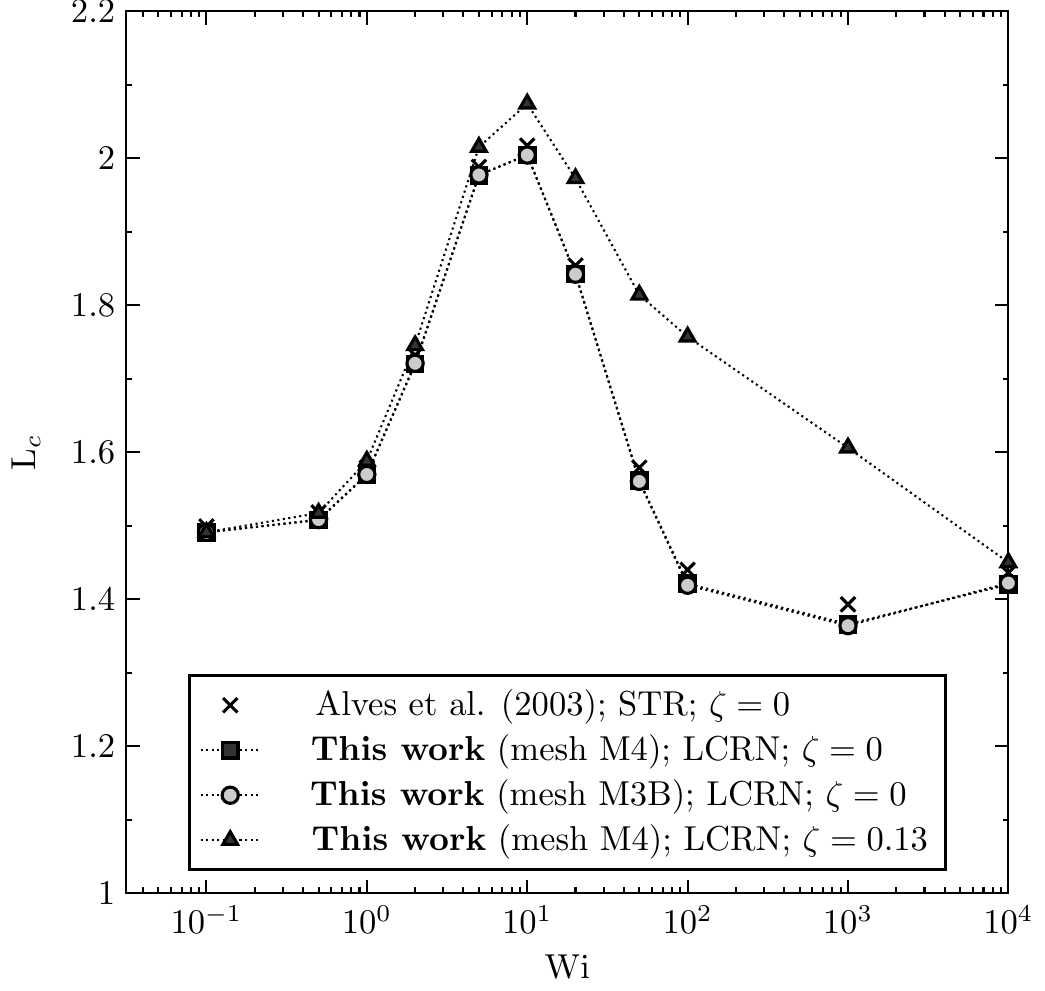}
        };
    \end{tikzpicture}
%%%%%%%%%%%%%%%%%%%%%%%%%%%%%%%%
%%%%%%%%%%%%%%%%%%%%%%%%%%%%%%%%
    \caption{Corner vortex size over $\operatorname{Wi}$ of an exponential PTT fluid in a planar 4:1 contraction. Comparison between published values and this work (mesh M4) up to $\operatorname{Wi} = 10^4$. A zero slip parameter and the stress tensor representation (STR) was used in the reference. The results in this work are shown for two sets of slip parameters: $\zeta = 0$ and $\zeta = 0.13$. They are obtained from the natural logarithm conformation representation (LCRN).}
    \label{fig:vortexSizePTT}
\end{figure}

\begin{figure}[h!]
\centering
    \begin{tikzpicture}[]
        \node[inner sep=0pt] (BLVC) at (0,0)
        {
            \includegraphics[width=265pt]{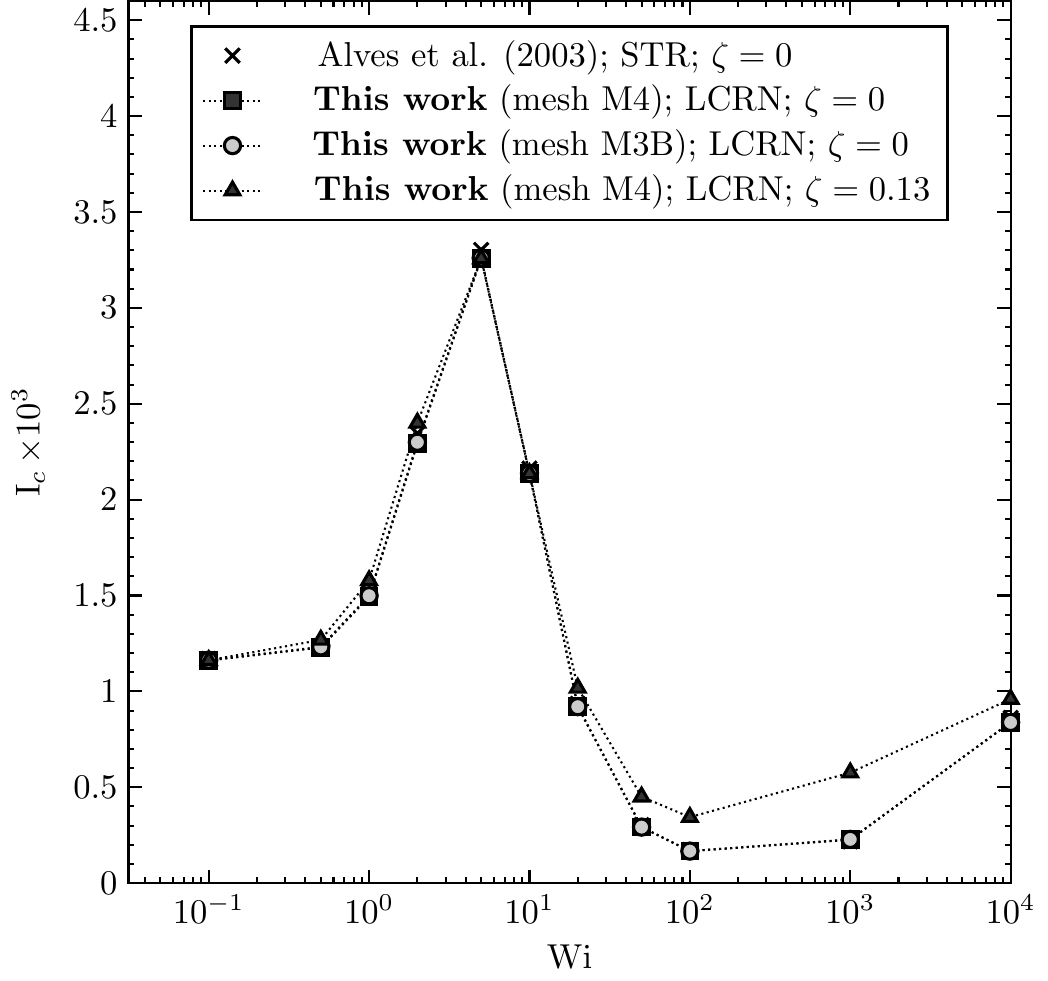}
        };
    \end{tikzpicture}
%%%%%%%%%%%%%%%%%%%%%%%%%%%%%%%%
%%%%%%%%%%%%%%%%%%%%%%%%%%%%%%%%
    \caption{Corner vortex intensity over $\operatorname{Wi}$ of an exponential PTT fluid in a planar 4:1 contraction. Comparison between published values and this work (mesh M4) up to $\operatorname{Wi} = 10^4$. A zero slip parameter and the stress tensor representation (STR) was used in the reference. The results in this work are shown for two sets of slip parameters: $\zeta = 0$ and $\zeta = 0.13$. They are obtained from the natural logarithm conformation representation (LCRN).}
    \label{fig:vortexIntensityPTT}
\end{figure}

The Phan-Thien Tanner (PTT) \cite{PhanThien1978, PhanThien1977} constitutive equation has two additional parameters, $\varepsilon$ and $\zeta$, compared to the Oldroyd-B equation; cf. Table \ref{tab:constitutiveEx}. Both parameters influence the shear-thinning properties of the fluid. We study two different parameter configurations:
\begin{itemize}
\item[1.] $\varepsilon = 0.25$ and $\zeta = 0$:
This set-up is sometimes called the simplified PTT fluid. Setting $\zeta$ to zero means that the motion is assumed to be affine. The shear-thinning is controlled only by $\varepsilon$. The value $\varepsilon = 0.25$ is frequently used in computational benchmarks. Reference data for the planar 4:1 contraction benchmark is available in literature, e.g.\ in \cite{Alves2003} accurate benchmark results were provided for the stress tensor representation (STR) and Weissenberg numbers from $0$ to $10000$. We compare this reference data to the results obtained with a change-of-variable representation in order to validate the numerical framework used in the present work.
\item[2.] $\varepsilon = 0.25$ and $\zeta = 0.13$:
The exponential PTT constitutive equations are often used to model concentrated polymer solutions or polymer melts by fitting the two parameters to some experimental data. This typically results in $\zeta$ being a small positive number. Our purpose is to study the influence of an increased $\zeta$ on the complex flow in the 4:1 contraction benchmark and to verify the validity of the generic numerical framework for a non-zero $\zeta$.
\end{itemize}
The results of both set-ups for the exponential PTT fluid are presented in Figure \ref{fig:vortexSizePTT} and Figure \ref{fig:vortexIntensityPTT}. The corresponding data is given in Table \ref{tab:PTTdata}. Unlike the Oldroyd-B fluid, the exponential PTT fluid does not form any lip vortex within the range of $\operatorname{Wi}$ from $0.1$ to $10000$. We observe one large vortex structure in the corner of the larger channel. Contrary to the Oldroyd-B fluid, the corner vortex of the exponential PTT fluid increases monotonically in size as the Weissenberg number is increased. Between $5 < \operatorname{Wi} < 10$, its spatial extent reaches a maximum. When the Weissenberg number is further increased, the corner vortex size decreases in magnitude and may even shrink below the size in the Newtonian limit at $\operatorname{Wi} = 0$, depending on the particular value of $\zeta$. Such a trend is observed when $\zeta$ is zero, where the extension of the corner vortex has a global minimum at $\operatorname{Wi} \approx 1000$. In \cite{Alves2003} it was argued that at certain $\operatorname{Wi}$ the shear thinnig of the PTT fluid becomes so strong that the apparent viscosity is reduced back to the level of the Newtonian limit, which is manifested by the reduction of the vortex size. For $\zeta = 0.13$, the corner vortex decreases slower but monotonically until the approximate size of the Newtonian limit is reached at $\operatorname{Wi} = 10000$. A similar evolution is observed in the vortex intensity. However, the variation between $\zeta = 0$ and $\zeta = 0.13$ is smaller in magnitude for $\operatorname{I}_c$. Contrary to the vortex size, the vortex intensity decreases below the value of the Newtonian limit in both cases, $\zeta = 0$ and $\zeta = 0.13$. The global minimum in $\operatorname{I}_c$ is reached at smaller Weissenberg numbers as in $\operatorname{L}_c$, i.e.\ at approximately $\operatorname{Wi} = 100$. Both quantities, $\operatorname{L}_c$ and $\operatorname{I}_c$, show a good agreement with the reference \cite{Alves2003}. In contrast to the Oldroyd-B fluid we observe a smaller mesh-sensitivity for the PTT fluids, especially at higher Weissenberg numbers.
%%%%%%%%%%%%%%%%%%%%%%%%%%%%%%%%%%%%%%%%%%%%%%%%%%%%%%%%%%%%%%%%
%%%%%%%%%%%%%%%%%%%%%%%%%%%%%%%%%%%%%%%%%%%%%%%%%%%%%%%%%%%%%%%%
%%%%%%%%%%%%%%%%%%%%%%%%%%%%%%%%%%%%%%%%%%%%%%%%%%%%%%%%%%%%%%%%
%%%%%%%%%%%%%%%%%%%%%%%%%%%%%%%%%%%%%%%%%%%%%%%%%%%%%%%%%%%%%%%%

%%%%%%%%%%%%%%%%%%%%%%%%%%%%%%%%%%%%%%%%%%%%%%%%%%%%%%%%%%%%%%%%
%%%%%%%%%%%%%%%%%%%%%%%%%%%%%%%%%%%%%%%%%%%%%%%%%%%%%%%%%%%%%%%%
\begin{table}[h!]
\begin{center}
\begin{tabular}{@{}lrrcrrc@{}}
\toprule
\quad &
\multicolumn{3}{c}{vortex size $\operatorname{L}_c$} &
%\multicolumn{4}{c}{relative deviation in $\operatorname{L}_c$ $[\%]$} &
\multicolumn{3}{c}{vortex intensity $\operatorname{I}_c$} \\ \cmidrule(l){2-4} \cmidrule(l){5-7}
%%%%%%%%%%%%%%%%%%%%%%%%%%%%%%%%%%%%%%%%%%%%%%%%%%%%%%%%%%%%%%%%
\quad &
\multicolumn{2}{c}{$\zeta = 0$} & \multicolumn{1}{c}{$\zeta = 0.13$} &
%\multicolumn{4}{c}{relative deviation in $\operatorname{L}_c$ $[\%]$} &
\multicolumn{2}{c}{$\zeta = 0$} & \multicolumn{1}{c}{$\zeta = 0.13$} \\ \cmidrule(lr){2-3} \cmidrule(l){4-4} \cmidrule(lr){5-6} \cmidrule(l){7-7}
%%%%%%%%%%%%%%%%%%%%%%%%%%%%%%%%%%%%%%%%%%%%%%%%%%%%%%%%%%%%%%%%
$\operatorname{Wi}$ \quad &
\multicolumn{1}{c}{M4} & \multicolumn{1}{c}{M3B} & \multicolumn{1}{c}{M4} &
\multicolumn{1}{c}{M4} & \multicolumn{1}{c}{M3B} & \multicolumn{1}{c}{M4} \\
\midrule
$0.1$ & $1.491$ & $1.491$ & $1.491$ \ & $1.161$ & $1.162$ & $1.162$ \ \\
$0.5$ & $1.508$ & $1.508$ & $1.517$ \ & $1.229$ & $1.232$ & $1.268$ \ \\
$1$ & $1.569$ & $1.570$ & $1.588$ \ & $1.493$ & $1.499$ & $1.579$ \ \\
$2$ & $1.720$ & $1.721$ & $1.745$ \ & $2.292$ & $2.299$ & $2.400$ \ \\
$5$ & $1.977$ & $1.977$ & $2.015$ \ & $3.258$ & $3.260$ & $3.257$ \ \\
$10$ & $2.005$ & $2.004$ & $2.074$ \ & $2.137$ & $2.137$ & $2.137$ \ \\
$20$ & $1.843$ & $1.842$ & $1.972$ \ & $0.922$ & $0.920$ & $1.018$ \ \\
$50$ & $1.561$ & $1.560$ & $1.814$ \ & $0.293$ & $0.292$ & $0.451$ \ \\
$100$ & $1.421$ & $1.419$ & $1.757$ \ & $0.167$ & $0.167$ & $0.344$ \ \\
$1000$ & $1.366$ & $1.364$ & $1.606$ \ & $0.226$ & $0.228$ & $0.576$ \ \\
$10000$ & $1.420$ & $1.422$ & $1.450$ \ & $0.839$ & $0.838$ & $0.960$ \ \\
\bottomrule
\end{tabular}
\end{center}
\caption{Benchmark results for the exponential PTT fluid, carried out with the natural logarithm conformation representation on the meshes M4 and M3B. The corner vortex size $\operatorname{L}_c$ and the corner vortex intensity $\operatorname{I}_c$ are listed for $\zeta = 0$ and $\zeta = 0.13$.}
\label{tab:PTTdata}
\end{table}
%%%%%%%%%%%%%%%%%%%%%%%%%%%%%%%%%%%%%%%%%%%%%%%%%%%%%%%%%%%%%%%%
%%%%%%%%%%%%%%%%%%%%%%%%%%%%%%%%%%%%%%%%%%%%%%%%%%%%%%%%%%%%%%%%

%%%%%%%%%%%%%%%%%%%%%%%%%%%%%%%%%%%%%%%%%%%%%%%%%%%%%%%%%%%%%%%%
%%%%%%%%%%%%%%%%%%%%%%%%%%%%%%%%%%%%%%%%%%%%%%%%%%%%%%%%%%%%%%%%
%%%%%%%%%%%%%%%%%%%%%%%%%%%%%%%%%%%%%%%%%%%%%%%%%%%%%%%%%%%%%%%%

\section{Summary and conclusions}
\label{sec:conclusions}
A generic FV framework for viscoelastic flow analysis on general unstructured meshes has been proposed, providing capabilities for a rigorous study of the convergence and the accuracy of different constitutive equation representations at high Weissenberg numbers. The validation of the methods and a detailed comparison of the results has been carried out for both Oldroyd-B and PTT fluids in the computational benchmark flow problem of a planar 4:1 contraction. It turns out that the change-of-variable representations are robust at high Weissenberg numbers beyond the critical limit and effectively alleviate the HWNP. However, this does not mean that the results are accurate. As a matter of fact, the methods are subject to severe requirements in order to avoid uncertainties in the predictions.

The accuracy of the results is found to be highly sensitive to the spatial resolution of the computational domain. When the mesh resolution is insufficient, the errors increase monotonically as the Weissenberg number is increased. The dependence of the required mesh-resolution to reach a certain degree of accuracy on the Weissenberg number causes additional limitations for highly elastic flows. Even though being able to carry out stable computations one might not meet the requirements to achieve a mesh-independent solution at high Weissenberg numbers, depending on the computational resources. Moreover, the mesh-sensitivity depends on the particular representation of the constitutive equations. Our results suggest that the results are more accurate when small root functions are used instead of the logarithm conformation representation, in particular on under-resolved meshes. On the other hand, a close consistency between the results from different change-of-variable representations is shown when enough spatial resolution is provided, indicating the convergence towards a unique mesh-independent solution.

The quality of the results is considerably improved by a local refinement of the stress boundary layers at the walls. A local refinement at the walls might also help to save computational costs. Yet, the issues of resolving the region in the direct neighborhood of the geometric singularity remain, regardless of any respective change in the constitutive variables. We do not find any convergence of the stress in a small region around the re-entrant corner. However, it seems that above a finite degree of resolution the local value of the stress at the geometric singularity has only a minor influence on the flow structure. We conclude that the change-of-variable representations are capable to predict accurately the complex flow in the contraction benchmark, provided that the steep stress boundary layers are well-resolved, while the problem of resolving the stress at geometric singularities remains unsolved.

The errors of the benchmark results in planar 4:1 contraction flows of Oldroyd-B fluids have been quantified in Table \ref{table:LCRdata}, demonstrating a high accuracy above the critical Weissenberg number limit. We intend to contribute the data to the computational rheology community for comparison and validation purposes.

\section*{Acknowledgements}
The authors, in particular \mbox{M.\ N.}, \mbox{H.\ M.}\ and \mbox{D.\ B.}, would like to acknowledge \mbox{BASF SE} for the financial support of this work, as well as Dr.\ Erik Wassner and Dr.\ Sebastian Weisse for their kind cooperation and several fruitful discussions. Extensive calculations on the Lichtenberg high performance computer of the Technische Universit{\"a}t Darmstadt were conducted for this research. The authors would like to thank Mr.\ Dipl.-Math.\ Thomas Opfer of HPC-Hessen, funded by the State Ministry of Higher Education, Research and the Arts, for helpful advice.

\clearpage

\bibliographystyle{amsplain}
\bibliography{LIT/bibliography}

\end{document}